\documentclass[fleqn,usenatbib]{mnras} 

\usepackage{newtxtext,newtxmath}
\usepackage{pdflscape}
\usepackage{ mathrsfs }

\usepackage[T1]{fontenc}
\usepackage{ae,aecompl}

\usepackage{graphicx}	
\usepackage{amsmath}	
\usepackage{amssymb}	

\newcommand{\mli}[1]{\mathit{#1}}
\newcommand{\zstar}{Z_*}
\newcommand{\agestar}{\mli{Age}_*}
\newcommand{\mur}{\mu_r}
\newcommand{\mustar}{\mu_*}
\newcommand{\SMA}{\mli{SMA}}
\newcommand{\sigmae}{\sigma_\text{e}}
\newcommand{\Reff}{\text{R}_\text{e}}
\newcommand{\Msun}{\text{M}_\odot}
\newcommand{\kms}{\text{km\,s}^{-1}}
\defcitealias{zibetti+17}{Z17}	
\defcitealias{RC3}{RC3}	

\title[Stellar populations of CALIFA ETGs]{Insights into formation scenarios 
of massive Early-Type galaxies from spatially resolved stellar population analysis in CALIFA}

\author[S. Zibetti et al.]
{Stefano Zibetti$^{1}$\thanks{e-mail: stefano.zibetti@inaf.it},
Anna R. Gallazzi$^{1}$,
Michaela Hirschmann$^{2,3}$,
\and Guido Consolandi$^{4}$,
Jes\'us Falc\'on-Barroso$^{5,6}$,
Glenn van de Ven$^{7}$,
Mariya Lyubenova$^{8}$
\\
$^{1}$INAF-Osservatorio Astrofisico di Arcetri, Largo Enrico Fermi 5, I-50125 Firenze, Italy\\
$^{2}$Institut d'Astrophysique de Paris, CNRS, Universit\'e Pierre \& Marie Curie, 98 bis Boulevard Arago, F-75014 Paris, France\\
$^{3}$DARK, Niels Bohr Institute, University of Copenhagen, Lyngbyvej 2, DK-2100 Copenhagen, Denmark\\
$^{4}$INAF-Osservatorio Astronomico di Brera, Via Brera 28, I-20121 Milano, Italy\\
$^{5}$Instituto de Astrof\'isica de Canarias, V\'ia L\'actea s/n, E-38205 La Laguna, Tenerife, Spain\\
$^{6}$Departamento de Astrof\'isica, Universidad de La Laguna, E-38205 La Laguna, Tenerife, Spain\\
$^{7}$Department of Astrophysics, University of Vienna, T\"urkenschanzstrasse 17, 1180 Vienna, Austria\\
$^{8}$European Southern Observatory, Karl-Schwarzschild-Stra{\ss}e 2, D-85748 Garching bei M\"unchen, Germany}

\date{Accepted 2019 October 30. Received 2019 October 15; in original form 2019 June 5}

\pubyear{2019}

\begin{document}
\label{firstpage}
\pagerange{\pageref{firstpage}--\pageref{lastpage}}
\maketitle

\begin{abstract}
We perform spatially resolved stellar population analysis for a sample of 69 early-type galaxies (ETGs) from the CALIFA
integral field spectroscopic survey, including 48 ellipticals and 21 S0's. We generate and quantitatively characterize
profiles of light-weighted mean
stellar age and metallicity within $\lesssim 2\Reff$, as a function of radius and stellar-mass surface density $\mustar$.
We study in detail the dependence of profiles on galaxies' global properties, including velocity dispersion
$\sigmae$, stellar mass, morphology.
ETGs are universally characterized by strong, negative metallicity gradients ($\sim -0.3\,\text{dex}$ per $\Reff$) within $1\,\Reff$,
which flatten out moving towards larger radii. A quasi-universal local $\mustar$-metallicity relation emerges, which displays
a residual systematic dependence on $\sigmae$, whereby higher $\sigmae$ implies higher metallicity at fixed $\mustar$. Age profiles 
are typically U-shaped, with minimum around $0.4\,\Reff$, asymptotic increase to maximum ages beyond $\sim 1.5\,\Reff$,
and an increase towards the centre. The depth of the minimum and the central increase anti-correlate with $\sigmae$.
A possible qualitative interpretation of these observations is a two-phase scenario. In the first phase, dissipative collapse 
occurs in the inner $1\,\Reff$, establishing a negative metallicity gradient. The competition between the outside-in quenching 
due to feedback-driven winds and some form of inside-out quenching, possibly caused by central AGN feedback or dynamical heating, 
determines the U-shaped age profiles. In the second phase, the accretion of ex-situ stars from quenched and low-metallicity
satellites shapes the flatter stellar population profiles in the outer regions.
\end{abstract}




\section{Introduction}\label{sec:intro}
Massive Early-Type Galaxies (ETGs, hereafter) have been a critical benchmark for models of galaxy formation and evolution
since several decades. The vast majority of these galaxies have been almost completely depleted of cold interstellar medium (ISM) and
lacking substantial star-formation activity for several Gyr. The degree of metal enrichment in their stars, typically at super-solar
level, implies that, relatively to the general galaxy population, massive ETGs have been able to reprocess their ISM more efficiently, 
with fewer losses of metals and in shorter times, as suggested by their relative enhanced ratio of $\alpha$ elements with respect 
to iron.
On the one hand, their global properties appeared similar to the natural outcome of a dissipative collapse regulated by stellar feedback 
\citep[historically dubbed as ``monolithic'' collapse scenario, e.g.][]{Eggen:1962,Larson:1974}  and, 
on the other hand, their scaling relations are not immediately reconcilable with na\"ive expectations from Cold Dark Matter (CDM) 
hierarchical models \citep[e.g.][and reference therein]{Renzini:2006}. 
According to these expectations, massive ETGs, as the most massive galaxies in the present-day Universe, should ``form'' later than
less massive galaxies (including late types), but the analysis of their stellar
content clearly indicates that most of their stars formed early on and little new stars were added in the last $5-8$ Gyr, as opposed to
less massive galaxies, which contain much younger stars or are even still currently forming stars.
This challenge to hierarchical models, often referred to as the ``anti-hierarchical nature of ETGs'', was mostly settled in the second
half of the years 2000's by a number of theoretical works \citep[most notably][]{DeLucia:2006,Neistein:2006}. They pointed
out the fundamental difference between the halo assembly history and the star formation history integrated over all progenitors. They
showed that while the former is obviously hierarchical in CDM models, so that more massive halos are formed later via merging of less
massive ones, the integrated star formation history of the progenitors is actually shifted back in time for more massive halos as 
a natural consequence of their progenitor halos forming earlier in higher density peaks \citep{Neistein:2006}.

At the same time as the hierarchical-monolithic debate was having its acme, semi-analytic models (SAMs) of galaxy formation and evolution
based on CDM N-body cosmological simulations had to face another major problem in the formation of massive ETGs. It was found that no
thermodynamical or stellar feedback mechanism is able to stop gas from being accreted onto massive halos/galaxies from the cosmic 
network, cooling down and forming stars. Radiative and mechanical feedback from the active galactic nuclei (AGN), powered by the 
super-massive black hole lurking in galaxy centres, was then identified as the responsible for the ``quenching'' and the inhibition of 
star formation in massive ETGs \citep[e.g.][]{Croton:2006}.
Later on, it was realized that the development of a massive/dense stellar spheroid can also stabilize the gas and prevent its
transformation into stars, thus giving rise to the so-called morphological quenching mechanism \cite[e.g.][]{Martig:2009},
which represents a possible alternative for AGN feedback, especially for galaxies less massive than 
$M_\text{halo}\sim 10^{12} \text{M}_\odot$.
So far, a detailed description of when, where, and how these mechanisms are at work still defies our theoretical understanding and 
observational tests.

In fact, we still lack a fully consistent theoretical framework that is able to account
for several other crucial (sets of) observations: \emph{(i)} the evolution of the mass-size relation across cosmic times, whereby ETGs at fixed
stellar mass are on average a factor $\sim 4$ more extended now than at redshift $\gtrsim 2$ \citep[e.g.][]{vanderWel:2014}; 
\emph{(ii)} the different properties of slow and fast rotators \citep[e.g.][]{Emsellem:2011}; 
\emph{(iii)} the internal structure and the variation of stellar populations across the spatial extent of the ETGs, which are the focus of this
paper; 
\emph{(iv)} the enhancement of $\alpha$ elements with respect to iron and its relation to total stellar mass or velocity dispersion 
\citep[e.g.][]{Trager:2000b,Gallazzi:2006};
\emph{(v)} the alleged variation of stellar initial mass function (IMF), with bottom-heavier IMF being predominant in more massive ETGs and in the
cores/densest regions \citep[e.g.][]{Conroy_vanDokkum:2012,Ferreras:2013,Martin-Navarro:2015}. 

It is arguable that these phenomena should essentially result from the interplay of the basic mechanisms we highlighted in this
introduction: the dissipative collapse of gas and the stellar feedback; 
the mergers, either as major mergers of similar-mass progenitors or as minor mergers, 
i.e. accretion of satellites; the feedback from the AGN; 
the morphological quenching that follows the formation of a massive and dense stellar spheroid.
When, where, and how these mechanisms take place determines spatial and temporal variations in the physical conditions in which
stars are formed and in the dynamics of the stars that are formed and/or accreted. The archaeological memory of these processes
is retained in the stellar population properties of present-day massive ETGs. Their spatial variations, in particular, can help
unravel the complex interplay of different mechanisms.

From a theoretical perspective, starting with the seminal work by \cite{DeLucia:2006}, a two phase scenario for the formation
of ETGs (and elliptical galaxies in particular) has increasingly gained support both from semi-analytic models and from cosmological
simulations. Quoting from \cite{Oser:2010}, who ran a set
of cosmological simulations and were the first to explicitly propose a ``two-phase scenario'', the formation of ETGs would consist of
``a rapid early phase at $z \gtrsim 2$ during which "in situ" stars 
are formed within the galaxy from infalling cold gas followed by an extended phase since $z \lesssim 3$ during which "ex situ" stars are 
primarily accreted''. As we will show in this paper, spatial variations of stellar population properties can actually test and
prove this scenario.


From the observational point of view, although variations of stellar populations in ETGs are evident already from colour gradients 
\citep[e.g.][]{deVaucouleurs:1961},
this kind of investigation requires spatially resolved spectroscopy at moderate
resolution, in order to track the spatial variation of age- and metallicity-sensitive absorption features across the galaxies.
Early works relied on long slit spectroscopy to trace absorption-feature strength variations along a radial direction 
\citep[e.g.][]{Carollo:1993,Mehlert:2003,Sanchez-Blazquez:2007}, and concluded that the absorption-strength gradients are essentially
due to metallicity and that the age of the populations is generally more homogeneous. The advent of integral field spectroscopy (IFS)
has opened a new era in this field, by empowering truly 2D-mapping capabilities in terms of stellar population properties.
A number of works on the radial variations of stellar population properties in ETGs have been published in the last decade from 
IFS surveys (see Sec. \ref{sec:discussion}), such as: SAURON \citep{deZeew:2002}, ATLAS$^\text{3D}$ \citep{Cappellari:2011}, 
SDSS-IV MaNGA \citep{Bundy:2015}, 
SAMI \citep{Bryant:2015}, and CALIFA \citep{Sanchez:2012aa}. Despite the wealth of measurements and the improved precision
of the available spectro-photometric data sets, a general \emph{ quantitative} consensus on the spatial distribution of the stellar
population properties (age and metallicity, in particular) of ETGs is still lacking. Significant systematic offsets persist among 
different estimates, in different data sets and/or obtained with different approaches, as it will be illustrated and discussed in
Sec. \ref{sec:discussion}. 

As we describe in detail in 
Sec. \ref{sec:SPanalysis}, in this paper we aim at further improving these measurements and reduce systematic uncertainties.   
To this goal, we adopt a bayesian method that takes into account the most robust constraints from both
spectroscopy and broad-band photometry \citep[see also][]{gallazzi+05,zibetti+17}. The inference of the stellar population properties
is then based on a vast suite of models that aims at covering the full possible complexity in terms of star-formation and chemical 
enrichment histories, as well as of dust attenuation, in order to fully account for degeneracies in physical parameter space at given 
observational constraints.

With our spatially resolved analysis we also aim at investigating the role of different scales in shaping the (spatial distribution
of the) stellar population properties of ETGs, in a sort of closer examination of the questions already addressed in our previous
work \citep{zibetti+17}:
\emph{(i)} Is it \emph{local} 
($\sim 1\,\text{kpc}$) scales what determines the local stellar population properties 
\citep[e.g.][]{Cano-Diaz:2016aa,Barrera-Ballesteros:2016aa,Gonzalez-Delgado:2014ab,Gonzalez-Delgado:2016aa}?
\emph{(ii)} Or is it a global parameter, such as mass \citep[e.g.][]{gavazzi_scodeggio96,scodeggio+02,Kauffmann:2003aa},
velocity dispersion \citep[e.g.][]{Bender:1993,Gallazzi:2006}, or overall age, what local properties mostly respond to?

The paper is organized as follows. Sec. \ref{sec:sample} introduces the sample of ETGs and the dataset used for the analysis.
Sec. \ref{sec:SPanalysis} provides full details on the methods and the data processing used to infer 2D maps of the stellar population
ages and metallicities. Sec. \ref{sec:SPprofiles} describes individual profiles of age and metallicity as a function of
radius and of surface brightness/mass density. Methods of extraction and uncertainties are presented and discussed,
as well as general trends.
Sec. \ref{sec:stack_SPprofiles} analyzes the average stellar population profiles for galaxies binned in classes of different global
properties, such as mass, velocity dispersion and E/S0 morphology. The dependence of the stellar population profiles on
global properties is quantified in Sec. \ref{sec:SPchars}. In Sec. \ref{sec:SPgradients} we focus on the descriptions of the profiles
in terms of gradients, as a convenient and popular way of compressing the information about the shape of the profiles. Correlations and
trends with global quantities are also analyzed. In Sec. \ref{sec:discussion} we discuss our results in the context of the vast
literature on the topic and propose a physical interpretation of our findings. Sec. \ref{sec:conclusions} summarizes and concludes this
paper.

\section{The CALIFA-SDSS ETG sample and dataset}\label{sec:sample}
This study is based on a sample of ETGs drawn from the main diameter-selected sample of the Calar Alto Legacy Integral
Field Area (CALIFA) survey \citep{Sanchez:2012aa,Walcher:2014aa} in its $3^\text{rd}$ and final
data release \citep[][DR3]{Sanchez:2016aa}. Apart from celestial coordinate constraints, these galaxies are selected
from the seventh data release of the Sloan Digital Sky Survey \citep{SDSS_DR7} requiring isophotal $r$-band diameter $45\arcsec < isoA_r < 79.2\arcsec$, $r$-band Petrosian magnitude $<20$ and available
redshift $0.005 <z< 0.03$ \citep[see][]{Walcher:2014aa}.

Galaxies are observed in integral-field spectroscopy at the 3.5 m telescope of the Calar Alto observatory with the Potsdam Multi Aperture Spectrograph, PMAS \citep{Roth:2005aa} in the PPAK mode \citep{Verheijen:2004aa,Kelz:2006aa}. The hexagonal field of view of $74\arcsec \times 64\arcsec$ is covered by a bundle of 331 science fibres, in three dithers that provide an effective filling factor close to 100\%.
Out of the 542 observed main sample galaxies, we consider only the 394 galaxies
that have been observed in both the blue ``V1200'' and red ``V500'' setups, combined into the so-called COMBO data-cubes\footnote{As in \citetalias{zibetti+17}, we exclude UGC\,11694 because of a very bright star near the centre, which contaminates a significant portion of the galaxy's optical extent, and UGC\,01123 because of problems in the noise spectra.}. The unvignetted spectral coverage extends from $3700\,$\AA ~to $7140\,$\AA, with a spatial sampling of $1\arcsec/\mathrm{spaxel}$ (effective spatial resolution $\sim 2.57\arcsec$ FWHM). These data-cubes typically reach a signal-to-noise ratio (SNR) of 3 per spectral resolution element and per spaxel at $\sim 23.4~\mathrm{mag~arcsec}^{-2}$ ($r$-band, see figure 14 of \citealt{Sanchez:2016aa}).

From this sample we select  morphologically classified ETGs, i.e. galaxies with morphological type earlier than S0a 
(S0a excluded), not classified as mergers \citep[see][]{Walcher:2014aa}.
After visual inspection we further discard three galaxies that are misclassified later types than E or S0: NGC\,693 
\citep[S0/a in][RC3, with evident nuclear starburst]{RC3}, IC\,3598 (SA(r)ab in \citetalias{RC3}), and UGC\,9629 (Sa 
in \citetalias{RC3}). These leaves us with a sample of 69 ETGs in total, including 48 E's and 21 S0's.

Total stellar masses for each galaxy are taken from \cite{Walcher:2014aa}. We adopt the estimates based on the fitting of the SDSS petrosian magnitudes only, excluding UV or NIR data. We recall here that these estimates are
based on the revised version of the \citet[BC03]{BC03} stellar population synthesis models indicated in the
literature as CB07, assuming a \cite{chabrier03} stellar initial mass function. Note that CB07 underestimates stellar 
masses by $\approx 0.1\,\text{dex}$ with respect to ``standard'' BC03 models \citep[see e.g.][]{ZCR09}.

We have independently analyzed the SDSS images of the sample and performed elliptical isophote fitting, from which 
we have derived average ellipticity $\epsilon$ and position angle PA with the procedure described in \cite{Consolandi:2016aa}. 
With this re-processing we were able to fix a few cases of apparently wrong estimates of PA and ellipticity reported in
the tables of \cite{Walcher:2014aa}.
From integrated photometry in elliptical apertures we 
have further derived total magnitudes and effective semi major axes, defined as the semi major axis (SMA) of the elliptical aperture enclosing half of the total flux and denoted by $R_\text{e}$.

Velocity dispersions, $\sigmae$ are available for 54 galaxies from \cite{Falcon-Barroso:2017aa}, while for the remaining 15 
galaxies measurements are computed in this work by JF-B. The velocity dispersions are
derived from integrated spectra within the $1\,R_\text{e}$ elliptical aperture, obtained in the V1200 setup 
(blue, high-resolution) of CALIFA. Hence these are effectively light-weighted mean velocity dispersions within $R_\text{e}$.

\begin{figure}
	\includegraphics[width=\columnwidth]{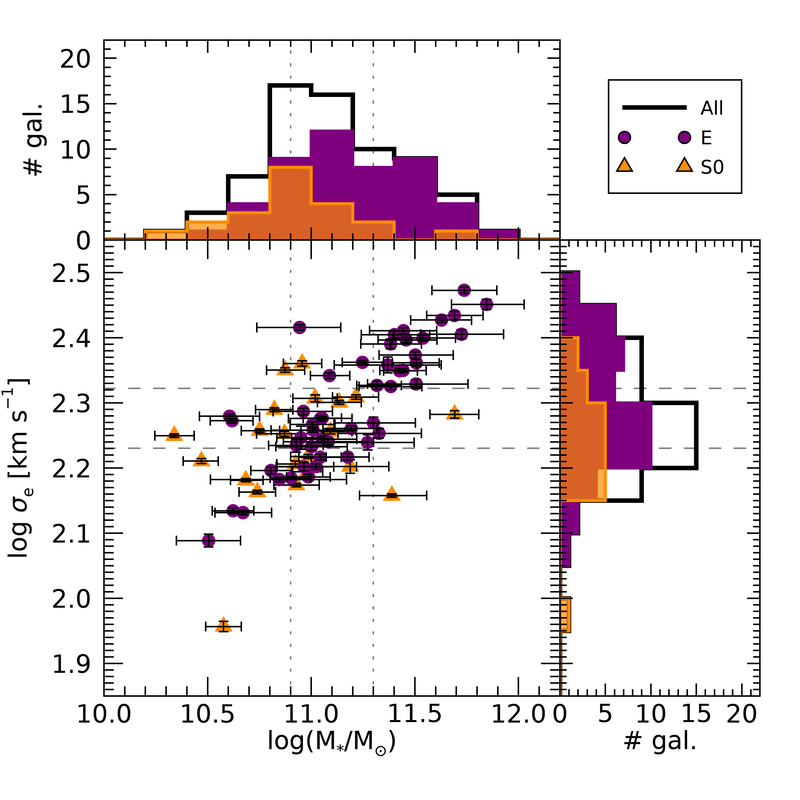}
    \caption{Distribution in $M_*$ and velocity dispersion $\sigmae$ within $1 R_\text{e}$ for the full ETG
    sample, and for the subsamples of Ellipticals and S0, in purple and dark orange, respectively.
    Vertical dotted lines (at  $M_*/\Msun=10^{10.9}=7.9\cdot 10^{10}$ and 
    $M_*/\Msun=10^{11.3}=2\cdot 10^{11}$) and horizontal dashed lines 
    (at $\sigmae/\kms=170 =10^{2.23}$ 
    and $\sigmae/\kms=210 =10^{2.32}$) 
    indicate the boundaries that define the three bins
    in stellar mass and velocity dispersion, respectively, used in the analysis.}
    \label{fig:mstar_sigma}
\end{figure}

Fig. \ref{fig:mstar_sigma} displays the distributions in $M_*$ and $\sigmae$ for the sample as a whole
and for the subsamples of ellipticals (E, in purple) and S0 (in orange). We span a range between $\sim100$ and  $
\sim 300\,\kms$ in $\sigmae$.  In terms of $M_*$ the
range covered spans from $2\cdot 10^{10}$ to $7\cdot 10^{11}\Msun$, thus extending a factor $\sim 3$ beyond the
limit of representativeness of CALIFA (in fact, galaxies with $M_*>2.5\cdot 10^{11}\,\Msun$ are under-represented 
in the main CALIFA sample). 
In Tab. \ref{tab:mstar_sigma_bins} we report the number of galaxies, the median stellar mass $M_*$,
and the median velocity dispersion $\sigmae$ for the full sample and for different subsamples, selected in 
morphology, $M_*$ or $\sigmae$. From both Fig. \ref{fig:mstar_sigma} and Tab. \ref{tab:mstar_sigma_bins}
it is apparent that E's and S0's span different ranges in $M_*$ and $\sigmae$. In particular S0's are biased low
in $M_*$ with respect to E's. Therefore we define a subsample of E's with a limit of $M_* <2\cdot 10^{11}\Msun$ 
(corresponding to the 90th percentile in the $M_*$ distribution of S0's) in order to control for $M_*$ 
when comparing S0's with E's.

While there is an obvious correlation between $M_*$ and $\sigmae$, the scatter is significant, especially for 
S0's. This justifies the distinct analysis of the dependence on the two parameters. In the following sections we 
will consider three bins in both $M_*$ and $\sigmae$, with the boundaries reported in Tab.  
\ref{tab:mstar_sigma_bins} and indicated by dashed lines in Fig. \ref{fig:mstar_sigma}.

\begin{table*}
 \caption{Characterization of the sample and the sub-samples in terms of stellar mass and velocity dispersion.}
 \label{tab:mstar_sigma_bins}
 \begin{tabular}{ccrrr} 
\hline
Sample & Boundaries & N & median $\log M_*$     & median $\sigmae$ \\
             &&    & $\log \Msun$ & $\kms$ \\
\hline
All         && 69 & 11.05 & 185\\
E           && 48 & 11.18 & 189\\
S0         && 21 & 10.93 & 178\\
E (mass-matched w/S0) &  $M_*/\Msun<2\cdot 10^{11}$ &28 &11.00 & 174\\
                  & $\log M_*/\Msun<11.3$ \\
\hline
All, low-$M_*$ & $M_*/\Msun<7.9\cdot 10^{10}$ & 16 & 10.68 & 162\\
                  & $\log M_*/\Msun<10.9$ \\[1ex]
All, mid-$M_*$ & $7.9\cdot 10^{10}\leq M_*/\Msun<2\cdot 10^{11}$ & 31 & 11.02 & 175\\
                  & $10.9\leq\log M_*/\Msun<11.3$ \\[1ex]
All, high-$M_*$ & $M_*/\Msun\geq2\cdot 10^{11}$ & 22 & 11.46& 236\\
                  & $\log M_*/\Msun\geq11.3$ \\
\hline
All, low-$\sigmae$ & $\sigmae/\kms < 170$ & 21 & 10.92 &154\\
All, mid-$\sigmae$ & $170\leq\sigmae/\kms < 210$ & 25 &11.02 & 182\\
All, high-$\sigmae$ & $\sigmae/\kms \geq 210$ & 23 & 11.44 & 236\\
                  
\hline
 \end{tabular}
\end{table*}

\section{Stellar population analysis in 2D}\label{sec:SPanalysis}
\subsection{Method}
We approach the analysis of the stellar populations in our galaxy sample by mapping their 2-dimensional
distribution. We follow the bayesian method already adopted in \citetalias{zibetti+17}, which builds on the original
work by \cite{gallazzi+05}, with a few modifications that will be highlighted in the next paragraphs.
At any given ``pixel'' of a given galaxy we measure a set of observables from the CALIFA IFS and the SDSS
imaging. The same observables are measured on an extensive suite of spectral models, each of them having a set
of associated physical parameters (e.g. light-weighted age, metallicity etc.). The likelihood of each set of real 
observables (with associated errors) for a given model $i$, is assumed to be
\begin{equation}
\mathcal{L}_i\propto\exp(-\chi_i^2/2),
\end{equation}
with the standard definition of $\chi^2$.
The (posterior) probability distribution function (PDF) of a physical parameter associated to the models is
derived by weighing the prior distribution of models in that parameter by their likelihood $\mathcal{L}_i$,
following Bayes' theorem.

In this paper we focus on two light-weighted mean properties of the stellar populations, specifically 
the $r$-band-light-weighted mean age $\agestar$ and metallicity $\zstar$ (see \citetalias{zibetti+17} sec. 2.3 and equations 4 and 6 
therein). Mean quantities are computed from the linear parameters\footnote{This is especially relevant to properly 
compare the present results with works in literature where log quantities are averaged. See also discussion in Appendix B (available online)
.}, i.e. age in Gyr and $Z$ as 
metal abundance ratio normalized to the solar value of $0.02$. We also derive the stellar mass surface density
based on the PDF of the scaling factor that one must apply to a 1-$\text{M}_\odot$ model spectrum in order to
match the SDSS photometry. 

The spectral library adopted in this study is the same as the one used in \citetalias{zibetti+17}, with the only exception
of a different prior on the dust attenuation parameters. 
It includes 500\,000 models generated from random star-formation 
histories (SFH), metal-enrichment histories, and effective dust attenuation. 
The base spectral library of simple stellar populations (SSPs) is the \citet{BC03} in the 2016 revised version (CB16),
which adopts the \citet{chabrier03} initial mass function, an updated treatment of evolved stars \citep{Marigo:2013},
and the MILES stellar spectral  library \citep{Sanchez-Blazquez:2006aa, Falcon-Barroso:2011aa}.
SFHs \'a la \cite{Sandage:1986aa}
are adopted for the continuous component: $\mli{SFR}_\tau(t)\propto\frac{t}{\tau}\exp\left(-\frac{t^2}{2\tau^2}\right)$.
A random burst component is also added on the top of it. Up to 6 burst can be added, with an intensity (i.e. fraction
of stars formed relative to the total formed in the continuum component) ranging between $10^{-3}$ and $2$. For
bursts with age $\mli{age}_\text{burst}<10^8\text{yr}$, the maximum fraction of stars formed is gradually decreased from $2$
to $10^{-2.5}$ at $\mli{age}_\text{burst}=10^5\text{yr}$, in order to avoid recent bursts that totally overshine the rest of the 
SFH.

A simple chemical enrichment history is also implemented. The metallicity of the stars formed at time $t$
increases from an initial 
value $Z_{*\,0}$ (randomly generated between 0.02 and 0.05 $\text{Z}_\odot$) to a final value
$Z_{*\,\text{final}}$ (also randomly generated between $Z_{*0}$ and $2.5~\text{Z}_\odot$) as a function of the time-integrated mass fraction, according to the law:
\begin{equation}
Z_*(t)=Z_*\left(M(t)\right)=Z_{*\,\text{final}}-\left(Z_{*\,\text{final}}-Z_{*\,\text{0}}\right)\left(1-\frac{M(t)}{M_\text{final}}\right)^\alpha,  \alpha > 0
\end{equation}
$\alpha$ is a random shape parameter that describes how quickly the enrichment occurs, from 
instantaneously ($\alpha \gg 1$) to delayed ($\alpha < 1$).

We mimic the stochasticity of the bursts by assigning each burst a metallicity $Z_{*\text{burst}}$ equal to the 
metallicity of stars formed in the continuous mode at the time of the burst, $Z_*(t=t_\text{obs}-\mli{age}_\text{burst})$, 
plus a random offset taken from a log-normal distribution with $\sigma=0.2\,\text{dex}$.

Dust attenuation is implemented following \cite{charlot_fall00}, who assume two components of dust: a diffuse ISM
with an effective attenuation law that goes as the wavelength $\lambda^{-0.7}$, and the dust in the birth cloud (BC), 
which embeds young  stars ($\text{age} \le 10^7~\text{yr}$) only, with an effective attenuation law that goes as $
\lambda^{-1.3}$.
Young stars, therefore, suffer attenuation from both components, yielding a total optical depth in $V$-band of 
$\tau_V$, with a fraction $\mu$ attributed to the diffuse ISM, and a fraction $1-\mu$ attributed to the BC. Older stars
are effectively attenuated only by the diffuse ISM, hence with a V-band optical depth of $\mu \tau_V$. The two free
parameters, $\tau_V$ and $\mu$ are randomly generated with probability distributions, which are flat at low
values and drop exponentially to $0$ between $\tau_V=4$ and $6$, and  between $\mu=0.5$ and $1$, respectively, 
as in \cite{dacunha+08}.

For this study we have allowed for a much larger fraction of dust-free models, i.e.  $90\%$, than the one adopted in 
\citetalias{zibetti+17}, $25\%$. This choice is justified by the restricted sample of ETGs analyzed here.
ETGs are known to have a much lower dust content than spirals. For instance, in the Herschel Reference Sample 
(HRS), \cite{Smith:2012aa} show that the ratio of dust over stellar mass is lower by a factor $50$ in ETGs with 
respect to spirals, on average, and the detection rate of ellipticals at $250\,\mu\text{m}$ is only $24\%$. With this 
prior we are able to provide tighter constraints (and lower residual biases) wherever dust is not required, by limiting 
the impact of dust on the dust-age-metallicity degeneracy. On the other hand, despite the small fraction of dusty 
models, we are able to correctly identify dust lanes and avoid significant biases in the (few) dusty regions. Although 
visual inspection has shown us that the extent of such regions is reduced with respect to what we get with the 
\citetalias{zibetti+17} prior, the number of pixels affected is small enough to produce negligible effects on the 
azimuthally averaged profiles of age and metallicity (see below).

It is important to stress that we do not aim at retrieving or fitting the full complexity of these parameters in the real
galaxies. Rather we want to include the maximum possible degree of complexity in our models, so to properly
take into account the parameter degeneracies on the estimates and uncertainties of the key physical quantities in 
which we are interested, namely \emph{the light-weighted mean age and metallicity of the stellar populations and the
stellar mass surface density}.

The key observables from which we derive the likelihood $\mathcal{L}_i$ are four stellar absorption indices and the
photometric fluxes in the five SDSS bands, $ugriz$. As absorption indices we use the Balmer indices, 
$\mathrm{H\beta}$ and $\mathrm{H\delta_A}+\mathrm{H\gamma_A}$, mainly age-sensitive, and two (mostly) metal-sensitive 
composite indices that show minimal dependence on $\alpha$-element abundance relative to iron-peak 
elements ($[\mathrm{Mg_2Fe}]$ and $[\mathrm{MgFe}]^\prime$). As opposed to previous works and to 
\citetalias{zibetti+17} in particular, we do not employ the $\mathrm{D4000_n}$ break, despite of its well proved 
sensitivity to age. The reason for this choice resides in the limited sensitivity and, most important, sky-subtraction 
accuracy, of the CALIFA dataset blue-ward of $4000\,$\AA. At fixed limiting surface brightness in $r$-band (or fixed 
limiting stellar mass surface density), ETGs display the lowest levels of surface brightness blue-ward of $4000\,$\AA~ 
with respect to the general galaxy population, due to their red spectral energy distribution and extreme $
\mathrm{D4000_n}$ break strength. Therefore even small residual pedestals from the sky subtraction can 
significantly affect the measurement of $\mathrm{D4000_n}$ in the outskirts of these galaxies, leading to biases that 
depend on surface brightness (radius). Since the main goal of this paper is to derive reliable and consistent stellar 
population profiles, we rather not use $\mathrm{D4000_n}$. It must be noted that the $u-g$ color is partly redundant 
with $\mathrm{D4000_n}$, so the information encoded in the break is only minimally lost.
For galaxies that do not display any apparent problem in the $\mathrm{D4000_n}$ map, we find that
age and metallicity maps that are obtained with and without $\mathrm{D4000_n}$ are very consistent with each 
other, with slightly larger uncertainties when $\mathrm{D4000_n}$ is excluded. On the other hand, when problems in 
the $\mathrm{D4000_n}$ map are apparent, differences in the physical parameters are seen, and associated 
uncertainties are larger when $\mathrm{D4000_n}$ is included.

Simulations of typical CALIFA-SDSS observations show that systematic biases at the level of a few 0.01 dex may be present
in both age and metallicity estimates, for the range of physical parameter relevant to ETGs. The largest biases are
expected for $\log Z_*/\text{Z}_\odot \gtrsim 0.3$: since in the models there is a hard boundary for $Z_*$ at 
$\log Z_*/\text{Z}_\odot=0.4$, the PDFs are skewed towards lower values and a bias is generated. As a consequence
of the age-metallicity degeneracy, an opposite bias is induced in the age estimates. Therefore, at the highest
stellar metallicity we expect to have underestimated metallicity by a few 0.01 dex up to 0.05 dex and, correspondingly,
overestimated ages by a few 0.01 dex up to 0.1 dex. A similar, although smaller, ``boundary'' effect is observed at the largest 
ages ($\agestar \gtrsim 8$~Gyr) for $\log Z_*/\text{Z}_\odot\lesssim 0.1$. In this regime, ages are underestimated by up 
to 0.05 dex with a corresponding overestimate of $Z_*$ by up to $0.05-0.07$dex. 
We will discuss the implication of these residual biases on the stellar population profiles in Sec. \ref{sec:profs_bias}.

\subsection{CALIFA-SDSS data processing}\label{sec:2Dproc}
The first step to study the dependence of stellar population properties on radial galactocentric distance and on
surface brightness/stellar-mass density, is to create 2D maps of age and metallicity as well as of stellar mass
surface density, $\mustar$. In order to achieve this,
we create maps of surface brightness in the 5 SDSS bands, and of index strength for the set defined in the previous 
section. These maps are matched in terms of sampling and effective resolution, following the procedure detailed in 
\citetalias{zibetti+17}. More specifically, we degrade the native resolution of the SDSS images to match the spatial
resolution of the CALIFA data-cubes (PSF $\text{FWHM}\sim2.57\arcsec$). Given the redshift distribution of
our sample, this angular resolution translates into a typical physical resolution of $\lesssim 1~\text{kpc}$. In terms of
effective radius, we typically resolve $\lesssim 0.1 R_\text{e}$.
More quantitatively, we consider as PSF radius the half width at half maximum (HWHM, i.e. 0.5 FWHM) of the PSF. 
The median ratio of PSF radius to $\Reff$ is 0.08; 70\% of the sample have this ratio $<0.1$, and the remaining 30\% between
0.1 and 0.2 (see the full distribution in Fig. \ref{fig:PSFdistr} of Appendix \ref{app:PSF}). Hence we conclude that we have 
sufficient spatial 
resolution to resolve the stellar population trends down to at least $0.2\,\Reff$ for the full sample, and down to $0.1\,\Reff$
for a representative majority of galaxies.

The stellar population analysis requires moderately high signal-to-noise ratio (SNR) in order to keep
uncertainties below $0.2-0.3\,\text{dex}$: a typical SNR of $\sim 15$ per \AA~($20$~per spectral pixel) is sufficient to this goal 
for CALIFA COMBO spectra, as we verified both on simulations and on real data. Since the SNR actually delivered 
by CALIFA is typically lower than that for most galaxies at galactocentric distances beyond $1 R_\text{e}$, we apply 
a spatially-adaptive smoothing of the cubes, following the approach of \texttt{adaptsmooth} 
\citep{ZCR09,adaptsmooth}. As in \citetalias{zibetti+17} we choose a target SNR of 20 and a maximum kernel radius
of 5\arcsec\footnote{In practice, smoothing is only applied at $\SMA\gtrsim 1\,\Reff$, with a kernel radius that increases
radially following the declining surface brightness. The spatial resolution is therefore not affected in the inner regions, but only
in the outer regions where gradients are already intrinsically milder.}.
We further restrict the analysis to spaxels with $r$-band surface brightness 
$\mu_r \leq 22.5~\mathrm{mag~arcsec}^{-2}$, as determined on the matched SDSS images, in order to define a
highly complete set of regions (completeness $>90\%$) over a well defined range in surface brightness 
\citepalias[see][]{zibetti+17}.

The next step in the processing is to derive the kinematic parameters (line-of-sight velocity $v$ and velocity 
dispersion $\sigma$) at every spaxel and decouple possible nebular emission lines from the underlying stellar
continuum. This is performed using an iterative procedure based on \texttt{pPXF}
\citep{Cappellari_pPXF} and \texttt{GANDALF} \citep{Sarzi:2006aa}.
We subtract the best-fit emission lines that are detected with an amplitude-over-noise ratio
larger than 2, from the original spectrum. Spectral absorption indices are measured on this ``clean'' spectrum in the
precise rest-frame defined by $v$, without applying any correction for $\sigma$. The effect of $\sigma$-broadening
on the indices is taken into account by directly modelling it in the models. In fact, in order to compute the $\chi^2$ of
each model, the observed indices are compared to model indices measured on model spectra that have previously 
been convolved to match the effective resolution and $\sigma$ in the observations \citep[see][]{gallazzi+05}.

The broad-band SDSS photometric fluxes are cleaned by the emission line contributions determined with 
\texttt{GANDALF}. In the $\chi^2$ computation, these fluxes are compared with the synthetic fluxes extracted from
the model spectra using properly shifted filter response functions that match the redshift and Doppler $v$-shift of
each spaxel.

From the posterior PDFs derived as described in the previous sections, we obtain maps of median-likelihood
stellar mass surface density $\mustar$, $r$-band-light-weighted age $\agestar$ and metallicity $\zstar$.
At each spaxel, the fiducial value of the quantity is taken as the median of the PDF, while the uncertainty is given
by half of the $16^\text{th}-84^\text{th}$ percentile range (corresponding to $\pm1\,\sigma$ in gaussian approximation).
It must be noted that this uncertainty includes both measurement errors as well as the intrinsic uncertainty due to the 
degenerate effect of different SFHs and chemical enrichment histories on the observable
quantities. For this reason, uncertainties on the estimates of light-weighted age and metallicity in individual spaxels
can hardly drop below $0.1\,\text{dex}$, no matter how much we shrink the error bars on the observable quantities.

\section{Stellar population profiles}\label{sec:SPprofiles}
\begin{figure*}
\includegraphics[width=0.8\textwidth]{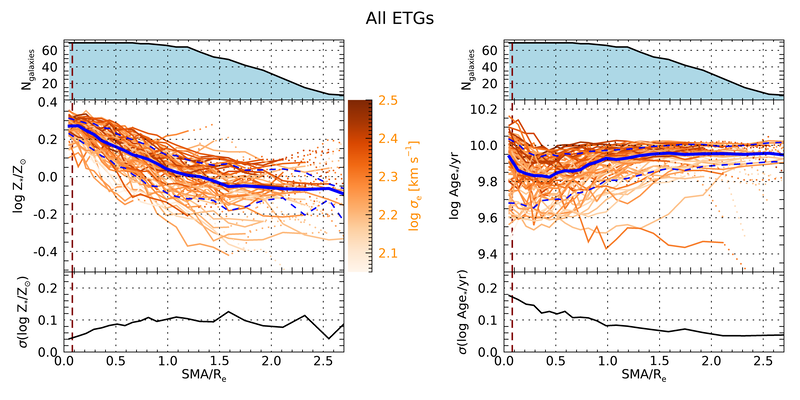}
\includegraphics[width=0.8\textwidth]{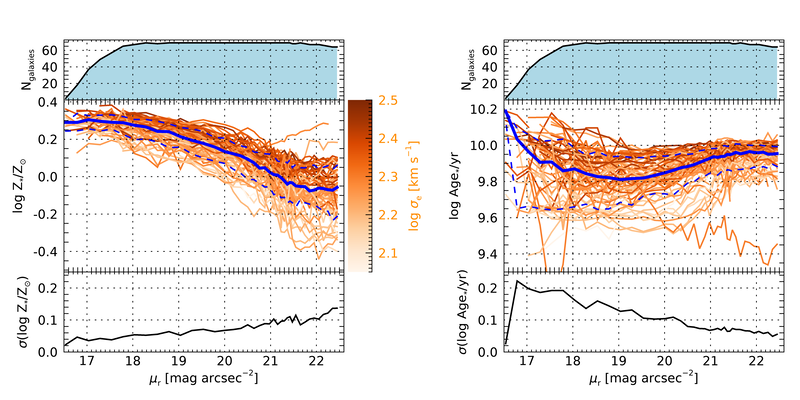}
\includegraphics[width=0.8\textwidth]{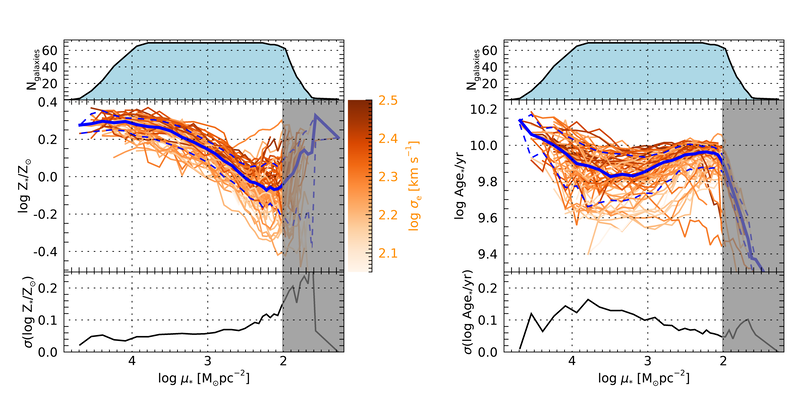}\caption{Profiles of 
light-weighted stellar metallicity $\zstar$ (\emph{left column}) and stellar age $\agestar$ 
(\emph{right column}) as a function of different quantities for the full sample of 69 ETGs. In the \emph{top row} quantities
are plotted as a function of the elliptical semi-major axis ($\mli{SMA}$) normalized by the half-light semi-major axis 
$\Reff$, as a function of the $r$-band surface brightness $\mur$ in the \emph{mid row}, and as a function of the
stellar mass surface density $\mustar$ in the \emph{bottom row}. The main panel of each plot displays the profile of each 
individual galaxy, color-coded according to the velocity dispersion $\sigmae$ within $1\,\Reff$ (see side colorbar).
In the top row, dotted lines indicate completeness $<0.67$, the vertical dashed lines mark the median PSF radius (HWHM). 
The blue lines represent the sample median (solid line) and the
$16^\mathrm{th}$ and $84^\mathrm{th}$ percentiles (dashed lines), only profiles with completeness $>0.67$ contribute. In the 
top and bottom panels of each plot we report the number of contributing galaxies and the scatter of the quantity on $y$-axis 
around the median, respectively.}
\label{fig:stacked_profs}
\end{figure*}

In the following subsections we describe three different kinds of profiles for stellar population parameters, 
which are shown in Fig.~\ref{fig:stacked_profs}, with $\zstar$ in the left column and $\agestar$ in the right one:
azimuthally-averaged elliptical radial profiles (top row), profiles as a function of $r$-band surface brightness 
$\mur$ (mid row), and profiles as a function of stellar-mass surface density $\mustar$ (bottom row). 
Each orange line corresponds to a galaxy, with its hue, ranging from light to dark orange, displaying the 
light-weighted average velocity dispersion within $1\,\Reff$, $\sigmae$. The blue solid line represents the
median of all galaxies at any given abscissa bin, the dashed blue lines are the corresponding 16th and 84th percentiles. Half of this percentile range is plotted in the bottom panels and
represents the scatter of the sample. The top panel of each plot displays the number of galaxies contributing with 
their profile at any given abscissa bin.

From the analysis of the PDF, typical uncertainties on age and metallicity in individual spaxels are both approximately 
$0.15\,\text{dex}$, including random measurement errors and systematic contributions inherent to the modelling.
An independent measurement of the uncertainty is provided by the scatter in the estimates for individual spaxels inside the bins 
used to create the profiles. For age determination the scatter is distributed with a median of $0.08\,\text{dex}$, between 
$0.04\,\text{dex}$ and $0.12\,\text{dex}$. For $\zstar$ determinations the median scatter is $0.1\,\text{dex}$ and varies between 
$0.05\,\text{dex}$ and $0.15\,\text{dex}$ approximately. In both cases, the scatter is less than the estimated error in individual
spaxels. This can be understood as a consequence of spaxel correlations and, most important, of systematic uncertainties being included 
in the error estimate based on our Bayesian analysis. If we make a rough evaluation (neglecting spaxel covariance) of the random
uncertainty in each bin as the rms around the median divided by the square root of the number of spaxels, we end up with estimates 
of the order of a few $0.01\,\text{dex}$ at most, thus well below our systematic uncertainties.

\subsection{Azimuthally-averaged elliptical profiles}\label{sec:elliptical_profs} 
Azimuthally-averaged elliptical profiles are obtained by binning the maps according to the semi major axis 
($\mli{SMA}$) of
elliptical annuli centred on the galaxy's nucleus, with ellipticity $\epsilon$ and position angle PA as determined
in Sec. \ref{sec:sample}. In each annulus we consider the median value of the stellar population parameter 
($\zstar$ and $\agestar$, respectively). This is plotted against the average (midpoint) value of $\mli{SMA}$ 
normalized to the $\Reff$. Because
of the limited field of view or of masked spaxels (due, e.g., to foreground stars or artefacts), only a portion of
spaxels may be available in a given elliptical annulus. If the representativeness drops below $2/3$, the profile
is drawn with a dotted line and those radial bins are not considered for the computation of the median and
percentiles of the sample (blue lines). Within $1\,\Reff$ we are highly complete, with $>65/69=94\%$ of galaxies contributing
in this range. The completeness drops to $51/69=74\%$ at $1.5\,\Reff$ and then to $32/69=46\%$ at $2\,\Reff$.

The stellar metallicity $\zstar$ monotonically decreases as a function of $\SMA$ in a very consistent way for all
galaxies (top left panel of Fig. \ref{fig:stacked_profs}).
The gradient is steeper within $1\,\Reff$, then the profiles flatten out beyond that radius. $\zstar$ decreases
by $\sim 0.3\,\text{dex}$ (roughly a factor $2$) going from the nucleus to $1\,\Reff$. The scatter of the sample around
the median is remarkably small, typically $0.1\,\text{dex}$ ($\sim 25\%$,) and decreases to $\sim 0.05$ ($\sim 12\%$)
from $0.5\,\Reff$
towards the centre. Note that such a scatter is smaller than expected from the systematic uncertainties in our simulations, 
which further indicates a strong regularity (universality) in the metallicity profiles of ETGs.
We also note a systematic tendency for the profiles of higher-$\sigmae$ galaxies (darker orange hue) to lay above 
those of lower-$\sigmae$ galaxies (lighter orange hue). We will quantify this effect better in 
Sec. \ref{sec:stack_SPprofiles}.

In terms of light-weighted age $\agestar$, profiles are overall flat (top right panel of 
Fig. \ref{fig:stacked_profs}). The median profile spans a range of $\sim 0.15\,\text{dex}$ only, between $6.8$ and 
$8.9\,\text{Gyr}$. Remarkably, the median age profile of the sample is \emph{not} monotonic, rather U-shaped. All galaxies display the
largest ages beyond $1\,\Reff$. This maximum age of $\sim 8.9\,\text{Gyr}$ is roughly constant for all galaxies, with a
sample r.m.s. of $\lesssim 0.07\,\text{dex}$.
Age decreases from $1\,\Reff$ inward to $0.5\,\Reff$. Below $0.5\,\Reff$ age profiles display a 
larger degree of diversity, as witnessed by the scatter, which increases to $\sim0.1\,\text{dex}$ (up to $0.2\,\text{dex}$ 
in the centre).
On average, moving to the centre, galaxies get as old as in the outskirts, although this trend is highly variable on a 
galaxy-to-galaxy basis and correlates with global quantities such as $\sigmae$, as we will show in Sec. \ref{sec:stack_SPprofiles}. A dependence of the age profiles on $\sigmae$ is already visible by looking at the dominant
hue of the lines, indicating that galaxies with higher velocity dispersion tend to have overall larger ages and typically
flatter profiles than galaxies with lower velocity dispersion.

The U-shape of the age profile is indeed a common feature to the majority of galaxies. In fact, from visual inspection of the 
individual profiles, we find: 28 galaxies that are fully consistent with the U-shape having a minimum at $0.4\,\Reff$; 
11 galaxies with U-shape but minimum inside $0.3\,\Reff$; 6  galaxies with U-shape but minimum outside $0.4\,\Reff$; 
3 galaxies with an extended plateau around the minimum; 3 galaxies with a noisy profile that is consistent with the 
median U-shape; the remaining 18 galaxies not showing any evidence for U-shape or inconsistent with that. In summary, 
51 out of 69 galaxies display U-shaped age profiles, with some variations in the position of the minimum.

\subsection{Profiles in surface brightness and stellar-mass surface density}\label{sec:sb_mu_profs}
Profiles in $r$-band surface brightness (SB, $\mur$) and stellar-mass surface density ($\mustar$) are obtained by binning
the spaxels in $\mur$ and $\mustar$, respectively. In Fig. \ref{fig:stacked_profs}, the median value of the stellar population
property inside the bin is plotted against the median $\mur$ (middle row) and $\mustar$ (bottom row), with metallicity in 
the left column and age on the right column.

As a function of $\mur$ (mid row), essentially all galaxies are represented for 
$\mur>18\,\text{mag arcsec}^{-2}$ down to the selection limit of $22\,\text{mag arcsec}^{-2}$. A decreasing number of
galaxies reach $\mur$ as bright as $\sim 16.5-17 \,\text{mag arcsec}^{-2}$, as a consequence of 
the different shapes of the surface brightness profiles of the ETGs in our sample. 

As a function of $\mustar$ (bottom row), we note that the cut-off at low stellar-mass
surface density is less abrupt than at low SB, due to errors in $M/L$.
Since we a apply a sharp selection cut at $\mur=22\,\text{mag arcsec}^{-2}$, which corresponds on average to
$\log\mustar\sim 2$, the tail of the distribution below this value is contributed (mainly) by spaxels whose
$M/L$ is under-estimated due to errors. Hence spaxels with $\log\mustar\lesssim 2$ are characterized by biased
estimates of stellar population properties. In particular, since errors in $M/L$ are correlated with errors in $\agestar$ 
and, in turn, errors in $\zstar$ are anti-correlated with errors in $\agestar$, points below the limiting $\mustar$ of $10^2\mathrm{M_\odot\,pc^{-2}}$ are 
severely biased also in $\agestar$ (down-turning profiles) and $\zstar$ (up-turning profiles). For this reason, that 
entire region must be neglected and is shaded in grey in Fig. \ref{fig:stacked_profs}.

Profiles of $\zstar$ stay almost flat in the highest 
SB/density regions and then decrease with steeper and steeper derivative as we move to lower SB/density.
The scatter is around or slightly above $0.05\,\text{dex}$ in the (inner) higher-SB/density regions, over almost 2 orders
of magnitudes in SB/density, and increases to $0.1-0.15\,\text{dex}$ only in the (outer) low-SB/density regions. As already
noted for the radial metallicity profiles, the scatter (especially in the inner, higher-SB/density regions)
is tiny compared to possible systematic uncertainties and points
to a high degree of universality in the dependence of $\zstar$ on radius and on $\mur$ or $\mustar$. Looking at the profiles
of the individual galaxies, it is apparent that there is a significant dependence of the $\zstar$ profiles on the velocity dispersion
$\sigmae$, which is much more evident than in the case of radial elliptical profiles. There is in fact an average shift
of the profiles of galaxies with higher $\sigmae$ towards larger $\zstar$, over a range that is comparable with
the scatter around the median profile.

Stellar light-weighted age profiles display a U-shape, even more evident than what is seen in radial profiles,
with a minimum corresponding to $\sim 6.8\,\text{Gyr}$ at $\mur\sim 19\,\text{mag arcsec}^{-2}$ or 
$\log\mustar/\text{M}_\odot \text{pc}^2\sim 3.5$. The scatter is typically larger than the one displayed in $\zstar$, 
and decreases from $\sim 0.15\,\text{dex}$ in the brightest regions to $\lesssim 0.1\,\text{dex}$ when we move to regions 
fainter than $20\,\text{mag arcsec}^{-2}$ 
or $\log\mustar/\text{M}_\odot \text{pc}^2 < 3$. Overall this is consistent with the radial profiles,
although we note that by binning in SB/density, the brightest/densest of the central regions reach older ages than fainter
or less dense ones, even older than the outer regions, and exceed $10\,\text{Gyr}$. We note also a trend for profiles
of higher-$\sigmae$ galaxies to display overall larger ages and a less deep minimum (i.e. flatter shape).

Apart from the above-mentioned difference at the dim end due to measurement effects, profiles as a function of $\mur$ 
and of $\mustar$ mirror each other very closely. This is not surprising if one considers that $M/L$ ratios span a small 
dynamical range for old and metal-rich stellar population like those in ETGs \citep[e.g.][their figure 1]{BC03}, and 
therefore $\mur$ and $\mustar$ trace each other very well. For the rest of the paper we will no longer discuss profiles
in $\mur$ and refer instead to profiles in $\mustar$, the latter being a more fundamental physical quantity.

\subsection{Relating azimuthally averaged elliptical profiles and profiles in stellar-mass surface density}\label{subsec:sma_vs_mustar}
\begin{figure}
\includegraphics[width=\columnwidth]{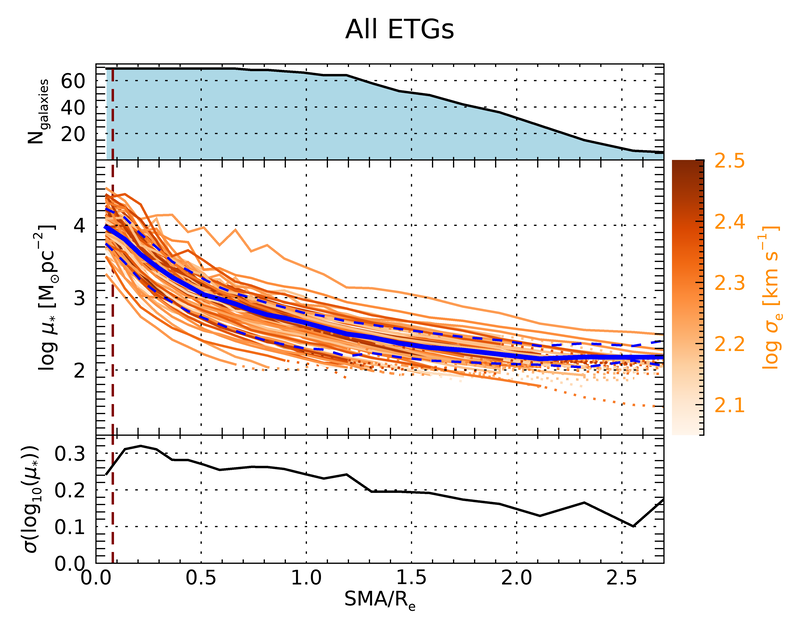}\caption{Profiles of stellar mass surface
density $\mustar$ as a function of the elliptical semi-major axis ($\mli{SMA}$) normalized by the half-light semi-major axis 
$\Reff$, for the full sample of 69 ETGs. The main panel displays the profile of each individual galaxy, color-coded according 
to the velocity dispersion $\sigmae$ within $1\,\Reff$ (see side colorbar). The blue lines represent the sample median (solid 
line) and the $16^\mathrm{th}$ and $84^\mathrm{th}$ percentiles (dashed lines). In the top and bottom panels we report the 
number of contributing galaxies and the scatter of $\log\mustar$ around the median, respectively. 
The vertical dashed line marks the median PSF radius (FWHM).
For a reference, at 
$\mli{SMA}=\Reff$ the median $\mustar(\Reff)$ is $10^{2.65} \Msun \text{pc}^{-2}$, and a $\mustar$ of $10^{3}\Msun\text{pc}^{-2}$ corresponds to 
$0.57~\Reff$.}\label{fig:stacked_mustarprof}
\end{figure}

Radial elliptical profiles and profiles in $\mustar$ are obviously related one to the other via radial elliptical profiles
of stellar surface mass density. We plot them in Fig. \ref{fig:stacked_mustarprof} with the same graphic format as for
Fig. \ref{fig:stacked_profs}.

All profiles but a few display similar shapes, i.e. the typical cuspy \cite{deV} profiles. As a result, at first order
approximation, radial elliptical profiles translate into profiles in $\mustar$ that are more stretched in the inner,
brighter parts, and more compressed in the outer, faint parts. This mere ``coordinate'' transformation explains the
basic difference in shape between these two kinds of profiles in Fig. \ref{fig:stacked_profs}.

The normalization of the profiles, on the other hand, exhibits a significant scatter of 
$\sim 0.25-0.3\,\text{dex}$ inside $1\,\Reff$. Because of the cut we apply in surface brightness, we note that we miss
an increasing number of galaxies as we move beyond $1\,\Reff$ and the sample becomes more and more biased towards
the galaxies of higher average surface brightness/mass density. The tight $\mustar-\zstar$ relation presented in the bottom
left plot of Fig. \ref{fig:stacked_profs}, which is unaffected by selection biases, hence implies that the parts of the
radial metallicity profiles missing at $\mli{SMA}>1\,\Reff$ are preferentially low-metallicity. In turn, this may \emph{(i)}
bias the median radial metallicity profile of the sample beyond $1\,\Reff$ to appear flatter than it is in reality and
\emph{(ii)} artificially decrease the scatter. On the other hand, the \emph{individual} profiles that extend far enough 
display a similar flattening as the median profile, hence reassuring about its real nature. 

Concerning the age profiles, we note that beyond $1\,\Reff$ $\log\mustar$ gets smaller than $3$, a regime where we observe a mild
anti-correlation between $\agestar$ and $\mustar$. Therefore, the outer radial age profiles miss preferentially larger ages 
and may be biased low. However, since the derivative of $\agestar$ with respect to $\mustar$ approaches $0$ as we move to low
surface mass density, we do not expect this bias to significantly alter the shape of the median radial profile.

Fig. \ref{fig:stacked_profs} highlights the existence of both a $\mli{SMA}-\zstar$ relation and of a 
$\mustar-\zstar$ relation. Both relations are remarkably tight, especially in the inner/high-surface-density regions. Still
it makes sense to investigate whether one is more ``fundamental'' than the other. We consider the scatter around the median
relations in a range where we are highly complete and the scatter is roughly constant, that is $0.5\,\Reff<\mli{SMA}<\Reff$
corresponding to $3.05>\log\mustar>2.65$ (see Fig. \ref{fig:stacked_mustarprof}). In these regions
the scatter around the $\mli{SMA}-\zstar$ relation is $\sim 0.10\,\text{dex}$ while the scatter around the $\mustar-\zstar$ 
relation is $\sim 0.07\,\text{dex}$. A clearly lower scatter in the $\mustar-\zstar$ relation is apparent even if we extend
the range to include the inner/higher-surface-density regions (down to $0.1\,\Reff$ or up to $4$ in $\log\mustar$):
the typical scatter in the $\mustar-\zstar$ relation is always around $\sim 0.06\,\text{dex}$, while the scatter in
the $\mli{SMA}-\zstar$ relation drops below $0.075\,\text{dex}$ only inside $0.3\,\Reff$.
In other words, $\mustar$ is a better predictor of the local $\zstar$ than the radial distance from the centre, thus
supporting the idea that the $\mustar-\zstar$ relation is the driving one, with the $\mli{SMA}-\zstar$ relation being
a consequence of the former one and of the quasi-universal shape of the $\mli{SMA}-\mustar$ profiles. In fact, 
one can work out that the larger
scatter in the $\mli{SMA}-\zstar$ relation with respect to the $\mustar-\zstar$ relation, within $1\,\Reff$,
is quantitatively consistent with this hypothesis.


\subsection{Impact of biases in stellar population profiles}\label{sec:profs_bias}
As mentioned at the end of Sec. \ref{sec:SPanalysis}, biases in the inferences of stellar population parameters may
arise as one approaches the physical limits of the parameter space covered by the models. In the actual profiles we
may be possibly biased
in the central ($\mli{SMA}<0.3-0.4\,\Reff$) and most dense regions ($\log\mustar\gtrsim 3.3$), where 
$\log Z_*/\text{Z}_\odot \gtrsim 0.2-0.3$. Due to the hard limit in $\log Z_*/\text{Z}_\odot = 0.4$ present in our library, 
in those regions, we might be underestimating the true metallicity by a few
$0.01\,\text{dex}$ up to $0.05\,\text{dex}$. We might also correspondingly overestimate the true age by a few $0.01\,\text{dex}$ up to 
$0.1\,\text{dex}$, although for ages as high as $\sim8-10\,\text{Gyr}$ the effect is expected to be even milder. 
As a consequence, metallicity profiles in the central/densest regions might be steeper in reality; in particular, the stark
flattening observed in the profiles vs. $\mustar$ might be partly an artefact. On the contrary, the central age cusp
might be enhanced with respect to the reality. 

On the other hand, in the less dense regions, typically beyond $1\,\Reff$, for ages $\gtrsim 8 \,\text{Gyr}$ and relatively low metallicity we might be 
biased low in age by a few $0.01\,\text{dex}$ (see last paragraph of Sec. \ref{sec:SPanalysis}) and, correspondingly, we might be biased high in $\zstar$ by a few $0.01\,
\text{dex}$. As a consequence, the age ``plateau'' at large distances/low densities might be slightly higher in reality, and
the metallicity profiles somewhat steeper. Note that a negative correction (i.e. to steeper slopes) to the 
radial derivative of metallicity is also expected from the surface brightness cut discussed in the previous section.

Considering the maximum amplitude of these biases, we do not expect significant changes in the shapes of the profiles plotted
in Fig. \ref{fig:stacked_profs}, rather
just small offsets and changes of slopes. In particular, all considerations and conclusions about the qualitative shapes of 
the profiles, the scatter and the existence of tight (quasi-)universal relations are robust against the possible biases of 
the stellar population analysis.

More systematic effects related to the choices of averaging linear quantities rather than their logarithm to estimate
$\agestar$ and $\zstar$, and of using a fixed universal IMF from \cite{chabrier03} are illustrated in Appendix \ref{app:systematics}.
Although different choices/assumptions in these respects may change our results \emph{quantitatively}, the \emph{qualitative}
picture and the trends that emerge from our analysis are robust.

\section{Averaged stellar population profiles}\label{sec:stack_SPprofiles}
In this section we analyze how stellar population profiles (both in $\mli{SMA}$ and in $\mustar$) depend on global
galaxy properties, namely on the stellar velocity dispersion within $\Reff$, $\sigmae$, on the total stellar mass, $M_*$,
and on the morphology (E vs. S0). To this goal, we bin galaxies in different classes and, in each of them, we proceed to 
compute the median averaged profiles and percentiles, as we did for the full sample in Fig. \ref{fig:stacked_profs}.
In particular, individual profiles contribute only as long as spaxel completeness is larger than $2/3=0.67$.

The different subsamples are defined in Table \ref{tab:mstar_sigma_bins} and are plotted in different colors in Fig. 
\ref{fig:comp_ellprof_stack} (radial profiles) and \ref{fig:comp_mustarprof_stack} (profiles in $\mustar$), according to the 
corresponding legends. As a reference, all plots report the median profiles of the unbinned sample (all ETGs in the
top three plots, all Es in the bottom one, repsectively) as solid blue line, with
shaded blue regions covering the $16^\mathrm{th}-84^\mathrm{th}$ percentile range. The same two percentiles are shown for
the subsamples as dashed lines in the corresponding color. Green-shaded regions indicate the range where less than $1/3$ of the
galaxies contribute.

\subsection{Averages in $M_*$ and $\sigmae$ bins}
\begin{figure*}
	\includegraphics[width=0.85\textwidth]{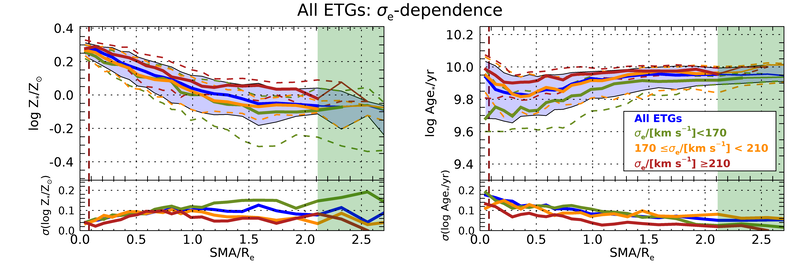}
	\includegraphics[width=0.85\textwidth]{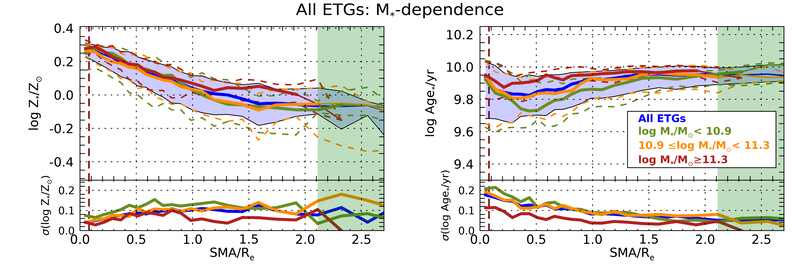}
	\includegraphics[width=0.85\textwidth]{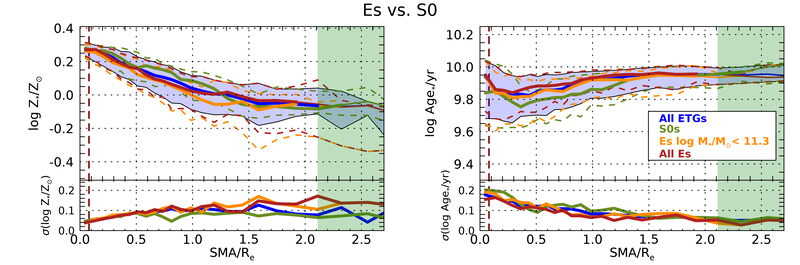}
	\includegraphics[width=0.85\textwidth]{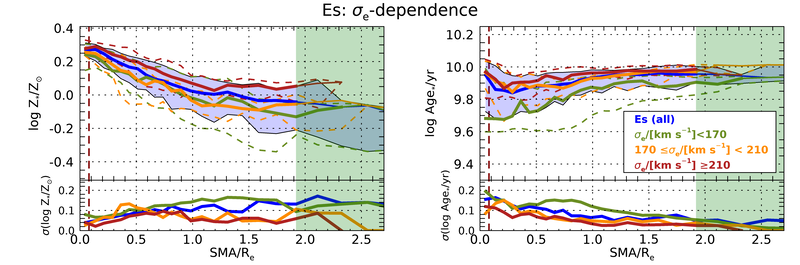}
    \caption{Azimuthally-averaged radial profiles of light-weighted stellar metallicity $\zstar$ (\emph{left column}) and
     stellar age $\agestar$ (\emph{right column}) for different galaxy subsamples. From the top row to the bottom: 
     all ETGs binned according to their $\sigmae$; all ETGs binned according to their $M_*$; Es vs S0 vs Es in the corresponding
     $M_*$ range of S0; only Es binned according to their $\sigmae$. The main panel of each plot displays the median profile of each 
     bin (solid lines) and the $16^\text{th}$ and $84^\text{th}$ percentiles (dashed lines), color-coded according to the legend.
     The blue line and shaded area correspond to the median and inter-percentile range of the full unbinned sample 
     (all ETGs for the first three rows, all Es only for the bottom row), and are reported
     for reference in all plots. Green-shaded regions indicate the range where less than $1/3$ of the galaxies in the unbinned
     sample contribute. Line representing less than $1/3$ of galaxies in a given subsample are thin. In the bottom panels of 
     each plot we report the scatter of individual profiles around the median of each subsample.
     The vertical dashed lines mark the median PSF radius (HWHM). \emph{Most notably, we observe that
     $\sigmae$ is the main parameter that modulates the shape of the profiles, while morphology plays a secondary role and
     $M_*$ appears to control the shape of the profiles only via its correlation with $\sigmae$}.}\label{fig:comp_ellprof_stack}
\end{figure*}

Radial profiles display a clear dependence on $\sigmae$ for both age and metallicity, as one can see in the top panels of 
Fig. \ref{fig:comp_ellprof_stack} (full sample of ``All ETGs'' in bins of $\sigmae$). 
At low and intermediate $\sigmae$ the median metallicity profiles are very similar, but are
significantly different from the metallicity profile of galaxies with $\sigmae \geq 210\, \kms$. High-$\sigmae$ galaxies share with 
lower-$\sigmae$ galaxies very similar $\zstar$ in the central regions, but their decrease of $\zstar$ with $\SMA$ is slower
and results in a difference of $\sim0.1\,\text{dex}$  in $\zstar$ at $\sim1.5\,\Reff$ with respect to lower-$\sigmae$ galaxies.
The effect of $\sigmae$ is particularly dramatic on age profiles. All galaxies share a very similar old age of $\sim 8.9~\text{Gyr}$
beyond $\sim1.5\,\Reff$, yet with a small but significant age offset correlated with $\sigmae$. Inside $\sim1.5\,\Reff$, high-$\sigmae$ 
galaxies display almost flat profiles, with an inflection around $\sim 0.4\,\Reff$; intermediate-$\sigmae$ galaxies reproduce very
closely the median profile for the full sample and are characterized by a U-shape with a minumum of $\sim 6.8\,\text{Gyr}$ at
$\sim0.4\,\Reff$; finally, the low-$\sigmae$ galaxies display a monotonic age decrease toward the centre.
Looking at the plots of scatter (both in $\zstar$ and age) we note that the scatter in the $\sigmae$ subsamples is generally lower 
than in the full sample (except for the $\zstar$ of low-$\sigmae$ galaxies at large radii). This is a further indication that
$\sigmae$ alone induces clear systematics effects on the stellar population profiles.

In the second row of panels of Fig. \ref{fig:comp_ellprof_stack} we plot the profiles in bins of $M_*$, including all ETG galaxies. 
We observe qualitatively very similar trends as with $\sigmae$. We note minor differences in the metallicity profiles, whereby
at $\SMA<1\,\Reff$ the three mass bins overlap almost perfectly. The age profile of low-mass galaxies is more noisy than for the 
corresponding low-$\sigmae$ bin. Finally we note that the scatter in the bins is typically as large as in the general sample, except
for the high-mass bin (which almost coincides with the high-$\sigmae$ bin). There is thus an indication that $M_*$ has systematic
effects on the profiles similar to $\sigmae$, but the correlation is weaker and possibly inherited via the $\sigmae-M_*$ correlation.

\begin{figure*}
	\includegraphics[width=0.85\textwidth]{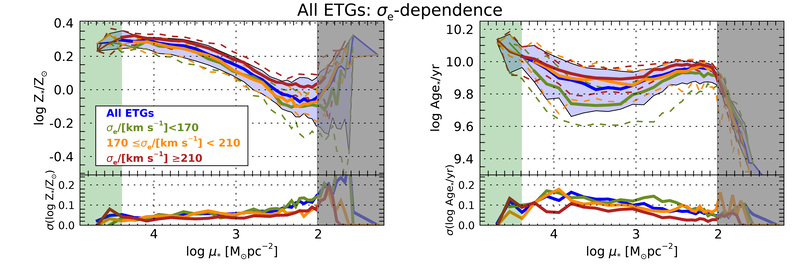}
	\includegraphics[width=0.85\textwidth]{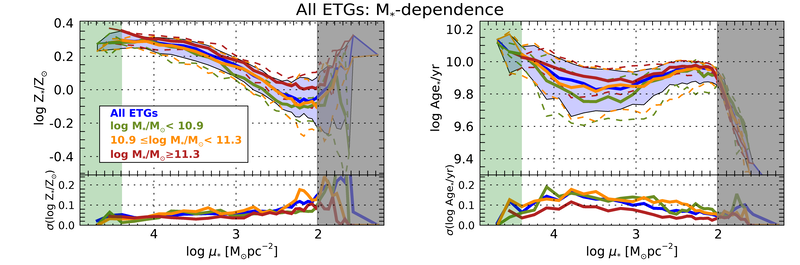}
	\includegraphics[width=0.85\textwidth]{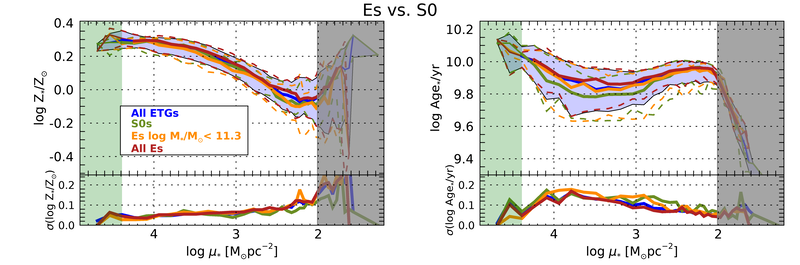}
	\includegraphics[width=0.85\textwidth]{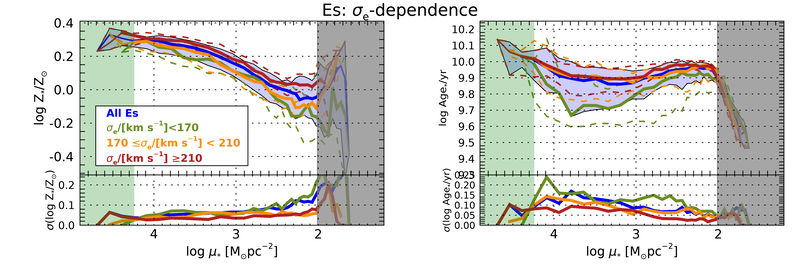}
    \caption{Profiles of light-weighted stellar metallicity $\zstar$ (\emph{left column}) and
     stellar age $\agestar$ (\emph{right column}) as a function of $\log\mustar$ for different galaxy subsamples. 
     From the top row to the bottom: 
     all ETGs binned according to their $\sigmae$; all ETGs binned according to their $M_*$; Es vs S0 vs Es in the corresponding
     $M_*$ range of S0; only Es binned according to their $\sigmae$. The main panel of each plot displays the median profile of each 
     bin (solid lines) and the $16^\text{th}$ and $84^\text{th}$ percentiles (dashed lines), color-coded according to the legend.
     The blue line and shaded area correspond to the median and inter-percentile range of the full unbinned sample 
     (all ETGs for the first three rows, all Es only for the bottom row), and are reported
     for reference in all plots. Green-shaded regions indicate the range where less than $1/3$ of the galaxies in the unbinned
     sample contribute. Line representing less than $1/3$ of galaxies in a given subsample are thin. Grey-shaded regions
     mask the $\mustar$ range where quantities are biased due to the surface brightness cut (see Sec. \ref{sec:sb_mu_profs}). 
     In the bottom panels of each plot we report the scatter of individual profiles around the median of each subsample.
     \emph{Most notably, a quasi-universal $\mustar$--$\zstar$ relation emerges, whose zero-point is modulated by
     $\sigmae$}.}
    \label{fig:comp_mustarprof_stack}
\end{figure*}

By comparing these binned profiles as a function of radius with the corresponding profiles as a function of stellar-mass
surface density in Fig. \ref{fig:comp_mustarprof_stack}, we observe that most of the profiles and relative trends are qualitatively 
consistent, after taking into account the different ``stretch'' caused by the change of variable in abscissa. 
The visual impression is that
different bins separate better, especially in $\zstar$, when profiles as a function of $\mustar$ are used instead of radial profiles.
We will better quantify this impression in the next section \ref{sec:SPchars}.

We note that the $\agestar-\mustar$ profile is U-shaped also for the low-$\sigmae$ subsample, contrary to the monotonically increasing 
behaviour observed in the corresponding radial profile. This is a consequence of the scatter in $\mustar$ in the central regions
and on its dependence on $\sigmae$ (see Fig. \ref{fig:stacked_mustarprof}). As most of the low-$\sigmae$ galaxies do not reach $\mustar$ 
as high as $10^{3.8}\Msun\text{pc}^{-2}$, the median inner \emph{radial} profiles are dominated by the points at lower $\mustar$, 
hence at lower age. However,
the top-right panel of Fig. \ref{fig:comp_mustarprof_stack} shows that even in low-$\sigmae$ galaxies there is a reversal of age 
gradients, provided that large enough densities are reached.

\subsection{Profile dependence on E vs. S0 morphology}
In the third and fourth rows of plots in Fig. \ref{fig:comp_ellprof_stack} and \ref{fig:comp_mustarprof_stack} we investigate the
impact of morphology on the stellar population profiles. 
The plots labeled ``Es vs. S0s'' in the third row of the two figures display the comparison between S0's (in green) and Ellipticals (full sample, in red). We also plot the median profiles for Ellipticals in the same mass range as S0's 
($\log M_*/\text{M}_\odot < 11.3$, in orange), in order to check to what extent differences in the profiles are induced by the different
mass range spanned by the two morphological classes. 

The median metallicity of S0's is systematically higher than the one in E's by $0.05-0.1\,\text{dex}$ for a substantial part
of the radial extent, between $\sim0.6$ and $\sim1.3\Reff$. Conversely, the median light-weighted age of the stellar
populations in S0 galaxies is systematically lower than in E's by $0.05-0.1\,\text{dex}$, over the same radial range.
By restricting the comparison to E's matching the mass range of S0's, we observe qualitatively the same effects, although the 
difference is marginally smaller on average in $\zstar$ and larger in age. This systematic variation between E's full sample and
the mass-matched sub-sample stems from the trend with stellar mass observed in the second row of panels 
in Fig. \ref{fig:comp_ellprof_stack}. 

Contrary to radial profiles, as a function of $\mustar$, the metallicity profiles of E's
and S0's are hardly distinguishable, a fact that further stresses the fundamental nature of the relation between $\zstar$ and $\mustar$,
which is insensitive to the morphology of the galaxy\footnote{The universality of $\zstar(\mustar)$ on one hand and the dependence of
$\zstar(\mli{SMA})$ on morphology on the other hand are a consequence of the stellar mass surface density 
profiles $\mustar(\mli{SMA})$ changing systematically with the morphology.}.
In terms of their $\agestar(\mustar)$ profiles, we observe systematic
differences between E's and S0's, with the latter having younger minima by some $0.1\,\text{dex}$, even when compared to the
mass-matched subsample of E's. 

The fourth row of panels in Fig. \ref{fig:comp_ellprof_stack} and \ref{fig:comp_mustarprof_stack} repeat the same analysis of the
respective top rows, i.e. the average profiles for different bins of $\sigmae$, but now excluding S0's. We find indeed very
similar profiles and trends. The most notable variations occur in the lowest-$\sigmae$ bin, whose difference relative to 
higher-$\sigmae$ bins appears amplified when S0's are excluded. In particular, the offset of the age profile of the low-$\sigmae$ bin
to younger values with respect to the general sample is more significant when S0's are excluded from the analysis. A possible explanation
for this might be that $\sigmae$ is a low-biased indicator of the dynamical support for S0's with respect to E's and therefore
the binning in $\sigmae$ for the general sample produces more heterogeneous subsamples than for the pure E sample.

The morphological classification into E and S0 is nowadays often regarded as a primitive tool to separate ``pressure supported'' from
``rotation supported'' systems, despite the morphological classification having its own peculiarities that are not captured
by a kinematic classification. A full characterization in terms of kinematics would allow us to properly separate 
the so-called ``slow rotators'' from the ``fast rotators'' \citep{Emsellem:2011}. Unfortunately we have this characterization available
only for the 54/69 galaxies in \cite{Falcon-Barroso:2017aa}, so we cannot perform a complete analysis here. However, from \cite{Falcon-Barroso:2015} we can easily see that S0's are an almost pure sample of fast rotators, while Es are a mixed bag of fast rotators and
slow rotators, with Es at $M_*>10\cdot 10^{11.3} \text{M}_\odot$ being almost only slow rotators. So, already from the plots
in Fig. \ref{fig:comp_ellprof_stack} and \ref{fig:comp_mustarprof_stack} we can infer that slow rotators tend to have flatter
age profiles than fast rotators, which, in turn, present stronger age minimum and profile inflection at $\sim 0.4 \Reff$. 
Massive slow rotators tend to slightly flatter radial profiles also in metallicity. 

\subsection{Characterization of profiles by reference values}
In order to provide a more quantitative characterization of the profiles, for each galaxy we evaluate $\agestar$ and $\zstar$ 
at different reference radial ($\SMA$) distances and stellar surface mass densities ($\mustar$).
We define the following set of reference radial distances:
\begin{itemize}
\item[-] ``center'' $(\SMA<0.1\,\Reff)$
\item[-] $0.2\,\Reff\,(0.15\leq\SMA/\Reff<0.25)$ 
\item[-] $0.4\Reff\,(0.35\leq\SMA/\Reff<0.45)$ 
\item[-] $\Reff\,(0.95\leq\SMA/\Reff<1.05)$
\item[-] $2\,\Reff\,(1.95\leq\SMA/\Reff<2.05)$
\end{itemize} 
In brackets we report the discrete $\SMA$ 
ranges used to compute the characteristic stellar population parameters. 
We introduced the $0.2\,\Reff$ distance because it is generally more robust both from a 
statistical point of view (more contributing spaxels) and from a observational/physical point of view (insensitive to residual
PSF mismatches between photometry and IFS, and to possible nuclear sources, e.g. AGN) with respect to the ``center'' region,
yet it is a fair representation of the innermost regions. The $0.4\,\Reff$ reference is chosen as the
approximate location of the age minimum.
Similarly, we define a set of three reference stellar mass surface densities as follows:
\begin{itemize}
\item[-] ``center'' $(\log \mustar = 4.0)$
\item[-] ``mid'' $(\log \mustar = 3.1)
$\item[-] ``outer'' $(\log \mustar = 2.3)$
\end{itemize}
For each reference $\mustar$, the characteristic stellar population
parameters are computed considering only the spaxels having $\mustar$ within $\pm 0.1\,\text{dex}$ from the reference.
For each galaxy subsample, we compute the median and the $16^\text{th}$ and $84^\text{th}$ percentiles
of the distribution of $\agestar$ and $\zstar$ in the spaxels bins defined above, and report them in 
tables \ref{tab:char_stelpop_sma} and \ref{tab:char_stelpop_mustar}, in the form
of $\text{median}^{+(\text{p84}-\text{median})}_{-(\text{median}-\text{p16})}$, along with the number of contributing galaxies.
The distributions of the characteristic stellar populations are represented in form of histograms in the right-hand side panels of
Fig. \ref{fig:stelpop_char_sigma} (and \ref{fig:stelpop_char_mstar}, identical), where the median values are highlighted by
arrows. These histograms and arrows clearly display the systematic shape of the stellar population profiles described in this
Section.

\section{Trends in stellar population profiles with global galaxy properties}\label{sec:SPchars}
In this section we further examine the dependence of stellar population profiles on global galaxy properties (e.g. $M_*$,
$\sigmae$, etc.), by studying the trends between these properties and $\agestar$ and $\zstar$ evaluated at reference
radial ($\SMA$) distances and stellar surface mass densities ($\mustar$), as defined in the previous section.

\begin{figure*}
\centerline{\large \textsf{Characteristic stellar population properties vs. $\sigma_e$}}
	\includegraphics[width=0.45\textwidth]{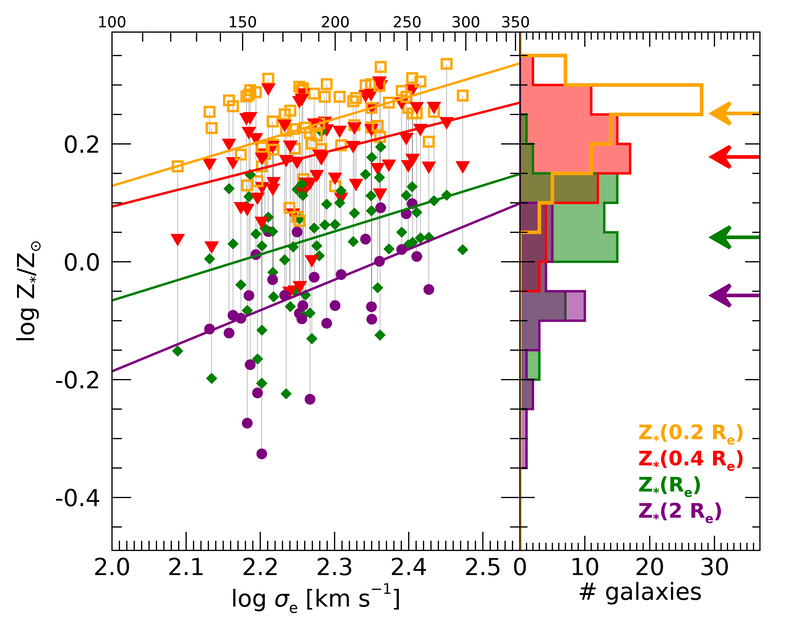}
	\includegraphics[width=0.45\textwidth]{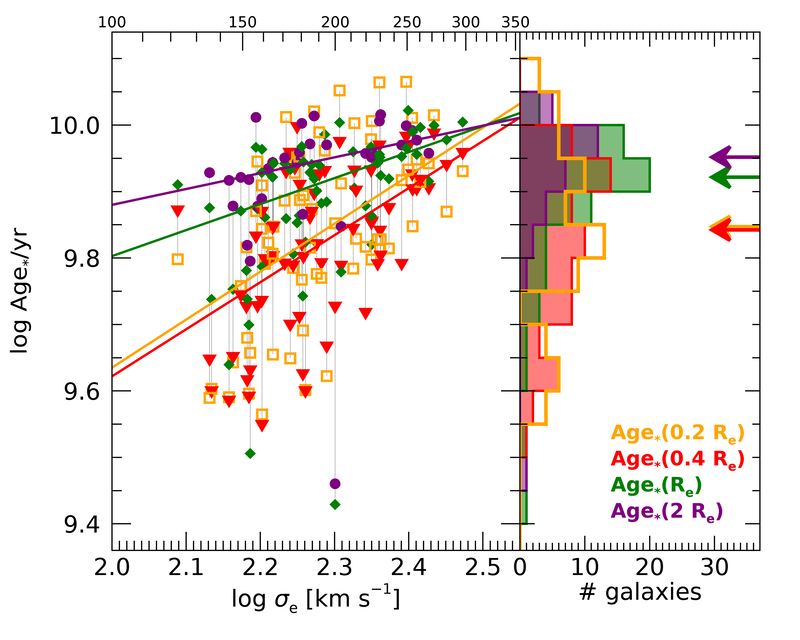}\\
	\includegraphics[width=0.45\textwidth]{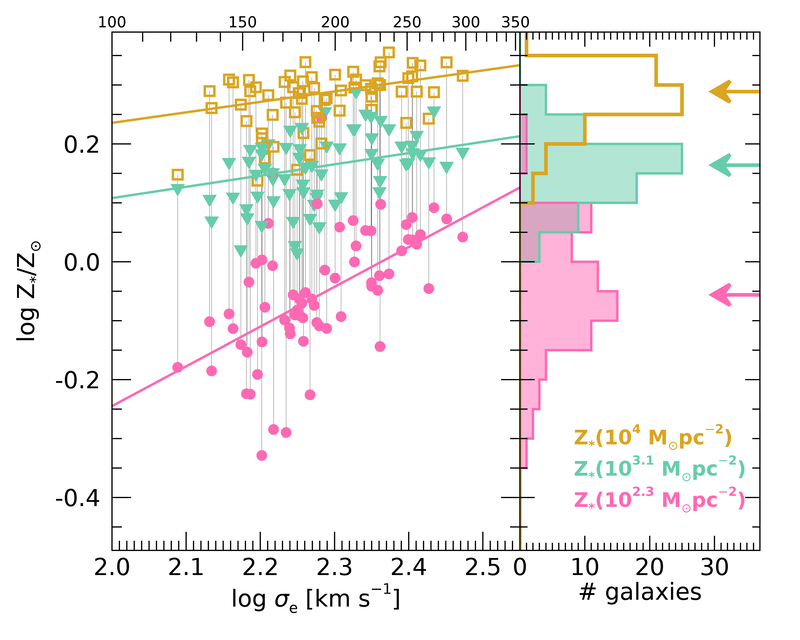}
	\includegraphics[width=0.45\textwidth]{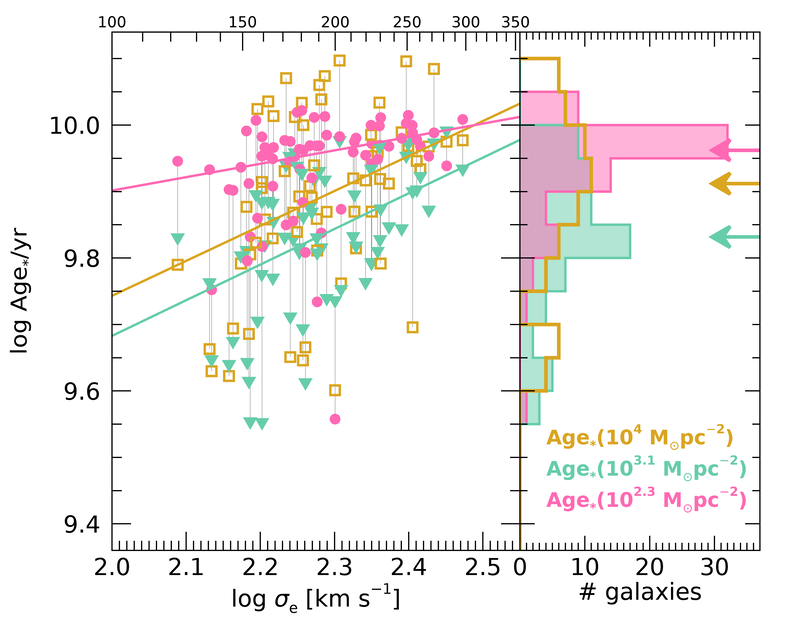}
    \caption{Characteristic stellar population properties ($\zstar$, \emph{left column}, and $\agestar$, \emph{right column}, resp.)
    are plotted as a function of the velocity dispersion within $\Reff$, $\sigmae$, in the main panel of each plot.
    The \emph{right-side} panel displays the histogram of the distribution, with arrows marking the median. In the \emph{top}
    row of plots stellar population properties are evaluated at some reference $\SMA$ in units of $\Reff$, while in the \emph{bottom}
    row they are evaluated at some reference $\mustar$. Different colors identify different reference $\SMA$ and $\mustar$ (see
    legends). Each thin vertical grey line connects points relative to the same galaxy. We plot in color the robust linear
    regression lines obtained from a least absolute deviation algorithm. \emph{Notably, all stellar population properties at
    characteristic $\mli{SMA}$ and $\mustar$ are significantly and positively correlated with $\sigmae$}.}\label{fig:stelpop_char_sigma}
\end{figure*}

In Fig. \ref{fig:stelpop_char_sigma} we show how $\zstar$ (left column) and $\agestar$ (right column), at different reference $\SMA$ 
(top row) and reference $\mustar$ (bottom row), respectively, correlate with the stellar velocity dispersion $\sigmae$. 
The main panel of each plot displays
the points for individual galaxies in different colours for the different reference quantities. Points for the same galaxy are
connected by vertical thin lines. The thick lines are obtained from robust linear regression via least absolute deviation minimization.
The coefficients of the fits, the mean absolute deviation (MAD), the Spearman's rank correlation coefficient $C_\text{Spearman}$,
and the resulting probability for null correlation $P_\text{null}$ are reported in columns 5 to 9 of tables 
\ref{tab:corr_char_stelpop_sma} and \ref{tab:corr_char_stelpop_mustar}. 
The right-side panels of the plots display the number distribution in $\zstar$ and $\agestar$ for the different reference
quantities. Arrows mark the position of the median of the distributions (see also tables \ref{tab:char_stelpop_sma} and \ref{tab:char_stelpop_mustar}). 

All characteristic stellar population properties are positively correlated with $\sigmae$, at $>99\%$ confidence level, 
according to a simple Spearman's rank correlation
test. In other words, at any radius or stellar mass surface density we find a trend for both $\zstar$ and $\agestar$ to increase at
increasing velocity dispersion. 

The trends of metallicities at different reference $\SMA$ (Fig. \ref{fig:stelpop_char_sigma}, top left panel) have all a very similar 
slope, thus
implying that the effect of $\sigmae$ on the radial metallicity profiles is essentially a vertical shift, by about $0.04~\text{dex}$
per $0.1\,\text{dex}$ in $\sigmae$, corresponding to about $0.15\,\text{dex}$ over the $\sigmae$ range spanned by our sample.
The only exception occurs in the very center (``center'' region, only reported in the tables, not shown in the plots), where
all profiles appear to converge. There we measure a weaker correlation with $\sigmae$ and all galaxies share the same estimated
$\log\zstar/\text{Z}_\odot\sim 0.3$ within a few $0.01\,\text{dex}$. Note that this central convergence may be partly an effect 
of saturation towards the highest metallicity allowed in the spectral library.

At fixed $\mustar$, the trends of $\zstar$ with $\sigmae$ (Fig. \ref{fig:stelpop_char_sigma}, bottom left panel)
are flatter than at fixed $\SMA$ in the high and intermediate
density regions, but significantly steeper in the low density regions, with an increase of $\sim0.07\,\text{dex}$ in metallicity
per $0.1\,\text{dex}$ in $\sigmae$. This is a quite remarkable effect as one would naively expect that $\sigmae$ (which traces
the dynamics in the densest regions) would affect mostly the densest regions of the galaxies. What we see, instead, is a correlation
between the inner dynamical state (possibly a tracer of the depth of the gravitational potential well) with the metallicity of
the low density regions, which simulations indicate being composed by a significant fraction of stars accreted from satellites 
\citep[e.g.][]{Hirschmann:2015,Rodriguez-Gomez:2016}. This, in turn, highlights a strong link between halo mass (whose $\sigmae$ is a 
proxy) and the surrounding environment in terms of the properties of stellar populations of its satellite galaxies.

Characteristic ages show steeper positive trends with $\sigmae$ for the inner reference radii ($0.4$, $0.2\,\Reff$) than for the
outer ones ($1$ and $2\,\Reff$). This is consistent with $\sigmae$ driving the scatter in the radial profiles of $\agestar$ and with the 
substantial decrease of the galaxy-to-galaxy scatter as one moves outwards. Going from $\sim 130$ to 
$\sim 300\,\text{km}\,\text{s}^{-1}$
galaxies roughly double their age in the inner $\sim 0.5\,\Reff$, but increase their age by $\sim 25\%$ only beyond $1\,\Reff$.
From the convergence of the different trend lines we also see that age profiles become essentially flat for 
$\sigmae\gtrsim 300\,\text{km}\,\text{s}^{-1}$. The trends of characteristic $\agestar$ at different $\mustar$ (bottom-right
plot of Fig. \ref{fig:stelpop_char_sigma}) confirm this picture by displaying flat slopes at low $\mustar$ and steeper slopes
at higher $\mustar$, with a substantial convergence at $\sim 300\,\text{km}\,\text{s}^{-1}$. We can summarize these trends by
saying that high-$\sigmae$ ETGs are homogeneously and maximally old, while at lower $\sigmae$ they become increasingly younger,
more so in the inner $\sim 0.5\,\Reff$.

\begin{figure*}
\centerline{\large \textsf{Characteristic stellar population properties vs. $M_*$}}
	\includegraphics[width=0.45\textwidth]{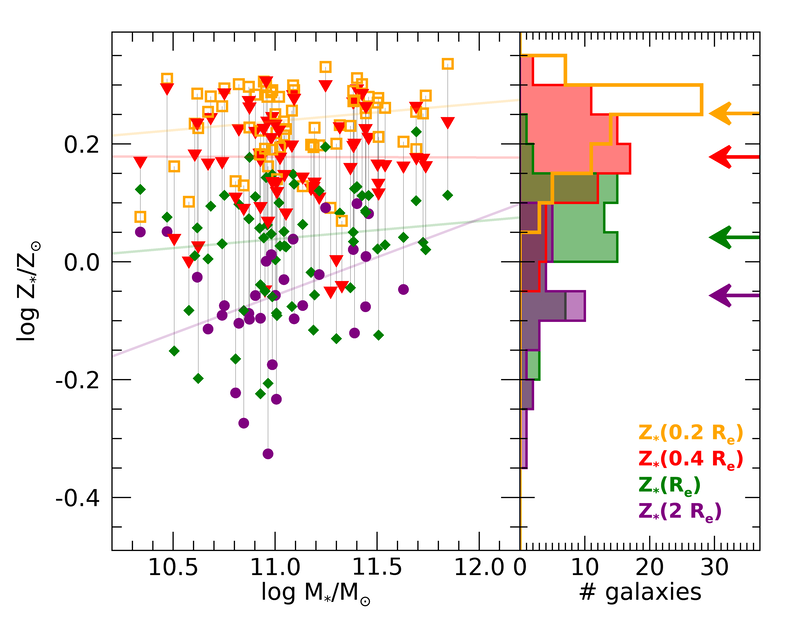}
	\includegraphics[width=0.45\textwidth]{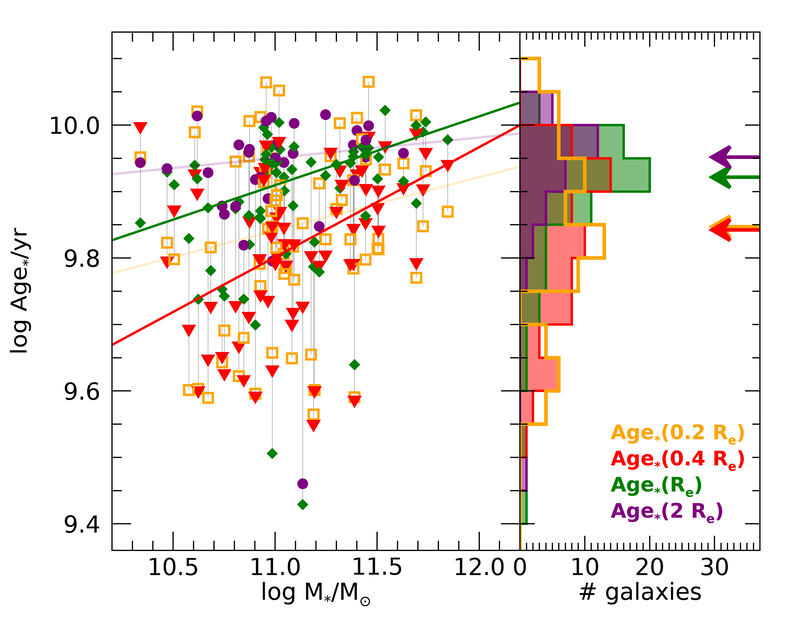}\\
	\includegraphics[width=0.45\textwidth]{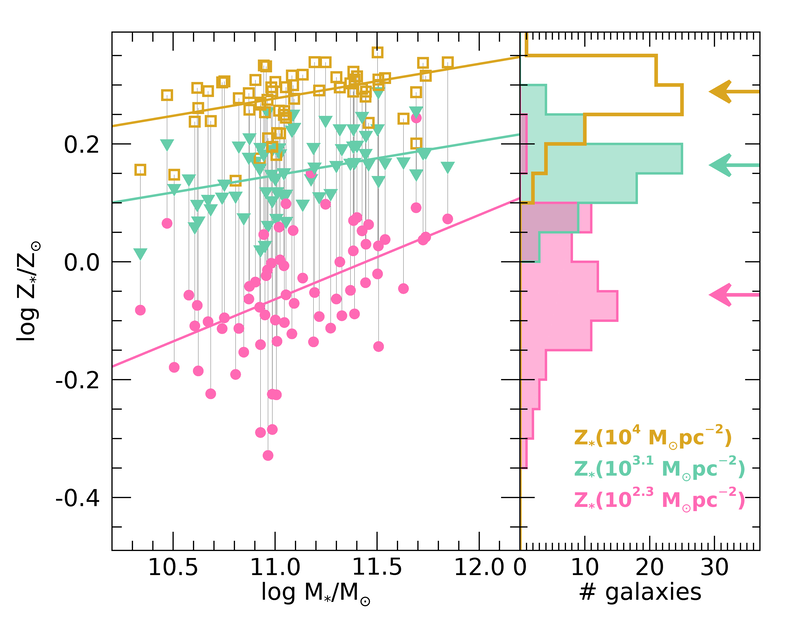}
	\includegraphics[width=0.45\textwidth]{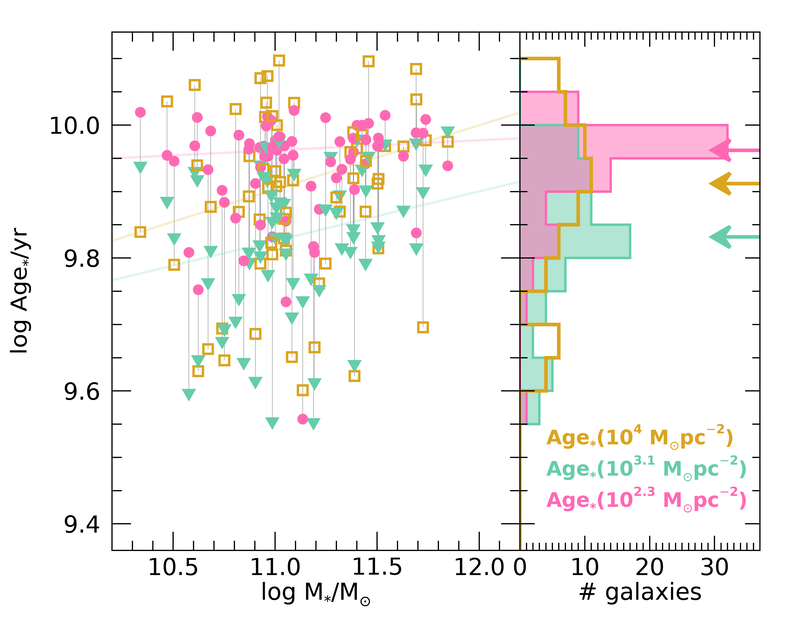}
    \caption{Characteristic stellar population properties ($\zstar$, \emph{left column}, and $\agestar$, \emph{right column}, resp.)
    are plotted as a function of the total stellar mass $M_*$ in the main panel of each plot.
    The \emph{right-side} panel displays the histogram of the distribution, with arrows marking the median. In the \emph{top}
    row of plots stellar population properties are evaluated at some reference $\SMA$ in units of $\Reff$, while in the \emph{bottom}
    row they are evaluated at some reference $\mustar$. Different colors identify different reference $\SMA$ and $\mustar$ (see
    legends). Each thin vertical line connects points relative to the same galaxy. We plot in color the robust linear
    regression lines obtained from a least absolute deviation algorithm. Thin lines are used for correlations that are significant
    with a probability less than $99\%$, according to the Spearman's rank correlation test. \emph{While positive correlations with $M_*$
    are found for all quantities, they are weaker than those with $\sigmae$ and, in several cases, not statistically significant. This 
    is an indication that $M_*$ drives stellar population profile indirectly, through the correlation with $\sigmae$}.}\label{fig:stelpop_char_mstar}
\end{figure*}

In Fig. \ref{fig:stelpop_char_mstar} we repeat the analysis of Fig. \ref{fig:stelpop_char_sigma}, but considering the stellar mass
$M_*$ as the variable against which to check for trends, instead of $\sigmae$. In this figure, regression lines are drawn as thick or
thin lines depending on whether the hypothesis of null correlation can be excluded with a probability of more of $99\%$ or less,
respectively. Although qualitatively we obtain similar trends of characteristic $\zstar$ and $\agestar$ increasing with $M_*$,
as opposed to the correlations with $\sigmae$ (all significant), only $6/14$ correlations with $M_*$ are significant. This can
be understood if we interpret the correlations with $\sigmae$ as ``primary'' correlations and the ones with $M_*$ as second
order correlations, ``inherited'' from the correlation between $M_*$ and $\sigmae$ (see Fig. \ref{fig:mstar_sigma}).
In particular, we note that none of the correlations with $\zstar$ at fixed radii retains its significance, whereas correlations
with $\zstar$ at fixed $\mustar$ do. This is further indication of the fundamental role of $\mustar$, rather than radial distance,
in determining the metallicity of the stars in an ETG.

Significant age trends with $M_*$ are only observed at $0.4\,\Reff$ and $1\,\Reff$. Those are also the most significant trends
against $\sigmae$ (see Tab. \ref{tab:char_stelpop_sma}) and this observation confirms the hypothesis that correlations with 
$M_*$ are actually second order correlations, ``inherited'' from the correlation between $M_*$ and $\sigmae$. 

We further explore possible correlations of the shape of the stellar population profiles 
with the ``global'' parameters represented by the metallicity and the age evaluated at $1\,\Reff$, 
$\zstar(\Reff)$ and $\agestar(\Reff)$, and with different concentration indices for the light distribution.
The choice of $\zstar(\Reff)$ and $\agestar(\Reff)$ as global parameters is justified because they can be considered as 
reasonable proxies for the galaxy-integrated stellar metallicity and age, respectively \citep[see also][]{Gonzalez-Delgado:2015aa}.
From Tab. \ref{tab:corr_char_stelpop_sma} and \ref{tab:corr_char_stelpop_mustar} we note that most correlations of $\zstar(\Reff)$ 
and $\agestar(\Reff)$ are with characteristic $\zstar$ and $\agestar$, respectively. These correlations are partly built-in,
given the continuity of the profiles. However, they also indicate that changes in the profile shapes, if any, correlate with their
normalization at $1\,\Reff$.

We also note some weak or marginally significant positive correlations between characteristic $\zstar$ and $\agestar(\Reff)$. 
Vice versa, we observe some tentative positive correlations between characteristic $\agestar$ and $\zstar(\Reff)$, although 
they are milder and much less significant ($P_\text{null}\gtrsim 0.1$). 
We interpret these correlations as second order effects, deriving from the primary 
correlations with $\sigmae$. Interestingly, there is virtually no correlation between $\zstar(\Reff)$
and $\agestar(\Reff)$.

Finally, we do not observe any significant correlation between characteristic stellar population parameters and concentration
index of the light profiles. In Tab. \ref{tab:char_stelpop_sma} and \ref{tab:char_stelpop_mustar} for illustration we only
report on the correlation analysis with the $C_{31}$ index, which is defined as the ratio between the radii enclosing
$75\%$ and $25\%$ of the total light, $R_{75}$ and $R_{50}$, respectively. We get consistent results also for other concentration
indices, namely for $C_{95}\equiv\frac{R_{90}}{R_{50}}$ and for $C_{21}\equiv\frac{R_{50}}{R_{25}}$, which are sensitive in
different degree to different parts of the profile, although obviously correlated (and equivalent in case of ideal S\'ersic profiles).
This lack of  measurable correlation may partly arise from the small dynamic range of the concentration parameters in our sample. 
In fact, \cite{Zhuang:2019} analysed the stellar metallicity profiles for a sample of CALIFA galaxies spanning the full
morphological range and found a clear dependence of the radial metallicity profile on the S\'ersic index.

\section{Stellar population gradients and their trends with global properties}\label{sec:SPgradients}

In this Section we provide some more quantitative estimates on the shape of the stellar population profiles, that can
be useful for comparison with other observations and with models (e.g. Hirschmann in prep.).

\subsection{Radial stellar population gradients}
We quantify the radial variations of stellar population properties inside ETGs by means of linear and logarithmic
radial gradients, that we define as:
\begin{equation}\label{eq:lingrad}
\nabla_{\text{lin}} X \equiv \frac{\partial \log X}{\partial \SMA}
\end{equation}
\begin{equation}\label{eq:loggrad}
\nabla_{\text{log}} X\equiv \frac{\partial \log X}{\partial \log \SMA}
\end{equation}
respectively, where $X$ is either $\zstar$ or $\agestar$. Despite of the ill-defined nature of a global gradient for profiles that are 
neither pure exponential nor power-law, as we showed in the previous sections, this kind of profile characterization is quite common in
the literature \citep[e.g.][]{kuntschner+10,Martin-Navarro:2018} and may be helpful for comparison purposes and to compress the 
information.
Considering the ``curvature'' of the profiles we define the radial gradients over two distinct ranges: an inner range from 
$0.2\,\Reff$ to $1\,\Reff$, and an outer range from $1\,\Reff$ to $2\,\Reff$. We exclude the very inner region to avoid biases in the
estimate due to low-number of spaxel statistics and PSF smearing effects. 
For each galaxy, the gradient values are computed as the ratio of finite differences between the values at the boundaries of the ranges.
In particular, in these differences we consider $\zstar$ and $\agestar$, respectively, evaluated as the median in the spaxels 
within $0.1$-$\Reff$-wide annuli centered at the respective boundaries of the ranges. In Tab. \ref{tab:grad_stelpop_sma} we report
the values of the radial (linear and logarithmic) gradients for the different galaxy subsamples, both for the inner and the outer 
regions. The reference value in each column is the median of the sample, the plus-minus values represent the 
$16^\text{th}-84^\text{th}$ percentile range.

Metallicity gradients are all negative but in two cases, where we measure slightly positive inner gradients (indeed very close to flat). 
We observe a flattening from inner to outer gradients in linear scale, and vice versa in logarithmic scale. 
This is consistent with the profiles observed in Fig. \ref{fig:stacked_profs} and \ref{fig:comp_ellprof_stack}
and with the different stretch when logarithmic radial scale is used instead of linear. The significance of this effect is evident from
the histograms in the side-panel of the top left plot of Fig. \ref{fig:stelpop_grads_sigma}. These distributions are clearly 
inconsistent with constant-slope profiles at more than $3\,\sigma$ significance.
From Tab. \ref{tab:grad_stelpop_sma}, we see very little variations among the different samples. The inner gradients in the various 
subsamples
depart from the median inner gradient of the full sample ($-0.27\,\text{dex}\,\Reff^{-1}$ or $-0.31\,\text{dex}\,\text{dex}^{-1}$)
by no more than $0.04\,\text{dex}\,\Reff^{-1}$ (or $0.05\,\text{dex}\,\text{dex}^{-1}$), which is half of the overall
galaxy-to-galaxy scatter. If anything, the most massive or high-$\sigmae$ galaxies tend to have slightly flatter inner $\zstar$ 
profiles than average. A similar hint is seen also in the outer gradients. However, as we will show in Sec. \ref{sec:SPgrads_vs_prop}, 
these trends are not statistically significant.

Inner age gradients are on average positive, with a typical increase of $20-25\%$ per $\Reff$ between the inner 
$0.2\,\Reff$ and $1\,\Reff$.
The scatter, however, is large, so that $19/65$ galaxies ($29\%$) have measured negative gradients. This results from the
diverse U-shapes observed in the age profiles within $1\,\Reff$. We just note a marginal indication for low-$\sigmae$
galaxies to display steeper positive age gradients, which is consistent with the top right plot of Fig. \ref{fig:stacked_profs}
As already observed in the previous sections, the age profiles become less scattered beyond $1\,\Reff$, where we still observe
mildly positive gradients but with a much less scattered distribution.

\subsection{Stellar population gradients along $\mustar$}
Similarly to radial gradients, in order to quantify the slopes of the variations of stellar population properties as a function 
of $\mustar$, we introduce the quantity (``gradient along $\mustar$''):
\begin{equation}\label{eq:mustargrad}
\nabla_{\mustar} X\equiv \frac{\partial \log X}{\partial \log \mustar}
\end{equation}
where $X$ is either $\zstar$ or $\agestar$. Also in this case, we define an inner range including the spaxels with 
high stellar mass surface density $3.1<\log\frac{\mustar}{\Msun\,\text{pc}^{-2}}<4$, and an outer range with
$2.3<\log\frac{\mustar}{\Msun\,\text{pc}^{-2}}<3.1$. The break point between the two regimes is arbitrary located
visually close to the inflection point in the median age profile in Fig. \ref{fig:stacked_profs} (bottom right plot).
$\nabla_{\mustar}$ gradients are obtained as the ratios of finite difference between the extremes of the range and by adopting as 
$\zstar(\mustar)$ and $\agestar(\mustar)$ the median $\zstar$ and median $\agestar$, respectively, in all spaxels with 
surface mass density within $\log\mustar-0.1$ and $\log\mustar+0.1$. The values for the $\nabla_{\mustar}$ gradients
for different subsamples are reported in Tab. \ref{tab:grad_stelpop_mustar}. The reference value is the median over each (sub)sample
and the plus-minus values correspond to the $84^{\text{th}}$ and the $16^{\text{th}}$ percentile of the distribution, respectively.
The distributions of gradient values are also represented in form of histograms in the side panels of the plots in the bottom
row of Fig. \ref{fig:stelpop_grads_sigma}. Given the monotonically decreasing nature
of $\mustar(\SMA)$ profiles, positive $\nabla_{\mustar}$ gradients correspond to negative radial gradients, and vice versa. 
For this reason, in the bottom panels of Fig. \ref{fig:stelpop_grads_sigma} the vertical axis is 
flipped (positive to negative) compared to the
top panels (negative to positive), in order to ease the visual comparison with the distribution of radial gradients.

Concerning metallicity gradients, we observe very similar distributions as for the radial case. Both inner and outer 
$\nabla_{\mustar}(\zstar)$ distribute in the positive range of values (i.e. decreasing $\zstar$ going to lower $\mustar$),
with just a couple of exceptions. The gradients are flatter
in the high $\mustar$ regime and steeper at low $\mustar$, as already noted in Sec. \ref{sec:SPprofiles} and from Fig. 
\ref{fig:stacked_profs}. The galaxies in the highest mass bin and in the highest $\sigmae$ tend 
to have flatter inner gradients, although the significance of the effect is rather low. 

Concerning age gradients, inner gradients present a broad distribution around $0$, with a slightly positive median of $0.06$ dex per dex
(i.e. decreasing $\agestar$ going to lower $\mustar$).
This distribution is the consequence of the inner range embracing the region around the age dip, with a net effect of an overall
flat profile. The distribution of outer age gradients is instead well defined negative (i.e. increasing $\agestar$ going to lower 
$\mustar$), with just a minor tail of galaxies extending
towards 0 and positive values. A trend for steeper slopes at smaller masses and $\sigmae$ is also apparent.

\subsection{Dependence of gradients on global properties}\label{sec:SPgrads_vs_prop}
In this Section we investigate possible dependencies of stellar population gradients on global quantities, in a similar way
as we did in Sec. \ref{sec:SPchars} for stellar population properties at fixed characteristic $\SMA$ or $\mustar$. In this way
we can isolate variations in the shape of the profiles from the changes in the overall normalization.

\begin{figure*}
\centerline{\large \textsf{Radial gradients vs. $\sigma_e$}}
	\includegraphics[width=0.45\textwidth]{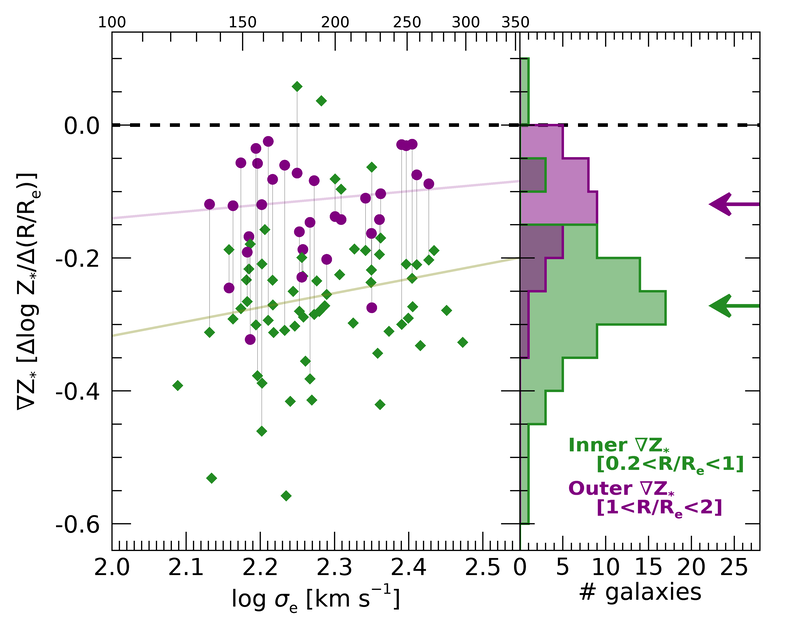}
	\includegraphics[width=0.45\textwidth]{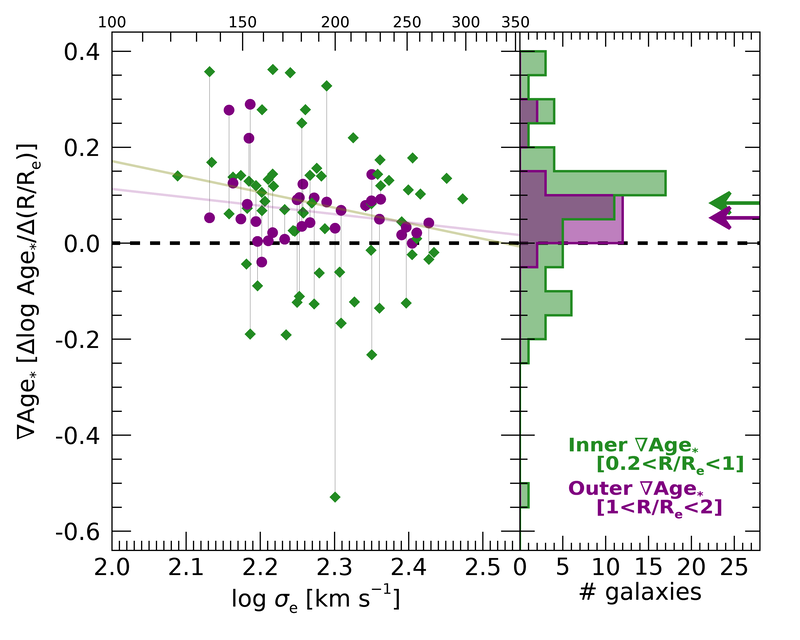}
\centerline{\large \textsf{Gradients along $\mustar$\, vs. $\sigmae$}}
	\includegraphics[width=0.45\textwidth]{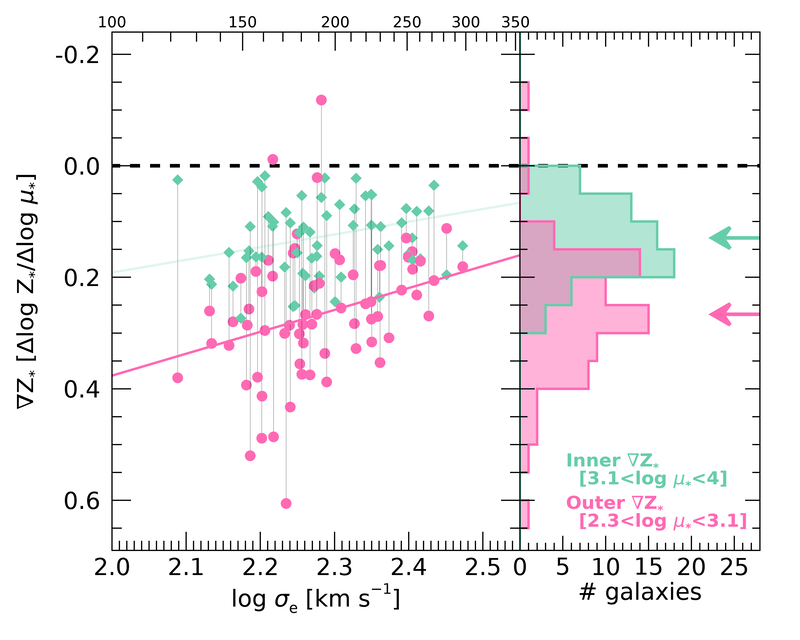}
	\includegraphics[width=0.45\textwidth]{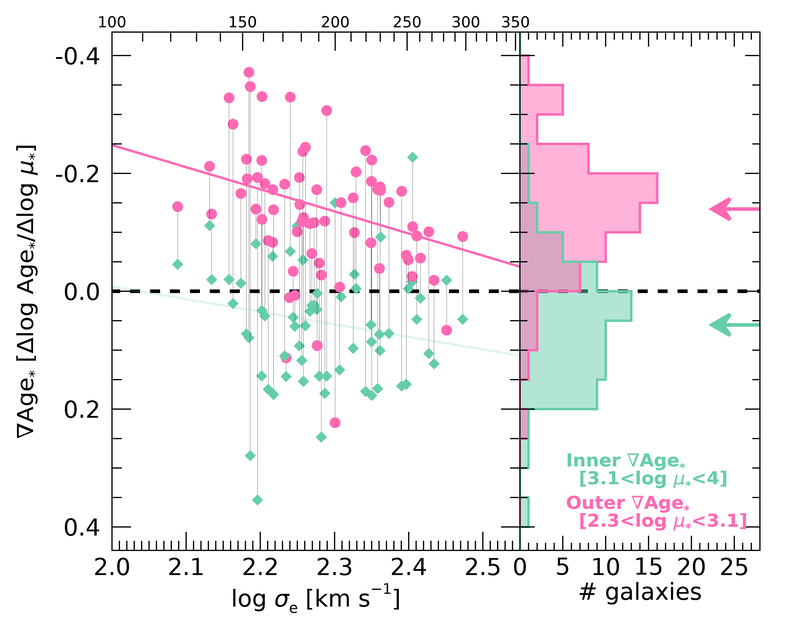}
    \caption{Dependence of various stellar population gradients on the velocity dispersion $\sigmae$. 
    Radial gradients are displayed in the \emph{top row}, gradients along $\mustar$ in the
    \emph{bottom row}; $\zstar$ gradients are presented in the \emph{left column} plots, $\agestar$ gradients
    in the \emph{right column}. The main panel of each plot reports the points for individual galaxies, with
    different colors for the different ranges of $\SMA$ or $\mustar$ (see legends). Points relative to the same galaxy are connected
    with vertical thin lines. Lines in color represent the robust fit to the data; a thin transparent style is used
    when the correlation is deemed as not significant (see text). The side panels to the \emph{right} of the main ones display
    the histograms of the gradient values. Arrows mark the median of the distributions. Note that the $y$-axis for the gradients
    along $\mustar$ is reversed, so to ease the comparison with radial gradients, whereby points below the zero line
    (dashed horizontal line) indicate trends of decreasing property going from the center towards the outskirts.}\label{fig:stelpop_grads_sigma}
\end{figure*}

We start off analyzing the dependence of gradients on the stellar velocity dispersion $\sigmae$, in Fig. \ref{fig:stelpop_grads_sigma}.
The main panels in these plots display the gradients (radial in the top row and along $\mustar$ in the bottom row) of metallicity 
(left column) and of age (right column) as a function of $\sigmae$, in various colours for different ranges, as indicated in the legends.
We performed robust linear regression via least absolute deviation minimization. The results are displayed by the lines in different
colours, matching the colour of the points. In tables \ref{tab:grad_rad_corr} and \ref{tab:grad_mustar_corr}
the coefficients of the fits, the mean absolute deviation (MAD), the Spearman's rank correlation coefficient 
$C_\text{Spearman}$, and the resulting probability for null correlation $P_\text{null}$ are reported in the last five columns,
respectively. If $P_\text{null}<0.01$ we deem the correlation as significant: the corresponding row in the table is checked and the
regression line is drawn with thick line; vice versa, the regression is shown with a thin transparent line.

Although we note general trends for profiles to become flatter as $\sigmae$ increases, we find significant statistical correlations
only between $\sigmae$ and outer gradients along $\mustar$, both in metallicity and in age. If we take this result together with 
the ubiquitous significant correlations that we found between properties at fixed characteristic $\SMA$ and $\mustar$ (see Fig. 
\ref{fig:stelpop_char_sigma}), we can conclude that $\sigmae$ drives systematic variations in the stellar population profiles, although
these variations do not affect the \emph{shape} of the profiles chiefly, rather their \emph{normalization}. In other words,
by increasing $\sigmae$ from $120$ to $300\,\text{km}\,\text{s}^{-1}$ we mainly shift age and metallicity profiles to higher
values, and in second place we produce flatter profiles. Note, however, that the gradients, as actually defined, do not capture 
the age dip around $0.4\,\Reff$, which is strongly correlated with $\sigmae$.

We repeat the correlation analysis of gradients against total stellar mass $M_*$, instead of $\sigmae$, and report the results 
in Tab. \ref{tab:grad_rad_corr} and \ref{tab:grad_mustar_corr}. Despite 
some very marginal indications for analogous trends as with $\sigmae$, none of them is significant at more than $90\%$ level.
This is yet another evidence that $\sigmae$ is the main driver of systematic changes in stellar population profiles and correlations
with $M_*$ are just inherited via the correlation between $\sigmae$ and $M_*$.

\begin{figure*}
\centerline{\large \textsf{Age gradients vs. $\mli{Age}(R_e)$}}
	\includegraphics[width=0.45\textwidth]{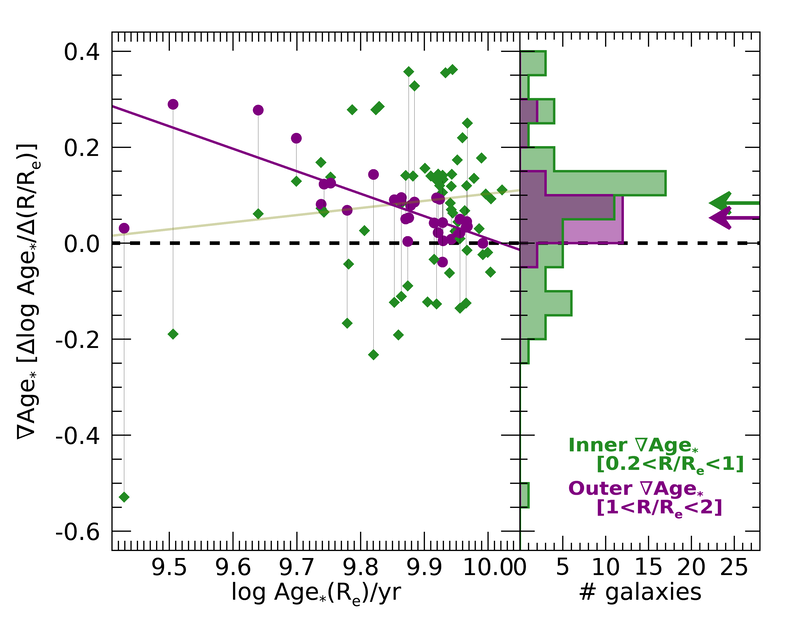}
	\includegraphics[width=0.45\textwidth]{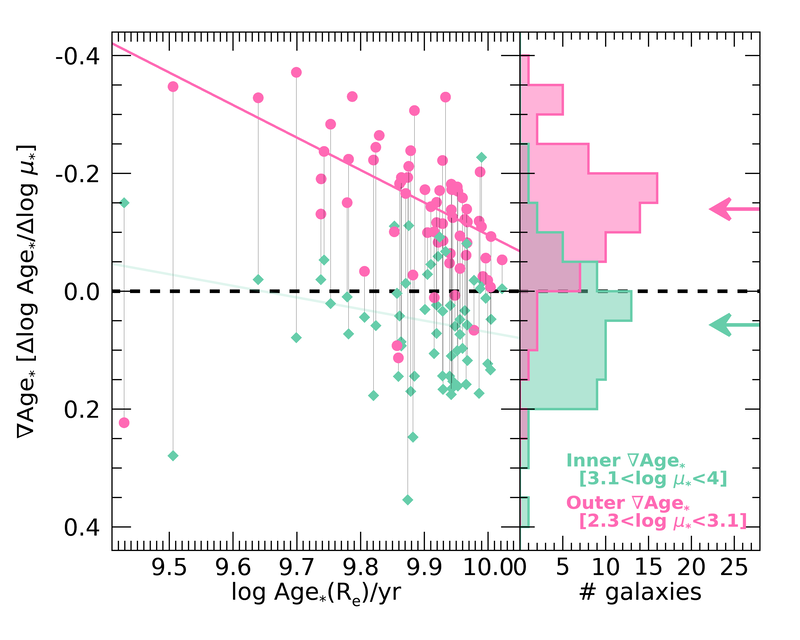}
    \caption{Dependence of various age gradients on the characteristic age at the effective radius, $\agestar(\Reff)$. 
    Radial gradients are displayed in the \emph{left-side} plot, gradients along $\mustar$ in the
    \emph{right-side} one. The main panel of each plot reports the points for individual galaxies, with
    different colors for the different ranges of $\SMA$ or $\mustar$ (see legends). Points relative to the same galaxy are connected
    with vertical thin lines. Lines in color represent the robust fit to the data; a thin transparent style is used
    when the correlation is deemed as not significant (see text). The side panels to the \emph{right} of the main ones display
    the histograms of the gradient values. Arrows mark the median of the distributions. Note that the $y$-axis for the gradients
    along $\mustar$ is reversed, so to ease the comparison with radial gradients, whereby points below the zero line
    (dashed horizontal line) indicate trends of decreasing $\agestar$ going from the center towards the outskirts.}
    \label{fig:agegrads_Agee}
\end{figure*}

\begin{figure*}
\centerline{\large \textsf{Metallicity gradients vs. $Z_*(R_e)$}}
	\includegraphics[width=0.45\textwidth]{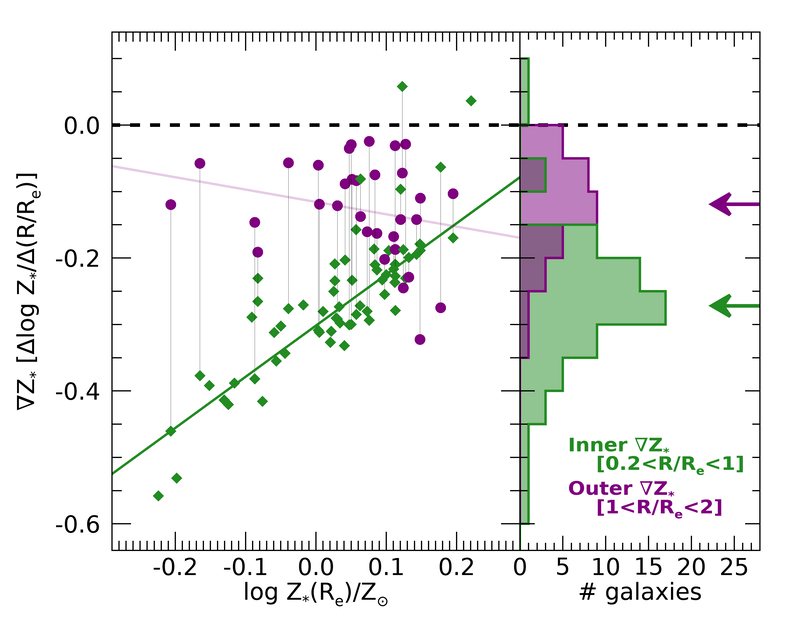}
	\includegraphics[width=0.45\textwidth]{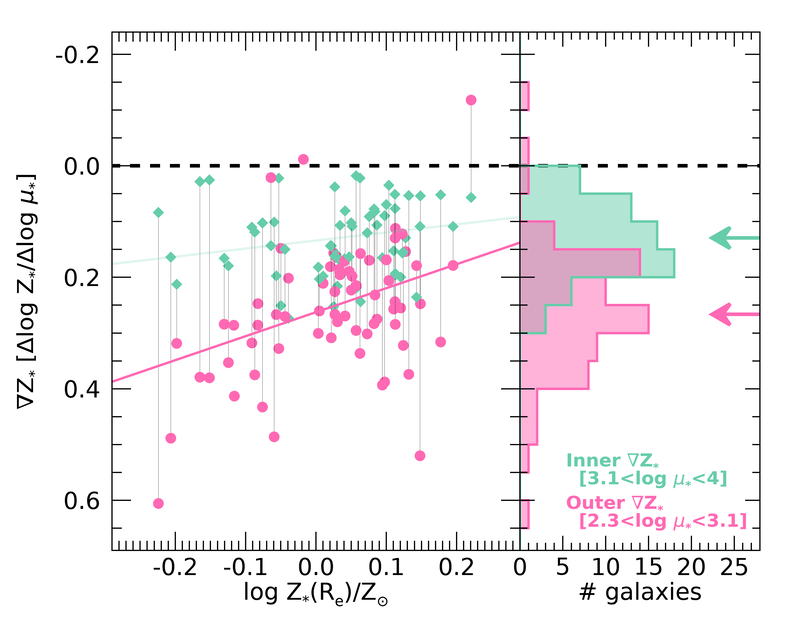}
    \caption{Dependence of various $\zstar$ gradients on the characteristic $\zstar$ at the effective radius, $\zstar(\Reff)$. 
    Radial gradients are displayed in the \emph{left-side} plot, gradients along $\mustar$ in the
    \emph{right-side} one. The main panel of each plot reports the points for individual galaxies, with
    different colors for the different ranges of $\SMA$ or $\mustar$ (see legends). Points relative to the same galaxy are connected
    with vertical thin lines. Lines in color represent the robust fit to the data; a thin transparent style is used
    when the correlation is deemed as not significant (see text). The side panels to the \emph{right} of the main ones display
    the histograms of the gradient values. Arrows mark the median of the distributions. Note that the $y$-axis for the gradients
    along $\mustar$ is reversed, so to ease the comparison with radial gradients, whereby points above the zero line
    (dashed horizontal line) indicate trends of increasing $\zstar$ going from the center towards the outskirts.}
    \label{fig:Zgrads_Ze}
\end{figure*}

\begin{figure*}
\centerline{\large \textsf{Metallicity gradients vs. $\agestar(R_e)$}}
	\includegraphics[width=0.45\textwidth]{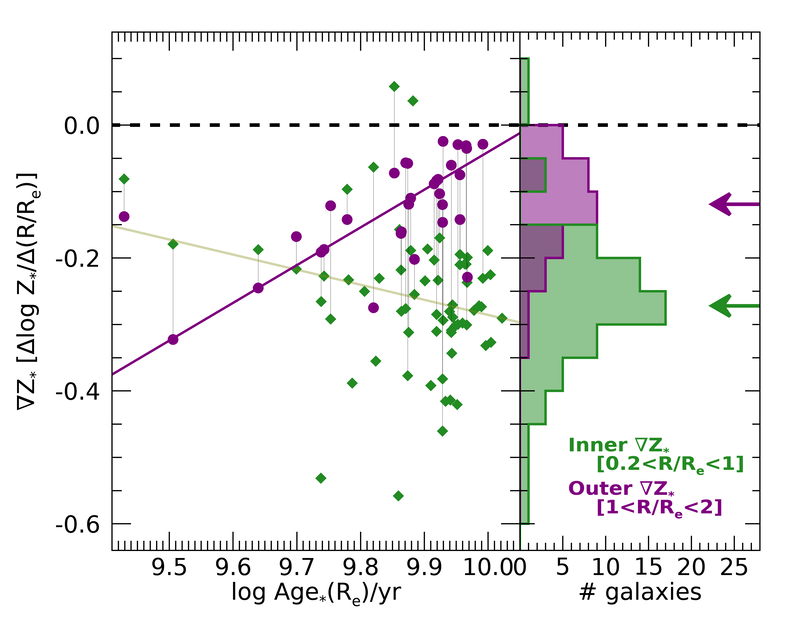}
	\includegraphics[width=0.45\textwidth]{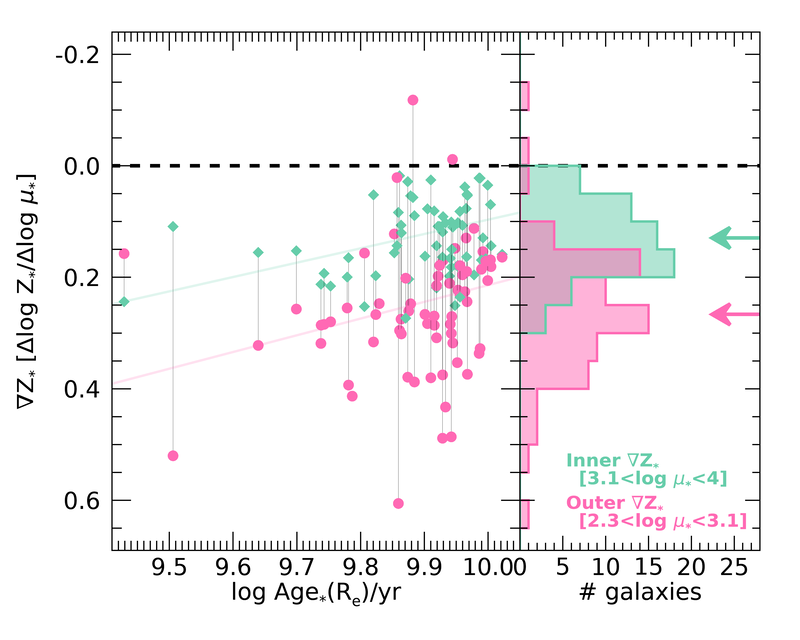}
    \caption{Dependence of various $\zstar$ gradients on the characteristic $\agestar$ at the effective radius, $\agestar(\Reff)$. 
    Symbols and lines as in Fig. \ref{fig:Zgrads_Ze}.}
    \label{fig:Zgrads_Agee}
\end{figure*}

We also investigate possible correlations between various gradients and the metallicity and age evaluated at 
the effective radius, as representative of the global metallicity and age of the galaxy. The results are reported in
Tab. \ref{tab:grad_rad_corr} and \ref{tab:grad_mustar_corr} and the main correlations are plotted in Fig.
\ref{fig:agegrads_Agee}, \ref{fig:Zgrads_Ze}, and \ref{fig:Zgrads_Agee}.

The strength of age gradients in the outer parts display anticorrelation trends with age at $1\,\Reff$.
These trends mainly result from the ages in the outer parts of galaxies being very uniform, so that the
variations in gradients essentially depend on variations in $\agestar(\Reff)$. The larger scatter in the innermost
regions results in non-significant correlations of the inner gradients with $\agestar(\Reff)$.
This can be seen in two panels of Fig. \ref{fig:agegrads_Agee}, for gradients in radial direction and
in $\mustar$ (left and right panel, respectively). Note that the trends are not driven nor made more 
significant by the two galaxies with significantly lower $\agestar(\Reff)$, as we verified by repeating the 
analysis with those galaxies excluded.

Age gradients do not display any strong correlation with $\zstar(\Reff)$. A marginally significant ($P_\text{null}<0.01$) 
anti-correlation is measured between the inner radial gradient of $\agestar$ and $\zstar(\Reff)$, whereby more metal-rich galaxies
tend to have flatter gradients. The slope of this relation, however, is quite flat and results in a
dynamic range for the gradients that is smaller than the scatter around the median value.

As a function of the metallicity at $1\,\Reff$, $\zstar(\Reff)$, we notice a significant trend of the inner radial metallicity 
gradients to become shallower as $\zstar(\Reff)$ increases, as shown in Fig. \ref{fig:Zgrads_Ze}. However, this trend is not seen 
for radial gradients relative to the outer regions. This is consistent with the \emph{central} metallicity being roughly uniform
while the scatter among the profiles keeps increasing while moving outwards up to approximately $1\,\Reff$~(see 
Fig. \ref{fig:stelpop_char_sigma} top left panel and, e.g., Fig. \ref{fig:stacked_profs} top left panel). 
Trends with $\zstar(\Reff)$ for metallicity gradients along $\mustar$ are very mild, although a marginally significant 
trend for outer gradients becoming flatter at larger $\zstar(\Reff)$ is measured.

In Fig. \ref{fig:Zgrads_Agee} we see that radial metallicity gradients beyond $1\,\Reff$ correlate with $\agestar(\Reff)$, going 
from $\sim-0.3\,\text{dex}\,\Reff^{-1}$ for the youngest ETGs to almost $0$ (flat) for the oldest ones. This might indicate
that metallicity gradients tend to be suppressed as galaxies age, possibly due to internal mixing processes or to external
accretion events that mostly affect the outer regions. On the other hand,
neither inner radial metallicity gradients nor metallicity gradients along $\mustar$ display any significant correlation with age.
This lack of strong trends for gradients along of $\mustar$ can be due to the quasi-universal shape for the
profiles of $\zstar(\mustar)$, with just a second order dependence on $\sigmae$ in the outer parts (see previous section and Fig. 
\ref{fig:stelpop_char_sigma} bottom left panel). The aforementioned mixing mechanisms or accretion events should affect $\mustar$
and $\zstar$ in a way that, somehow, preserves the $\zstar(\mustar)$ relation while suppressing the radial $\zstar$ gradients, i.e.
produce a slowly decreasing mass surface density profiles along with a mild radial decrease in metallicity.

Finally, we looked for possible correlations between stellar population gradients and light concentration indices (see end of Sec. 
\ref{sec:SPchars}), but found none. As already noted, the range in concentration indices in our sample of (massive) ETGs is
too small to make a statistically significant detection of correlations possible with a relatively small dataset like ours.
However, based on the results by \cite{Zhuang:2019}, it is tantalizing to think that stellar population gradients may actually 
depend on the light concentration index over a broader range, thus covering not only ETGs but also late type galaxies.

As a general conclusion, we can state that \emph{whenever a trend is visible (even if the statistical significance is low), 
it goes in the direction of older/more metal-rich/more massive/higher velocity-dispersion galaxies having flatter gradients.}

\section{Discussion}\label{sec:discussion}
Stellar population profiles and gradients in ETGs have been the subject of extensive studies
in the last decades, both observationally and theoretically. As already pointed out in the Introduction, the attention to
this topic is justified by the fundamental constraints that stellar population profiles can provide in order to discriminate
between different scenarios of formation and evolution of ETGs. In this section we discuss how our results 
position themselves in this context and suggest a possible evolutionary scenario for the ETGs that can explain our observations.

\subsection{Stellar population profiles: state of observations}\label{subsec:discuss_obs}
From an observational point of view, we can compare our results with various analyses of long-slit or integral field spectroscopic
observations, which mainly focus on radial profiles rather than on their dependence on $\mustar$ 
\citep[with the notable exception of][see below]{Gonzalez-Delgado:2014ab}. 

Age and metallicity estimates in the literature are based 
on a very diverse set of methods (absorption indices, full spectral fitting, photometry) and assumptions regarding the models. 
Popular ``fitting'' methods are based on different philosophies. Concerning the observational data, methods that focus on absorption 
indices (and colours) privilege the reliability of the models predictions over a limited set of features and wavelengths and the 
robustness of the measurements, while full spectral fitting methods privilege the statistical power given by the large number of pixel 
wavelengths at the expenses of possible model mismatches and flux calibration biases.
Differences are also found in the statistical approach, ranging from frequentist to bayesian, from parametric to fully non-parametric.

Concerning the models, there is a huge spectrum of approaches. Comparing with simple stellar population (SSP) models is still very
popular for ETGs, despite their SFH not being for sure a single burst of single metallicity. Among approaches based on composite
stellar populations, choices range from single parametric SFH, to parametric plus stochastic SFH, to fully non-parametric SFH.
Dust attenuation is also treated very differently by various authors: in some works it is assumed to be negligible and is not
modelled, in others it is treated in screen approximation, in others (like in this work) it is implemented in a mixed star-dust
geometry.

How these model assumptions affect (or possibly bias) estimates of ages and metallicity depend also on the statistical method adopted.
We stress that one of the key advantages of our bayesian method is to consider the full PDF of the physical quantities, over
the broadest range of theoretical models, therefore to account for complexity and intrinsic model degeneracy.
Last but not least, different works are based on different basic stellar population synthesis ingredients, such as evolutionary
tracks, isochrones, spectral libraries.

As a consequence, a direct comparison between different works, especially in quantitative terms, is often all but straightforward. 
A comprehensive review of all
these studies and of the reasons of their (dis)agreement is beyond the scope of this paper. It is interesting, however, to highlight
the points that persist among the different studies (and therefore can be considered ``robust'') and the novelties of our analysis
and results. In the following discussion we will consider only results for ETGs in the velocity dispersion range covered by
our present work, $100\lesssim \sigma/[\text{km\,s}^{-1}]\lesssim 300$, or equivalently in the stellar mass range above 
$\sim10^{10.3} \text{M}_\odot$.

\subsubsection{Radial metallicity gradients}
Most past works in the literature agree on the presence of negative metallicity gradients within $1\,\Reff$, whose amplitude 
varies from $-0.3\,\text{dex}$ per $\Reff$ (or equivalently $-0.3\,\text{dex}$ per $\text{dex}$ in radius) to $\sim-0.1$, 
hence broadly consistent with our findings. It is worth noting that different authors adopt different methods to measure gradients.
Some authors work with linear radial scale, some other with logarithmic radial scale (we report both).
Some measure the gradients via linear regression, while some others (including us) estimate the gradient via discrete difference ratio.

\cite{Mehlert:2003} analyzed
a sample of 35 ETGs (both E and S0) in the Coma cluster with long-slit spectroscopic observations that allowed for a minimum spatial
coverage of $1\,\Reff$. Based on stellar absorption indices and SSP predictions, they estimate typical central metallicity values of
$[\text{Z}/\text{H}]\sim 0.2$ and logarithmic radial gradients within $1\,\Reff$ of $-0.16\,\text{dex}$ per $\text{dex}$, with a
r.m.s. of $\sim 0.10$ -- $0.15$. A consistent result was found by \cite{Spolaor:2009} using a compilation of their own data and results
from \cite{Brough:2007,Reda:2007,Sanchez-Blazquez:2007}: their logarithmic radial gradients of metallicity are in the range 
$0$ -- $ -0.5$, with a clear trend for flatter profiles at larger velocity dispersions, provided that the brightest cluster/group 
galaxies are excluded.

\cite{kuntschner+10} analyse a sample of 48 ETGs as part of the IFS survey SAURON, also
based on stellar absorption indices interpreted both with SSP models and with composite stellar populations according to their 
reconstructed SFH. They obtain consistent results with the two approaches and found typical central metallicities of 
$[\text{Z}/\text{H}]\sim 0.1$ and logarithmic radial gradients within $1\,\Reff$ of $\sim -0.3$, very similar to our estimates. 
In their figure 14 one can also note a tentative trend for metallicity gradients to get shallower as $\sigmae$ increases.
A very similar result was found by \cite{Koleva:2011} based on a sample of 40 ETGs observed in long-slit spectroscopy.
Steep negative metallicity gradients, similar to those found in the works listed so far, can be inferred from the aperture-integrated
measurements published by \cite{McDermid:2015} based on ATLAS$^\text{3D}$ data.
 
Stellar population gradients have been derived for large samples of ETGs ($N\gg 100$) in the IFS survey MaNGA, based on different
inference methods. \cite{Goddard:2017b} adopted a full-spectrum fitting approach based on the {\sc firefly} code (see references in
their paper) and found typically mild logarithmic gradients in metallicity, both light- and mass-weighted, in the range $-0.13$ -- $
-0.06$, which is slightly shallower than our measurements. The generally poorer physical 
spatial resolution in MaNGA than in CALIFA (typically a factor 2 to 4 worse, depending on the bundle employed in MaNGA),
may explain part of this difference, as measured profiles appear shallower than in reality, 
as shown by the simulations by \cite{Ibarra-Medel:2019}. 
Interestingly, by mass-weighing they measure marginally flatter gradients than by light-weighing.
Considering the typical temporal evolution of $M/L$, which increases with time, and the almost perfect correspondence between
light- and mass-weighted metallicity in the centre, this effect suggests that at $R\gtrsim 1\,\Reff$ the metallicity
decreases in the youngest stars, contrary to expectations from a scenario of self-enrichment. In turn, this can be interpreted as a 
sign of ex-situ accreted stars or stars formed from external metal-poor gas.
Indeed, the accretion of a metal poor stellar envelope is also invoked by \cite{Oyarzun:2019} to explain the flattening of the
outer metallicity profile that they observe in the most massive ETGs in MaNGA, based on a simple prescription derived from the 
Illustris simulation \citep{DSouza_Bell:2018}.
While the strength of the metallicity gradient is almost unchanged in different mass bins, \cite{Goddard:2017b} show a significant
overall offset of $\sim 0.1\,\text{dex}$ in the two top mass bins ($10.55 < \log M_*/\text{M}_\odot < 11.05$ and 
$\log M_*/\text{M}_\odot > 11.05$). We recall that we find a similar trend with $\sigmae$, while our trend with $M_*$ is small
and not statistically significant.
 
\cite{Li_Mao_Cappellari:2018} performed full-spectrum fitting analysis on a sample of $\sim 2000$ galaxies in MaNGA, 
including $952$ ETGs, by using {\sc ppxf} to estimate light-weighted ages and metallicity.
They found negative metallicity gradients for the vast majority of massive / high-$\sigmae$ ETGs, with logarithmic slopes in the
range $-0.25$ -- $\sim0$, with a median of $\sim-0.1$. These gradients are somewhat shallower than what we measure, however
part of the disagreement might be due to their poorer spatial resolution, and part
to the upper limit of $[\text{Z/H}]<0.22$ built in their models, which does not allow their
metallicity profiles to go as high as ours in the centre of galaxies. In agreement with our findings, these authors also 
identify a clear $\sigma$-metallicity relation and a trend for metallicity gradient to get flatter at larger $\sigma$ (primarily)
and larger $M_*$. These trends are also confirmed by the detailed elemental abundance analysis by \cite{Parikh:2019}, based on  
a sample of 366 ETGs in MaNGA with a pure index fitting approach. The analysis of 303 MaNGA ellipticals 
by \cite{DominguezSanchez:2019}, where the IMF was allowed to vary among and within galaxies, also confirms steep metallicity
gradients for the most massive and high-velocity-dispersion galaxies.

Different analyses of stellar populations across the galaxy extent have been published for the CALIFA survey as well. 
\cite{Gonzalez-Delgado:2015aa} show the average radial profiles of mass-weighted $\log\,Z_*$ 
(as well as light-weighted $\log\,\mli{Age}$) for 41 Es and 32 S0s. Concerning the metallicity, they find remarkably flat profiles,
with logarithmic slopes within $1\,\Reff$ of $\sim 0$ for the ellipticals and $\lesssim 0.1$ for the S0s (as inferred from their
figure 17). This appears inconsistent with the majority of the analysis published in the literature, although the reason for this is
not obvious.
In fact, using a largely overlapping observational dataset, we find much steeper gradients\footnote{Note that the averaging in \cite{Gonzalez-Delgado:2015aa} is done on logarithmic quantities, whereas in the present work we average the linear quantities.
This may possibly lead to a relative negative bias with respect to our estimates and other works in literature 
(see also Appendix \ref{app:systematics}).
Another possible cause of relative bias may be that STARLIGHT's best fit solutions are mostly determined by the overall shape of
the SED (colours) and are somewhat biased towards larger ages, while other solutions of similar likelihood along the age-metallicity
degeneracy curve are discarded.}. Also \cite{Martin-Navarro:2018}
performed an analysis of SSP-equivalent age and metallicity profiles based on stellar absorption indices only, on a similar and
largely overlapping CALIFA dataset, and found relatively steep logarithmic slopes for the metallicity, in the range 
$\sim-0.25$ -- $\sim-0.15$, hence much closer to our estimates and the rest of the literature.

This brief review on radial profiles of stellar metallicity in ETGs can be summarized as follows: \emph{there is a general although not
unanimous consensus on relatively steep metallicity gradients, with $[Z_*/\text{Z}_\odot]$ values that are definitely super-solar in the
centre and decrease to sub-solar values already at $1\,\Reff$. Different results appear to be driven more by different modelling 
techniques rather than by the different observational datasets or statistical limitations.}

\subsubsection{Local $\mustar-\zstar$ relation}\label{subsec:massmetrel}
Most works in the literature focus on the dependence of metallicity on radius, yet we have shown that a very tight and
quasi-universal relation is established between the local surface mass density in stars and their metallicity, whereby this relation
is only modulated by $\sigmae$ (and by $M_*$) in second approximation.
In a previous work \cite{Gonzalez-Delgado:2014ab} have analyzed a sample of 300 CALIFA galaxies covering the full range of morphology
and stellar mass and concluded that both total stellar mass and local stellar mass density affect metallicity. In particular,
they claim that in spheroids the main driver in changing metallicity is global ($M_*$), while local effects ($\mustar$) just 
modulate the global trends.
Despite the direction of the trends is the same as ours, the importance of global-vs-local drivers is reversed. 
This is probably a consequence of their metallicity profiles in ETGs being much flatter than ours, a feature that is in contrast with 
large part of the previous literature.

On the other hand, it is interesting to note that \cite{Scott:2009} showed a significant relation between local metallicity and the
local escape velocity $v_{\text{esc}}$ in SAURON galaxies, which becomes even tighter if a combination of age and metallicity is
considered. However, this is only partially in agreement with our findings. In fact, these authors show that a tight relation between 
$v_{\text{esc}}$ and $\mustar$ exists \emph{in each individual galaxy}, and this implies, in turn, a strong $\mustar-\zstar$ relation.
However, their relation between $v_{\text{esc}}$ and $\mustar$ shifts significantly as a function of galaxy mass. 
Therefore their analysis would not reproduce a universal $\mustar-\zstar$ relation like the one we observe.

Finally, we point out that a significant correlation between surface density of dynamical mass and local $\zstar$ 
was found by \cite{Zhuang:2019} for a general sample of 244 CALIFA galaxies of all Hubble types, not only ETGs. Despite the
stellar metallicity being derived with the same method as in the present work, the results by \cite{Zhuang:2019} strengthen 
the significance of our findings by using an independent estimate of mass surface density, solely based on dynamics, and by extending 
the relation to all morphologies.

\subsubsection{Age profiles}
Age profiles and gradients have been concurrently investigated with metallicity profiles in most of the works cited in the previous
paragraphs. In general ETGs are found to have relatively flat age profiles, with overall variations within 
$\sim 0.1 \,\text{dex}\,(25\%)$ inside $1\,\Reff$. However, there is a broad diversity of results from different studies. 

The strongest negative age gradients are reported by \citet[][CALIFA]{Gonzalez-Delgado:2015aa}, who found that the typical age decrease 
from $\sim  10\,\text{Gyr}$ in the centre to $\sim5\,\text{Gyr}$ at $1\,\Reff$, with a logarithmic slope of $\sim -0.3$.
\citet[][MaNGA]{Li_Mao_Cappellari:2018} report age gradients broadly distributed around $\sim -0.08$, with a significant number of 
galaxies displaying positive gradients. Similarly, \cite{Parikh:2019} find mildly negative age gradients in their analysis of MaNGA
ETGs. Typically flat profiles are reported by both \cite{Koleva:2011} and \citet[][CALIFA]{Martin-Navarro:2018}. Interestingly,
\cite{Goddard:2017b} show that light-weighted and mass-weighted ages have opposite gradients in their analysis of MaNGA ETGs: while
the light-weighted age mildly decreases with radius, the mass-weighted age, which is overall larger by some $0.2$ -- $0.3\,\text{dex}$,
increases with radius. This could be interpreted if we make the hypothesis that the SFH of the integrated (in-situ plus accreted) 
stellar population in the outer parts had a peak further in the past but a longer extension to recent times, with respect to
the stellar populations in the inner parts. In yet another analysis of the MaNGA data, based on spectral indices, 
\cite{DominguezSanchez:2019} report positive age gradients, but only for the most luminous ETGs ($-22.5<M_\text{r}\text{[mag]}<-23.5$)
with the highest velocity dispersion ($2.40<\log(\sigma_0/\text{km\,s}^{-1}<2.50$), while negative age gradients are reported in all
other bins.
Finally, \citet[][SAURON]{kuntschner+10} report mildly positive age
gradients, with typical logarithmic slopes $\sim +0.05$ -- $+0.1$. Similar values can be inferred from the aperture-integrated
ages reported by \citet[][ATLAS$^\text{3D}$]{McDermid:2015}.

Positive age gradients have been reported for S0 galaxies by different authors, based either on absorption features 
\citep[e.g][]{Fisher:1996,Bedregal:2011,Silchenko:2012} or optical-NIR colours \citep{ProchaskaChamberlain:2011}.
They are typically interpreted as evidence of outside-in quenching .

Our U-shaped age profiles fall within the broad range of profile shapes and gradients that are reported in the literature.
Assuming that the U-shape we measure is the true one, this may help explain part of the diversity of previous results, whereby 
a single constant logarithmic slope is assumed to describe a profile whose gradient changes its sign along the radius. 
However, this is (or may be)
only part of the story: the different methods adopted in the different studies are certainly affected by different relative
systematic biases, which stem, e.g., from the different population synthesis models, assumed SFH and metal distributions, observational
constraints, and also different spatial resolution. These biases can easily exceed $0.1\,\text{dex}$ and change in different regimes of age/metallicity.

We stress, on the other hand, that we rely on a very extended yet clean set of observables, i.e. indices that do not suffer of flux 
calibration issues and are chosen to be as insensitive as possible to ingredients that we can poorly model (e.g. variable elemental 
abundance ratios), plus photometry from $3500$\,\AA\, to $9000$\,\AA. Moreover, the spectral library we adopt for the interpretation allows
for the broadest ignorance on the true SFH, chemical composition and dust attenuation. Although, due to the complex pattern of
mutual degeneracies, making a direct connection between the radial profiles of observed quantities and inferred physical 
parameters (age) is all but straightforward, in Appendix \ref{app:ind_col} we show that indeed the inflection of the age 
profiles corresponds to different changes of slopes in the colour and index profiles, which reassures us about the reality of 
the U-shape. Moreover,
the fact that the amplitude of the age minimum correlates with a formally completely independent physical parameter such as $\sigmae$
is a strong indication that there is a real underlying physical effect.
All this, together with the extensive testing we performed (see Sec. \ref{sec:SPanalysis}), makes us confident in the robustness of our 
results on the U-shape of the age profiles.

\subsection{Which evolutionary scenario for ETGs?}\label{subsec:discuss_scenario}
From the theoretical point of view, spatial variations of stellar population properties are a key benchmark for 
our understanding  of the formation and evolution of ETGs. 
Although the historical antithesis between the dissipative (``monolithic'') collapse scenario and the na\"ive rendition of 
the hierarchical merging
scenario, whereby an elliptical is the result of a spiral-spiral merger, is superseded by a much more complex picture where 
the typical physical mechanisms at work in both of them are deeply interlaced,
these two pictures can still be seen as extreme or archetypical scenarios, which can apply to different phases and dominate
in different galaxy regions.

\emph{Dissipative collapse} models that do not implement any other feedback than stellar winds result in steep metallicity gradients 
with logarithmic slopes of $\sim -0.3$ \citep[e.g.][]{Kobayashi:2004} and in positive age gradients as consequence of the outside-in 
quenching driven by stellar winds \citep[so-called ``outside-in formation scenario'', e.g.][]{Pipino:2008,Pipino:2010}
\footnote{Note that a similar effect of outside-in quenching may be produced also by AGN feedback, as recent simulations by 
\cite{Brennan:2018} indicate.}. 
Comparing these predictions with our results, we see that this can be
only part of the story. While our observed metallicity gradients match the dissipative collapse predictions within
$1\,\Reff$, positive age gradients occur only at $R\gtrsim 0.4\,\Reff$ and flatten out at $R\gtrsim 1\,\Reff$.
On the other hand, the effect of major mergers is to flatten the gradients significantly 
\citep[e.g.][]{Hopkins:2009a,Taylor_Kobayashi:2017} and to produce a large scatter.
Therefore the observed steep metallicity gradients, ubiquitous in our sample, and the very small galaxy-to-galaxy scatter 
rule out the possibility that (the inner parts of) ETGs are heavily affected by recent major mergers. 

One possible way to explain the observed reversal of
the age gradients inside $\sim0.4\,\Reff$ (hence the presence of an age minimum) is to invoke some mechanism of
\emph{quenching that acts from the inside outwards}. Such an inside-out quenching is indeed observed in massive star-forming galaxies
at $z\sim2$ \citep[e.g.][]{Tacchella:2015,Tacchella:2018}. \emph{Dynamical heating} due to the growing stellar density 
\citep[``morphological quenching'', e.g.][]{Martig:2009} or \emph{AGN feedback} could
both represent plausible physical mechanisms to shut down star-formation
progressively from the nucleus outwards. This would also qualitatively fit the observation that the age minimum is deeper for
lower-$\sigmae$ galaxies. In fact, in the case of AGN feedback, given the relation that exists between the mass of the 
central super-massive black hole and the velocity dispersion of galaxies \citep[e.g.][]{Gebhardt:2000}, we can envisage that 
galaxies with higher $\sigmae$ had a more powerful and effective
AGN, that was able to shut down the star formation more quickly and up to larger radii, while lower-$\sigmae$ galaxies were 
able to proceed in their outside-in dissipative-like formation for longer time. In fact, we do observe that the age profiles of the 
lowest-$\sigmae$ ETGs in our sample approach a radially almost-monotonic increasing behaviour. This scenario needs more testing,
especially in a quantitative sense, to be confirmed. High-resolution simulations are being analyzed by our group and results will
be presented in a future paper (Hirschmann et al. in preparation).

Alternatively, the shapes of the age profiles may be qualitatively explained by a scenario in which ETGs form the bulk of their
stars within $\sim 1 \Reff$ on timescales that would result in negligible age gradients as observed today.
On the top of these stars, more
recent gas accretion may produce episode(s) of so-called ``\emph{disk re-growth}'', which rejuvenates the stellar population 
\citep[][]{DeLucia+2011}. Such a process, however, would have a variable intensity as a function of radius and, in particular, the 
conversion of gas into stars may be prevented by the hot dynamical state in the inner regions. This would produce a minimum 
in the age profiles 
at intermediate radii as a sort of fossil record of such episodes. The dependence of the age gradient inversion on the 
velocity dispersion is also in qualitative agreement with this hypothesis, in that ETGs with higher velocity dispersion appear less
affected by this kind of rejuvenation than ETGs with lower $\sigmae$. 
Although plausible, also this second scenario is purely speculative
at this stage, and more testing (e.g. by checking against stellar kinematics) and simulations are needed.

Finally, the flattening of the gradients, both in age and metallicity, beyond $\sim1.5\,\Reff$, can be interpreted as the transition to 
a regime, which develops in a \emph{second phase} of galaxy formation, where \emph{stars accreted from satellites} start to represent a portion comparable to the one provided by the in-situ formed stars, 
according to simulations \citep[see, e.g.][]{Hirschmann:2015,Rodriguez-Gomez:2016}. Specifically, the lower mass of the accreted 
satellites would explain why the metallicity profiles flatten out to sub-solar values $\log\zstar/\text{Z}_\odot\sim-0.1$
\citep[see also][]{Oyarzun:2019}. 
The accreted satellites would have been quenched early on \citep[see][]{Pasquali:2010,Pasquali:2019}, and this would explain why 
the age profiles flatten out to maximally old ages of $\approx 10\,\text{Gyr}$. 
Support to such a second phase of ex-situ
stellar accretion is also lent by observations of the distinct chemical and kinematic properties of the stars at large radii
\citep[see e.g.][]{Coccato:2010,Pulsoni:2018}.

This \emph{two-phase formation scenario} for ETGs resembles very closely the one proposed by \cite{Oser:2010}, purely based on
cosmological simulations \citep[see also, e.g.,][]{DeLucia:2006,Navarro-Gonzalez:2013}.
Moreover, it provides a natural explanation for the evolution of high-redshift compact passive galaxies, which are 
commonly interpreted as the descendants of compact sub-mm galaxies that quickly formed stars in an intense dissipative episode and then
grow to present-day ETGs via minor mergers \citep[see e.g.][]{Toft:2014, Oser:2012, Choi:2018}. 
In this framework it is also possible to explain,
at least qualitatively, many of the open issues that we outlined in the Introduction. Chiefly this applies to the size growth, 
but also the enhancement in $\alpha$ elements and the systematic variations of IMF may be interpreted as a consequence of the
different density at which stars belonging to different (portions of) galaxies were formed, either in the main progenitor(s) or in the
accreted satellites.


\section{Summary and conclusions}\label{sec:conclusions}
In this paper we have analyzed the spatial variations of the $r$-band-light-weighted mean age and metallicity of the stellar
populations in 69 ETGs drawn from the CALIFA survey, including 48 ellipticals and 21 S0s. Our analysis is based on a bayesian
statistical approach, whereby a state-of-the-art suite of 500\,000 spectral synthesis models is compared to the observed maps of
stellar absorption indices and broad-band fluxes, which is so far unique in the literature. The large field of view and the
depth of the CALIFA-SDSS dataset is such that we can reliably cover the radial extent of our galaxies out to 1.5 $\Reff$ ($2 \Reff$)
for $74\%$ ($46\%$) of the sample with a median resolution of $0.08\,\Reff$, and we are essentially complete to 
stellar mass surface density of $100\,\text{M}_\odot\,\text{pc}^{-2}$.

We have shown that steep negative radial gradients in metallicity are ubiquitous and very consistent among galaxies over a relatively
broad range of velocity dispersion, stellar mass and morphology. The central regions of ETGs reach metallicities more
than twice solar and decrease below solar abundance between 1 and $1.5\,\Reff$. 
On the contrary, age profiles show relatively small radial variations, of the order of $40\%$ at most on average. The most striking
feature of the age profiles is that, on average, there is an inversion of slope at $\sim 0.3$ -- $0.4\,\Reff$, where a minimum age is reached, i.e. profiles are U-shaped.
All ETGs share maximally old ages at large galactocentric distances ($\mli{SMA}\gtrsim 1.5\,\Reff$). Within $\sim 1\,\Reff$ we note an
increased scatter, that mainly correlates with the global velocity dispersion of the galaxies, $\sigmae$. High-$\sigmae$ galaxies
display a shallow minimum at $\sim 0.3$ -- $0.4\,\Reff$ and older age in the center. For lower-$\sigmae$ galaxies the minimum age 
decreases and the inversion of the age profile towards the center becomes weaker.

Metallicity appears to be primarily determined by the local stellar mass surface density $\mustar$, 
as shown by the tiny galaxy-to-galaxy scatter in the $\zstar$ vs. $\mustar$ plot. There is, in fact, a \emph{quasi-universal
local mass-metallicity relation}, which is only mildly modulated by $\sigmae$, whereby the highest-$\sigmae$ galaxies 
($\sim 300\,\text{km\,s}^{-1}$) have their metallicity profiles offset to $\sim 0.15\,\text{dex}$ higher values than the
lowest-$\sigmae$ ones ($\sim 150\,\text{km\,s}^{-1}$). This result confirms in greater detail, yet over a more limited set of galaxies,
the findings by \cite{Zhuang:2019}. 

For both age and metallicity, we have analyzed in great detail possible dependencies of profile shapes, normalizations and 
gradients on global galaxy properties, such as $\sigmae$, stellar mass, morphology, light-concentration, global age and metallicity. 
We showed that the most significant correlations are found with $\sigmae$. Correlations with stellar mass are weaker and appear as 
largely inherited from the correlations with $\sigmae$ via the $\sigmae$-mass relation. All other parameters, including morphology
(i.e. E vs. S0), affect the profiles only in a minor way, if anything.

Finally, we have discussed how the observations reported in this work support a two-phase formation scenario for ETGs. 
The first phase
would be dominated by in-situ star formation in a dissipative collapse (series of) episode(s), which determines the (inner) strongly 
negative metallicity gradient and, due to outside-in wind-driven quenching, an initial positive age gradient. 
The onset of powerful AGN feedback (whose effectiveness scales with $\sigmae$, 
via the well known correlation between this quantity and the mass of the central black hole) or the development of strong dynamical 
heating would determine an inside-out quenching, 
hence the reversal in the age profiles that we observe at $0.3$ -- $0.4\,\Reff$. Alternatively, the age minimum at $\sim 0.4 \Reff$
might be the fossil record of disk regrowth in the inner regions, whose star formation efficiency is regulated (or suppressed) 
by the local velocity dispersion.
This phase would be mainly recorded in the properties of the inner $\sim 1\,\Reff$. Outside of this radius, a second phase dominated by
minor mergers and accretion events of early-on quenched satellites would then produce the flattening of the profiles around ages 
of $\sim 10\,\text{Gyr}$ and sub-solar metallicities.

In a forthcoming paper (Hirschmann et al., in preparation) we will investigate how cosmological SPH simulations at high-resolution 
and implementing AGN feedback can reproduce the current observations and possibly lend support to the proposed scenario.

\section*{Acknowledgements}
S.~Z. wishes to dedicate this work to the memory of Prof. Franco Colombo, whose skills and dedication in teaching and nurturing
young talents will never be forgotten.

We thank the anonymous referee for the constructive report that has lead us to a better assessment of uncertainties and systematics
and to improve the discussion on possible interpretations. We are also grateful to Gabriella De Lucia, Fabio Fontanot, 
Mariangela Bernardi, Sandro Tacchella, and Magda Arnaboldi for useful discussions and suggestions.

A.~R.~G. and S.~Z. are supported by the INAF PRIN-SKA 2017 program 1.05.01.88.04.ESKAPE-HI.\\
M.~H. acknowledges financial support from the Carlsberg Foundation via a 
Semper Ardens grant  (CF15-0384), and the support from the visiting program at INAF-O.A.Arcetri.\\
J.~F.-B. acknowledges financial support from grant AYA2016-77237-C3-1-P from the Spanish Ministry of 
Economy and Competitiveness (MINECO).\\
G.~v.~d.~V. acknowledges funding from the European Research Council (ERC) under the European Union's Horizon 2020 research 
and innovation programme under grant agreement No 724857 (Consolidator Grant ArcheoDyn).



\bibliographystyle{mnras}
\bibliography{all_mybibs} 




\appendix
\section{On the effective spatial resolution of the maps}\label{app:PSF}
\begin{figure}
\centerline{\includegraphics[width=0.45\textwidth]{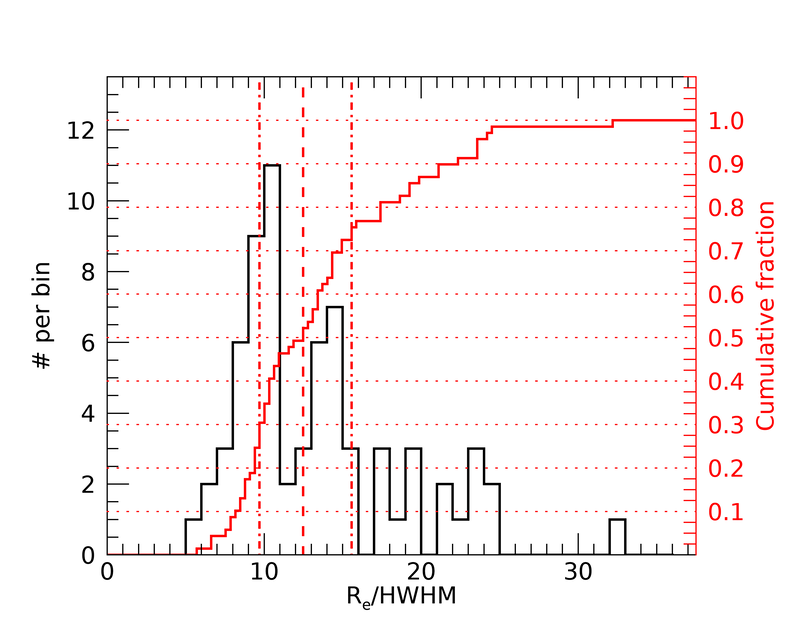}}
\caption{Differential (black line) and cumulative (red line) distribution of the effective radius $\Reff$ of the galaxies analyzed
in this study, normalized by the PSF radius, i.e. the Half-Width at Half Maximum (HWHM) of the PSF. The vertical dashed red line
indicates the median of the distribution, the dot-dashed red lines mark the lower and upper quartiles.}
\label{fig:PSFdistr}
\end{figure}

The effective physical resolution of our stellar population maps is quantified by the ratio of the effective radius $\Reff$ (in arcsec)
to the PSF radius, defined as the Half-Width at Half Maximum, i.e. $2.57/2=1.29$\,arcsec (median value for CALIFA datacubes).
Fig. \ref{fig:PSFdistr} presents both the differential (black lines) and the cumulative (red solid line) distribution of such ratio
$\Reff/\text{HWHM}$ for the full sample.
With a median value of 12.48 (dashed vertical line) and lower and upper quartiles of 9.70 and 15.56 respectively (dot-dashed
vertical lines), we can safely assert that we resolve the stellar population profiles down to 0.1 $\Reff$ for the majority of
the galaxies. Thus our results, especially those regarding the presence of a minimum of the age around $0.4\,\Reff$ are not
significantly biased by resolution effects.

\section{Systematic effects related to averaging scheme and IMF}\label{app:systematics}
In this work we have considered the $r$-band-light-weighted mean age $\agestar$ and metallicity $\zstar$, defined
from the linear parameter as
follows (see also \citetalias{zibetti+17} sec. 2.3):
\begin{equation}
\agestar=\frac{\int\limits_{t=0}^{t_0} \mathrm{d}t~(t_0-t)~\mathrm{SFR}(t)~\mathscr{L}^\prime(t)}{\int\limits_{t=0}^{t_0} \mathrm{d}t~\mathrm{SFR}(t)~\mathscr{L}^\prime(t)}\label{eq:lwage}
\end{equation}
\begin{equation}
\zstar=\frac{\int\limits_{t=0}^{t_0} \mathrm{d}t~\zstar(t)~\mathrm{SFR}(t)~\mathscr{L}^\prime(t)}{\int\limits_{t=0}^{t_0} \mathrm{d}t~\mathrm{SFR}(t)~\mathscr{L}^\prime(t)}\label{eq:lwage}
\end{equation}
where $t_0$ is the time elapsed since the start of the SFH and $\mathscr{L}^\prime(t)$ is the luminosity arising per unit \emph{formed} stellar mass from the ensemble of SSPs of age $t_0-t$.
Other works in
the literature adopt instead the average of logarithmic quantities \citep[e.g.][]{Gonzalez-Delgado:2015aa}. Here we want to briefly 
discuss how the two estimates should compare, based on the corresponding maps that we produce with our bayesian method for the two
averaging schemes.

Log-averaged quantities give more weight to young ages and low $\zstar$ and therefore should be biased in that sense with respect to linear-averaged quantities. From the direct comparison of our maps, we see indeed that log-averaged age maps are negatively biased, but
the bias is stronger for younger ages. In the oldest regions the age bias is a few $-0.01$ dex at most, while in the youngest ones
the bias is around $-0.1$ -- $-0.15$ dex. As a consequence, log-averaged age profiles not only are shifted to slightly younger ages overall, but are also amplified and have deeper minima with respect to the linear-averaged ones.
In $\zstar$ maps there is also a negative relative bias of the log-averaged values with respect to the linear-averaged ones. However it is more uniform across the range, going from some $-0.05$ -- $-0.1$ dex at super-solar metallicities to $-0.1$ -- $-0.15$ dex at sub-solar metallicities. As a consequence, log-averaged metallicity profiles are expected to be lower overall and slightly steeper than the
linear-averaged ones.

All in all, the adoption of a log-averaging scheme would not change the \emph{qualitative} picture emerging from the analysis presented 
in the paper, although the \emph{quantitative} characterization would be slightly affected.

Despite the growing evidence in the literature for a variable stellar IMF, even within the extent of individual galaxies 
\citep[e.g.][]{Martin-Navarro:2015}, in this work we have interpreted all observations assuming a universal \cite{chabrier03} IMF.
By comparison with a version of our own spectral library based on a \cite{Salpeter:1955} IMF, we have verified that the observable 
quantities (absorption indices and colours) used to infer stellar population properties display negligible dependence on the IMF at
fixed SFH and chemical enrichment history. Intensive quantities such as age and metallicity are therefore unaffected by IMF variations
in our analysis. 

As opposed, IMF variations do affect the M/L ratio in a very evident and systematic way. Going from a 
\cite{chabrier03} IMF to a \cite{Salpeter:1955} IMF the ratio M/L increases by a factor $1.75$ (i.e. $+0.24$\,dex) according to 
\cite{gallazzi+08}. Systematic galaxy-wide variations may affect the scatter in the relations between $\mustar$ and $\zstar$ and
between $\mustar$ and $\agestar$. If IMF variations occur following a radial trend within galaxies similar to those presented in, e.g.,
\cite{Martin-Navarro:2015}, they may change the shapes and, in particular, the slopes of such relations as well,
by making $\mustar(\SMA)$ profiles steeper and hence stretching the relations with $\mustar$. A shift by $\sim0.25$\,dex in $\mustar$, 
however, does not produce substantial qualitative changes in the relations presented in this work. On the other hand, IMF variations
that depend both on radius and on the luminosity/mass/velocity dispersion of the galaxy, as presented by \cite{DominguezSanchez:2019},
would modify the secondary dependence of the $\zstar-\mustar$ relation on $\sigmae$ (see Fig. \ref{fig:stelpop_char_sigma}, 
bottom-left panel), in the sense of reducing such dependence.

\section{Tables of stellar population properties at characteristic locations}
\newcommand\T{\rule{0pt}{2.6ex}}       
\newcommand\B{\rule[-1.2ex]{0pt}{0pt}} 
\begin{landscape}
\begin{table}
  \centering
 \caption{Stellar population properties at characteristic radii. The galaxy (sub)sample is indicated in column (1). The following
 groups of three columns each report metallicity and age (with related $16^\text{th}-84^\text{th}$ percentile ranges), and the number 
 of galaxies in the (sub)sample for the five radial regions, as labeled in the top headline. See main text for details.}
 \label{tab:char_stelpop_sma}
\scriptsize
 \begin{tabular}{c|rrc|rrc|rrc|rrc|rrc} 
\hline
Sample & \multicolumn{3}{c}{Center $(\mli{SMA}<0.1\,R_e)$}& \multicolumn{3}{c}{$\mli{SMA}=0.2\,R_e$}& \multicolumn{3}{c}{$\mli{SMA}=0.4\,R_e$}& \multicolumn{3}{c}{$\mli{SMA}=R_e$}& \multicolumn{3}{c}{$\mli{SMA}=2\,R_e$}\\
&$\log Z_*/Z_\odot$&$\log \mli{Age}_*/[\rm{yr}]$&$N$
&$\log Z_*/Z_\odot$&$\log \mli{Age}_*/[\rm{yr}]$&$N$
&$\log Z_*/Z_\odot$&$\log \mli{Age}_*/[\rm{yr}]$&$N$
&$\log Z_*/Z_\odot$&$\log \mli{Age}_*/[\rm{yr}]$&$N$
&$\log Z_*/Z_\odot$&$\log \mli{Age}_*/[\rm{yr}]$&$N$\\
(1) & (2) & (3) & (4) & (5) & (6) & (7) & (8) & (9) & (10) & (11) & (12) & (13) & (14) & (15) & (16)\\
\hline
All ETGs &$ 0.27^{+0.04}_{-0.04}$ & $ 9.94^{+0.10}_{-0.26}$ & $69$ & $ 0.25^{+0.04}_{-0.07}$ & $ 9.85^{+0.13}_{-0.19}$ & $69$ & $ 0.18^{+0.09}_{-0.07}$ & $ 9.84^{+0.09}_{-0.15}$ & $69$ & $ 0.04^{+0.08}_{-0.12}$ & $ 9.92^{+0.05}_{-0.13}$ & $65$ & $-0.06^{+0.10}_{-0.06}$ & $ 9.95^{+0.05}_{-0.07}$ & $32$ \B \T \\
$\log M_*/\rm{M_\odot} < 10.9 $&$ 0.25^{+0.05}_{-0.10}$ & $ 9.92^{+0.12}_{-0.29}$ & $16$ & $ 0.25^{+0.04}_{-0.12}$ & $ 9.82^{+0.17}_{-0.21}$ & $16$ & $ 0.18^{+0.09}_{-0.14}$ & $ 9.73^{+0.17}_{-0.10}$ & $16$ & $ 0.06^{+0.06}_{-0.21}$ & $ 9.86^{+0.06}_{-0.12}$ & $16$ & $-0.09^{+0.14}_{-0.13}$ & $ 9.93^{+0.04}_{-0.07}$ & $11$ \B \T \\
$10.9 \leq \log M_*/\rm{M_\odot} < 11.3 $&$ 0.27^{+0.04}_{-0.04}$ & $ 9.95^{+0.09}_{-0.30}$ & $31$ & $ 0.23^{+0.07}_{-0.04}$ & $ 9.84^{+0.12}_{-0.19}$ & $31$ & $ 0.18^{+0.10}_{-0.08}$ & $ 9.81^{+0.13}_{-0.10}$ & $31$ & $ 0.03^{+0.11}_{-0.11}$ & $ 9.93^{+0.04}_{-0.14}$ & $29$ & $-0.06^{+0.07}_{-0.12}$ & $ 9.95^{+0.06}_{-0.10}$ & $14$ \B \T \\
$\log M_*/\rm{M_\odot} \geq 11.3 $&$ 0.28^{+0.05}_{-0.03}$ & $ 9.95^{+0.08}_{-0.14}$ & $22$ & $ 0.26^{+0.03}_{-0.06}$ & $ 9.89^{+0.12}_{-0.09}$ & $22$ & $ 0.20^{+0.07}_{-0.06}$ & $ 9.91^{+0.05}_{-0.11}$ & $22$ & $ 0.08^{+0.03}_{-0.06}$ & $ 9.96^{+0.04}_{-0.05}$ & $20$ & $ 0.01^{+0.07}_{-0.09}$ & $ 9.97^{+0.02}_{-0.02}$ & $7$ \B \T \\
$\sigma_e/[\rm{km\,s}^{-1}] < 170 $&$ 0.26^{+0.04}_{-0.04}$ & $ 9.68^{+0.30}_{-0.08}$ & $21$ & $ 0.20^{+0.09}_{-0.04}$ & $ 9.76^{+0.09}_{-0.16}$ & $21$ & $ 0.17^{+0.05}_{-0.10}$ & $ 9.73^{+0.12}_{-0.13}$ & $21$ & $ 0.00^{+0.09}_{-0.16}$ & $ 9.87^{+0.07}_{-0.13}$ & $21$ & $-0.10^{+0.11}_{-0.18}$ & $ 9.92^{+0.03}_{-0.10}$ & $12$ \B \T \\
$170 \leq \sigma_e/[\rm{km\,s}^{-1}] < 210 $&$ 0.27^{+0.03}_{-0.03}$ & $ 9.95^{+0.11}_{-0.11}$ & $25$ & $ 0.23^{+0.07}_{-0.09}$ & $ 9.89^{+0.13}_{-0.12}$ & $25$ & $ 0.17^{+0.10}_{-0.09}$ & $ 9.82^{+0.14}_{-0.11}$ & $25$ & $ 0.06^{+0.06}_{-0.14}$ & $ 9.92^{+0.03}_{-0.11}$ & $22$ & $-0.07^{+0.05}_{-0.03}$ & $ 9.96^{+0.04}_{-0.11}$ & $10$ \B \T \\
$\sigma_e/[\rm{km\,s}^{-1}] \geq 210 $&$ 0.28^{+0.05}_{-0.03}$ & $ 9.99^{+0.09}_{-0.18}$ & $23$ & $ 0.27^{+0.03}_{-0.04}$ & $ 9.92^{+0.09}_{-0.10}$ & $23$ & $ 0.23^{+0.06}_{-0.06}$ & $ 9.91^{+0.06}_{-0.10}$ & $23$ & $ 0.08^{+0.06}_{-0.06}$ & $ 9.96^{+0.04}_{-0.05}$ & $22$ & $ 0.02^{+0.07}_{-0.10}$ & $ 9.98^{+0.03}_{-0.02}$ & $10$ \B \T \\
All Es&$ 0.27^{+0.04}_{-0.04}$ & $ 9.95^{+0.08}_{-0.23}$ & $48$ & $ 0.25^{+0.04}_{-0.06}$ & $ 9.87^{+0.12}_{-0.19}$ & $48$ & $ 0.18^{+0.09}_{-0.09}$ & $ 9.87^{+0.07}_{-0.15}$ & $48$ & $ 0.03^{+0.08}_{-0.12}$ & $ 9.94^{+0.04}_{-0.08}$ & $44$ & $-0.05^{+0.13}_{-0.19}$ & $ 9.96^{+0.04}_{-0.08}$ & $18$ \B \T \\
S0 &$ 0.28^{+0.03}_{-0.06}$ & $ 9.84^{+0.20}_{-0.20}$ & $21$ & $ 0.26^{+0.03}_{-0.08}$ & $ 9.82^{+0.14}_{-0.19}$ & $21$ & $ 0.20^{+0.08}_{-0.07}$ & $ 9.79^{+0.07}_{-0.14}$ & $21$ & $ 0.09^{+0.04}_{-0.13}$ & $ 9.86^{+0.09}_{-0.11}$ & $21$ & $-0.07^{+0.09}_{-0.02}$ & $ 9.94^{+0.06}_{-0.08}$ & $14$ \B \T \\
Es, $\log M_*/\rm{M_\odot} < 11.3 $&$ 0.26^{+0.04}_{-0.04}$ & $ 9.95^{+0.09}_{-0.30}$ & $28$ & $ 0.23^{+0.06}_{-0.07}$ & $ 9.83^{+0.13}_{-0.18}$ & $28$ & $ 0.17^{+0.06}_{-0.10}$ & $ 9.81^{+0.13}_{-0.17}$ & $28$ & $ 0.00^{+0.06}_{-0.16}$ & $ 9.92^{+0.02}_{-0.12}$ & $26$ & $-0.06^{+0.10}_{-0.22}$ & $ 9.94^{+0.07}_{-0.12}$ & $12$ \B \T \\
Es, $\sigma_e/[\rm{km\,s}^{-1}] < 170 $&$ 0.24^{+0.06}_{-0.09}$ & $ 9.68^{+0.31}_{-0.08}$ & $11$ & $ 0.20^{+0.09}_{-0.06}$ & $ 9.68^{+0.23}_{-0.08}$ & $11$ & $ 0.13^{+0.10}_{-0.09}$ & $ 9.73^{+0.14}_{-0.13}$ & $11$ & $-0.02^{+0.13}_{-0.18}$ & $ 9.88^{+0.07}_{-0.18}$ & $11$ & $-0.17^{+0.12}_{-0.10}$ & $ 9.89^{+0.04}_{-0.07}$ & $7$ \B \T \\
Es, $170 \leq \sigma_e/[\rm{km\,s}^{-1}] < 210 $&$ 0.26^{+0.03}_{-0.02}$ & $ 9.96^{+0.07}_{-0.10}$ & $16$ & $ 0.23^{+0.05}_{-0.09}$ & $ 9.89^{+0.12}_{-0.12}$ & $16$ & $ 0.15^{+0.09}_{-0.19}$ & $ 9.90^{+0.04}_{-0.11}$ & $16$ & $-0.05^{+0.08}_{-0.04}$ & $ 9.93^{+0.01}_{-0.07}$ & $13$ & $-0.06^{+0.03}_{-0.18}$ & $ 9.97^{+0.04}_{-0.02}$ & $3$ \B \T \\
Es, $\sigma_e/[\rm{km\,s}^{-1}] \geq 210 $&$ 0.28^{+0.05}_{-0.03}$ & $ 9.97^{+0.08}_{-0.16}$ & $21$ & $ 0.27^{+0.03}_{-0.04}$ & $ 9.91^{+0.09}_{-0.10}$ & $21$ & $ 0.23^{+0.06}_{-0.06}$ & $ 9.91^{+0.05}_{-0.10}$ & $21$ & $ 0.08^{+0.03}_{-0.06}$ & $ 9.96^{+0.04}_{-0.04}$ & $20$ & $ 0.04^{+0.05}_{-0.09}$ & $ 9.98^{+0.02}_{-0.02}$ & $8$ \B \T \\
\hline
 \end{tabular}
 \normalsize
\end{table}
\end{landscape}

\begin{landscape}
\begin{table}
  \centering
 \caption{Stellar population properties at characteristic stellar mass surface density $\mustar$. The galaxy (sub)sample is indicated in column (1). The following
 groups of three columns each report metallicity and age (with related $16^\text{th}-84^\text{th}$ percentile ranges), and the number 
 of galaxies in the (sub)sample for the three characteristic $\mustar$, as labeled in the top headline. See main text for details.}
 \label{tab:char_stelpop_mustar}
 \begin{tabular}{c|rrc|rrc|rrc} 
\hline
Sample & \multicolumn{3}{c}{Center ($\log \mu_*=4.0)$}& \multicolumn{3}{c}{Mid ($\log \mu_*=3.1)$}& \multicolumn{3}{c}{Outer ($\log \mu_*=2.3)$}\\
&$\log Z_*/Z_\odot$&$\log \mli{Age}_*/[\rm{yr}]$&$N$
&$\log Z_*/Z_\odot$&$\log \mli{Age}_*/[\rm{yr}]$&$N$
&$\log Z_*/Z_\odot$&$\log \mli{Age}_*/[\rm{yr}]$&$N$\\
(1) & (2) & (3) & (4) & (5) & (6) & (7) & (8) & (9) & (10)\\
\hline
All ETGs &$ 0.29^{+0.03}_{-0.05}$ & $ 9.91^{+0.11}_{-0.15}$ & $63$ & $ 0.16^{+0.05}_{-0.07}$ & $ 9.83^{+0.10}_{-0.12}$ & $69$ & $-0.06^{+0.11}_{-0.08}$ & $ 9.96^{+0.04}_{-0.10}$ & $69$ \B \T \\
$\log M_*/\rm{M_\odot} < 10.9 $&$ 0.28^{+0.02}_{-0.12}$ & $ 9.88^{+0.15}_{-0.21}$ & $14$ & $ 0.11^{+0.08}_{-0.04}$ & $ 9.79^{+0.12}_{-0.15}$ & $16$ & $-0.10^{+0.05}_{-0.08}$ & $ 9.95^{+0.04}_{-0.15}$ & $16$ \B \T \\
$10.9 \leq \log M_*/\rm{M_\odot} < 11.3 $&$ 0.28^{+0.04}_{-0.06}$ & $ 9.91^{+0.13}_{-0.14}$ & $28$ & $ 0.15^{+0.07}_{-0.07}$ & $ 9.83^{+0.11}_{-0.12}$ & $31$ & $-0.07^{+0.12}_{-0.15}$ & $ 9.95^{+0.04}_{-0.12}$ & $31$ \B \T \\
$\log M_*/\rm{M_\odot} \geq 11.3 $&$ 0.30^{+0.02}_{-0.02}$ & $ 9.96^{+0.03}_{-0.09}$ & $21$ & $ 0.19^{+0.04}_{-0.02}$ & $ 9.87^{+0.10}_{-0.06}$ & $22$ & $ 0.03^{+0.04}_{-0.09}$ & $ 9.98^{+0.02}_{-0.04}$ & $22$ \B \T \\
$\sigma_e/[\rm{km\,s}^{-1}] < 170 $&$ 0.26^{+0.04}_{-0.09}$ & $ 9.82^{+0.19}_{-0.16}$ & $17$ & $ 0.14^{+0.04}_{-0.07}$ & $ 9.77^{+0.11}_{-0.16}$ & $21$ & $-0.11^{+0.11}_{-0.11}$ & $ 9.93^{+0.03}_{-0.12}$ & $21$ \B \T \\
$170 \leq \sigma_e/[\rm{km\,s}^{-1}] < 210 $&$ 0.28^{+0.04}_{-0.06}$ & $ 9.89^{+0.17}_{-0.23}$ & $23$ & $ 0.13^{+0.06}_{-0.06}$ & $ 9.83^{+0.11}_{-0.09}$ & $25$ & $-0.09^{+0.08}_{-0.02}$ & $ 9.96^{+0.05}_{-0.11}$ & $25$ \B \T \\
$\sigma_e/[\rm{km\,s}^{-1}] \geq 210 $&$ 0.30^{+0.03}_{-0.02}$ & $ 9.95^{+0.04}_{-0.08}$ & $23$ & $ 0.20^{+0.05}_{-0.03}$ & $ 9.90^{+0.08}_{-0.09}$ & $23$ & $ 0.04^{+0.04}_{-0.08}$ & $ 9.98^{+0.02}_{-0.02}$ & $23$ \B \T \\
All Es&$ 0.29^{+0.02}_{-0.06}$ & $ 9.92^{+0.08}_{-0.13}$ & $44$ & $ 0.17^{+0.06}_{-0.07}$ & $ 9.86^{+0.08}_{-0.10}$ & $48$ & $-0.04^{+0.10}_{-0.12}$ & $ 9.96^{+0.04}_{-0.10}$ & $48$ \B \T \\
S0 &$ 0.28^{+0.03}_{-0.08}$ & $ 9.87^{+0.16}_{-0.18}$ & $19$ & $ 0.15^{+0.05}_{-0.05}$ & $ 9.81^{+0.12}_{-0.13}$ & $21$ & $-0.08^{+0.07}_{-0.06}$ & $ 9.96^{+0.04}_{-0.13}$ & $21$ \B \T \\
Es, $\log M_*/\rm{M_\odot} < 11.3 $&$ 0.27^{+0.05}_{-0.05}$ & $ 9.91^{+0.12}_{-0.22}$ & $25$ & $ 0.13^{+0.07}_{-0.06}$ & $ 9.83^{+0.09}_{-0.18}$ & $28$ & $-0.10^{+0.15}_{-0.09}$ & $ 9.95^{+0.02}_{-0.12}$ & $28$ \B \T \\
Es, $\sigma_e/[\rm{km\,s}^{-1}] < 170 $&$ 0.25^{+0.04}_{-0.10}$ & $ 9.81^{+0.11}_{-0.14}$ & $9$ & $ 0.13^{+0.06}_{-0.06}$ & $ 9.76^{+0.12}_{-0.15}$ & $11$ & $-0.15^{+0.16}_{-0.07}$ & $ 9.91^{+0.04}_{-0.12}$ & $11$ \B \T \\
Es, $170 \leq \sigma_e/[\rm{km\,s}^{-1}] < 210 $&$ 0.27^{+0.04}_{-0.04}$ & $ 9.93^{+0.13}_{-0.12}$ & $14$ & $ 0.12^{+0.08}_{-0.05}$ & $ 9.87^{+0.07}_{-0.06}$ & $16$ & $-0.09^{+0.04}_{-0.04}$ & $ 9.96^{+0.01}_{-0.11}$ & $16$ \B \T \\
Es, $\sigma_e/[\rm{km\,s}^{-1}] \geq 210 $&$ 0.30^{+0.03}_{-0.01}$ & $ 9.95^{+0.04}_{-0.08}$ & $21$ & $ 0.20^{+0.05}_{-0.03}$ & $ 9.90^{+0.08}_{-0.08}$ & $21$ & $ 0.04^{+0.03}_{-0.07}$ & $ 9.98^{+0.02}_{-0.02}$ & $21$ \B \T \\
\hline
 \end{tabular}
\end{table}
\end{landscape}
\let\T\undefined       
\let\B\undefined 

\newcommand\T{\rule{0pt}{2.6ex}}       
\newcommand\B{\rule[-1.2ex]{0pt}{0pt}} 

\begin{table*}
  \centering
 \caption{Correlations between stellar population properties at characteristic radii and global properties. (1) Checked if correlation is significant; (2)-(3) $y$ and $x$ variables; (4) region in which stellar population properties $y$ are computed; (5)-(6) coefficients of the linear fit
 $y=a+b\,x$; (7) mean absolute deviation (MAD); (8) Spearman rank coefficient; (9) probability of null correlation.}
 \label{tab:corr_char_stelpop_sma}
 \begin{tabular}{ccccrrrrrrrr} 
\hline
Significant  & $y$               &$x$               &Region                            &$a$       &$b$      &MAD      &$C_{\rm{Spearman}}$ & $P_{\rm{null}}$ \\
(1)          & (2)               &(3)               &(4)                               &(5)       & (6)     &(7)      & (8)                & (9) \\
  \hline
$\checkmark$&                      logZ&               $\log\sigma_\text{e}$&    Center $(\mli{SMA}<0.1\,R_\text{e})$ & $  -0.089$ & $  +0.158$& $   0.035$& $  +0.316$ & $\rm{ 8.205e-03}$\\
$\checkmark$&                      logZ&               $\log\sigma_\text{e}$&              $\mli{SMA}=0.2\,R_\text{e}$& $  -0.627$ & $  +0.378$& $   0.045$& $  +0.382$ & $\rm{ 1.287e-03}$\\
$\checkmark$&                      logZ&               $\log\sigma_\text{e}$&              $\mli{SMA}=0.4\,R_\text{e}$& $  -0.548$ & $  +0.321$& $   0.062$& $  +0.363$ & $\rm{ 2.335e-03}$\\
$\checkmark$&                      logZ&               $\log\sigma_\text{e}$&                $\mli{SMA}=1\,R_\text{e}$& $  -0.847$ & $  +0.390$& $   0.074$& $  +0.365$ & $\rm{ 2.757e-03}$\\
$\checkmark$&                      logZ&               $\log\sigma_\text{e}$&               $\mli{SMA}=2\,R_\text{e}$ & $  -1.223$ & $  +0.518$& $   0.064$& $  +0.563$ & $\rm{ 8.014e-04}$\\
$\checkmark$&                    logAge&               $\log\sigma_\text{e}$&    Center $(\mli{SMA}<0.1\,R_\text{e})$ & $  +7.641$ & $  +0.994$& $   0.113$& $  +0.475$ & $\rm{ 3.766e-05}$\\
$\checkmark$&                    logAge&               $\log\sigma_\text{e}$&              $\mli{SMA}=0.2\,R_\text{e}$& $  +8.191$ & $  +0.722$& $   0.093$& $  +0.536$ & $\rm{ 2.492e-06}$\\
$\checkmark$&                    logAge&               $\log\sigma_\text{e}$&              $\mli{SMA}=0.4\,R_\text{e}$& $  +8.206$ & $  +0.708$& $   0.075$& $  +0.589$ & $\rm{ 1.250e-07}$\\
$\checkmark$&                    logAge&               $\log\sigma_\text{e}$&                $\mli{SMA}=1\,R_\text{e}$& $  +9.020$ & $  +0.391$& $   0.065$& $  +0.576$ & $\rm{ 5.208e-07}$\\
$\checkmark$&                    logAge&               $\log\sigma_\text{e}$&               $\mli{SMA}=2\,R_\text{e}$ & $  +9.403$ & $  +0.238$& $   0.048$& $  +0.558$ & $\rm{ 8.986e-04}$\\
            &                      logZ&                          $\log M_*$&    Center $(\mli{SMA}<0.1\,R_\text{e})$ & $  -0.308$ & $  +0.052$& $   0.035$& $  +0.243$ & $\rm{ 4.449e-02}$\\
            &                      logZ&                          $\log M_*$&              $\mli{SMA}=0.2\,R_\text{e}$& $  -0.095$ & $  +0.030$& $   0.049$& $  +0.115$ & $\rm{ 3.525e-01}$\\
            &                      logZ&                          $\log M_*$&              $\mli{SMA}=0.4\,R_\text{e}$& $  +0.188$ & $  -0.001$& $   0.067$& $  +0.034$ & $\rm{ 7.858e-01}$\\
            &                      logZ&                          $\log M_*$&                $\mli{SMA}=1\,R_\text{e}$& $  -0.298$ & $  +0.031$& $   0.078$& $  +0.191$ & $\rm{ 1.271e-01}$\\
            &                      logZ&                          $\log M_*$&               $\mli{SMA}=2\,R_\text{e}$ & $  -1.483$ & $  +0.130$& $   0.075$& $  +0.279$ & $\rm{ 1.226e-01}$\\
            &                    logAge&                          $\log M_*$&    Center $(\mli{SMA}<0.1\,R_\text{e})$ & $  +9.614$ & $  +0.030$& $   0.128$& $  +0.104$ & $\rm{ 3.973e-01}$\\
            &                    logAge&                          $\log M_*$&              $\mli{SMA}=0.2\,R_\text{e}$& $  +8.959$ & $  +0.080$& $   0.106$& $  +0.157$ & $\rm{ 2.019e-01}$\\
$\checkmark$&                    logAge&                          $\log M_*$&              $\mli{SMA}=0.4\,R_\text{e}$& $  +7.983$ & $  +0.165$& $   0.086$& $  +0.348$ & $\rm{ 3.594e-03}$\\
$\checkmark$&                    logAge&                          $\log M_*$&                $\mli{SMA}=1\,R_\text{e}$& $  +8.771$ & $  +0.104$& $   0.068$& $  +0.444$ & $\rm{ 2.092e-04}$\\
            &                    logAge&                          $\log M_*$&               $\mli{SMA}=2\,R_\text{e}$ & $  +9.614$ & $  +0.031$& $   0.056$& $  +0.260$ & $\rm{ 1.509e-01}$\\
            &                      logZ&      $\log \mli{Age}_*(R_\text{e})$&    Center $(\mli{SMA}<0.1\,R_\text{e})$ & $  -1.216$ & $  +0.150$& $   0.036$& $  +0.264$ & $\rm{ 3.357e-02}$\\
$\checkmark$&                      logZ&      $\log \mli{Age}_*(R_\text{e})$&              $\mli{SMA}=0.2\,R_\text{e}$& $  -1.510$ & $  +0.177$& $   0.045$& $  +0.319$ & $\rm{ 9.647e-03}$\\
            &                      logZ&      $\log \mli{Age}_*(R_\text{e})$&              $\mli{SMA}=0.4\,R_\text{e}$& $  -0.728$ & $  +0.093$& $   0.062$& $  +0.240$ & $\rm{ 5.411e-02}$\\
            &                      logZ&      $\log \mli{Age}_*(R_\text{e})$&                $\mli{SMA}=1\,R_\text{e}$& $  +1.157$ & $  -0.112$& $   0.078$& $  +0.013$ & $\rm{ 9.155e-01}$\\
$\checkmark$&                      logZ&      $\log \mli{Age}_*(R_\text{e})$&               $\mli{SMA}=2\,R_\text{e}$ & $  -3.214$ & $  +0.321$& $   0.070$& $  +0.460$ & $\rm{ 8.124e-03}$\\
$\checkmark$&                    logAge&      $\log \mli{Age}_*(R_\text{e})$&    Center $(\mli{SMA}<0.1\,R_\text{e})$ & $  +0.209$ & $  +0.979$& $   0.115$& $  +0.442$ & $\rm{ 2.307e-04}$\\
$\checkmark$&                    logAge&      $\log \mli{Age}_*(R_\text{e})$&              $\mli{SMA}=0.2\,R_\text{e}$& $  +1.113$ & $  +0.880$& $   0.090$& $  +0.526$ & $\rm{ 6.898e-06}$\\
$\checkmark$&                    logAge&      $\log \mli{Age}_*(R_\text{e})$&              $\mli{SMA}=0.4\,R_\text{e}$& $  +0.264$ & $  +0.966$& $   0.068$& $  +0.714$ & $\rm{ 2.474e-11}$\\
$\checkmark$&                    logAge&      $\log \mli{Age}_*(R_\text{e})$&                $\mli{SMA}=1\,R_\text{e}$& $  -0.001$ & $  +1.000$& $   0.000$& $  +1.000$ & $\rm{ 0.000e+00}$\\
$\checkmark$&                    logAge&      $\log \mli{Age}_*(R_\text{e})$&               $\mli{SMA}=2\,R_\text{e}$ & $  +4.693$ & $  +0.532$& $   0.036$& $  +0.775$ & $\rm{ 1.967e-07}$\\
            &                      logZ&              $\log Z_*(R_\text{e})$&    Center $(\mli{SMA}<0.1\,R_\text{e})$ & $  +0.276$ & $  +0.060$& $   0.035$& $  +0.271$ & $\rm{ 2.890e-02}$\\
$\checkmark$&                      logZ&              $\log Z_*(R_\text{e})$&              $\mli{SMA}=0.2\,R_\text{e}$& $  +0.241$ & $  +0.384$& $   0.034$& $  +0.586$ & $\rm{ 2.905e-07}$\\
$\checkmark$&                      logZ&              $\log Z_*(R_\text{e})$&              $\mli{SMA}=0.4\,R_\text{e}$& $  +0.171$ & $  +0.569$& $   0.039$& $  +0.769$ & $\rm{ 7.280e-14}$\\
$\checkmark$&                      logZ&              $\log Z_*(R_\text{e})$&                $\mli{SMA}=1\,R_\text{e}$& $  -0.000$ & $  +1.000$& $   0.000$& $  +1.000$ & $\rm{ 0.000e+00}$\\
            &                      logZ&              $\log Z_*(R_\text{e})$&               $\mli{SMA}=2\,R_\text{e}$ & $  -0.116$ & $  +0.814$& $   0.059$& $  +0.417$ & $\rm{ 1.764e-02}$\\
            &                    logAge&              $\log Z_*(R_\text{e})$&    Center $(\mli{SMA}<0.1\,R_\text{e})$ & $  +9.941$ & $  +0.060$& $   0.131$& $  +0.156$ & $\rm{ 2.140e-01}$\\
            &                    logAge&              $\log Z_*(R_\text{e})$&              $\mli{SMA}=0.2\,R_\text{e}$& $  +9.840$ & $  +0.221$& $   0.109$& $  +0.211$ & $\rm{ 9.203e-02}$\\
            &                    logAge&              $\log Z_*(R_\text{e})$&              $\mli{SMA}=0.4\,R_\text{e}$& $  +9.821$ & $  +0.199$& $   0.096$& $  +0.156$ & $\rm{ 2.135e-01}$\\
            &                    logAge&              $\log Z_*(R_\text{e})$&                $\mli{SMA}=1\,R_\text{e}$& $  +9.923$ & $  -0.027$& $   0.075$& $  +0.013$ & $\rm{ 9.155e-01}$\\
            &                    logAge&              $\log Z_*(R_\text{e})$&               $\mli{SMA}=2\,R_\text{e}$ & $  +9.933$ & $  +0.210$& $   0.052$& $  +0.262$ & $\rm{ 1.467e-01}$\\
            &                      logZ&                            $C_{31}$&    Center $(\mli{SMA}<0.1\,R_\text{e})$ & $  +0.156$ & $  +0.018$& $   0.036$& $  +0.199$ & $\rm{ 1.015e-01}$\\
            &                      logZ&                            $C_{31}$&              $\mli{SMA}=0.2\,R_\text{e}$& $  +0.147$ & $  +0.016$& $   0.049$& $  +0.036$ & $\rm{ 7.686e-01}$\\
            &                      logZ&                            $C_{31}$&              $\mli{SMA}=0.4\,R_\text{e}$& $  +0.231$ & $  -0.009$& $   0.067$& $  -0.123$ & $\rm{ 3.183e-01}$\\
            &                      logZ&                            $C_{31}$&                $\mli{SMA}=1\,R_\text{e}$& $  +0.093$ & $  -0.007$& $   0.078$& $  -0.066$ & $\rm{ 5.997e-01}$\\
            &                      logZ&                            $C_{31}$&               $\mli{SMA}=2\,R_\text{e}$ & $  -0.030$ & $  -0.005$& $   0.077$& $  +0.036$ & $\rm{ 8.437e-01}$\\
            &                    logAge&                            $C_{31}$&    Center $(\mli{SMA}<0.1\,R_\text{e})$ & $  +9.782$ & $  +0.026$& $   0.126$& $  +0.164$ & $\rm{ 1.785e-01}$\\
            &                    logAge&                            $C_{31}$&              $\mli{SMA}=0.2\,R_\text{e}$& $  +9.714$ & $  +0.020$& $   0.107$& $  +0.121$ & $\rm{ 3.266e-01}$\\
            &                    logAge&                            $C_{31}$&              $\mli{SMA}=0.4\,R_\text{e}$& $  +9.741$ & $  +0.014$& $   0.096$& $  +0.067$ & $\rm{ 5.890e-01}$\\
            &                    logAge&                            $C_{31}$&                $\mli{SMA}=1\,R_\text{e}$& $  +9.717$ & $  +0.033$& $   0.071$& $  +0.192$ & $\rm{ 1.248e-01}$\\
            &                    logAge&                            $C_{31}$&               $\mli{SMA}=2\,R_\text{e}$ & $  +9.922$ & $  +0.005$& $   0.056$& $  +0.090$ & $\rm{ 6.228e-01}$\\
   \hline
 \end{tabular}
\end{table*}

\begin{table*}
  \centering
 \caption{Correlations between stellar population properties at characteristic $\mu_*$ and global properties. (1) Checked if correlation is significant; (2)-(3) $y$ and $x$ variables; (4) region in which stellar population properties $y$ are computed; (5)-(6) coefficients of the linear fit
 $y=a+b\,x$; (7) mean absolute deviation (MAD); (8) Spearman rank coefficient; (9) probability of null correlation.}
 \label{tab:corr_char_stelpop_mustar}
 \begin{tabular}{ccccrrrrrrrr} 
\hline
Significant  & $y$               &$x$               &Region                            &$a$       &$b$      &MAD      &$C_{\rm{Spearman}}$ & $P_{\rm{null}}$ \\
(1)          & (2)               &(3)               &(4)                               &(5)       & (6)     &(7)      & (8)                & (9) \\
  \hline
$\checkmark$ & $\log Z_*$        &               $\log\sigma_\text{e}$&Center ($\log \mu_*=4.0)$         & $-0.120$ & $+0.178$&$0.034$ & $+0.398$ & $1.239\rm{e}-03$ \\
$\checkmark$ & $\log Z_*$        &               $\log\sigma_\text{e}$&Mid ($\log \mu_*=3.1)$            & $-0.274$ & $+0.191$&$0.044$ & $+0.468$ & $5.016\rm{e}-05$ \\
$\checkmark$ & $\log Z_*$        &               $\log\sigma_\text{e}$&Outer ($\log \mu_*=2.3)$          & $-1.594$ & $+0.675$&$0.064$ & $+0.602$ & $4.355\rm{e}-08$ \\
$\checkmark$ & $\log\mli{Age}_*$ &               $\log\sigma_\text{e}$&Center ($\log \mu_*=4.0)$         & $+8.691$ & $+0.526$&$0.090$ & $+0.415$ & $7.100\rm{e}-04$ \\
$\checkmark$ & $\log\mli{Age}_*$ &               $\log\sigma_\text{e}$&Mid ($\log \mu_*=3.1)$            & $+8.610$ & $+0.536$&$0.073$ & $+0.529$ & $2.996\rm{e}-06$ \\
$\checkmark$ & $\log\mli{Age}_*$ &               $\log\sigma_\text{e}$&Outer ($\log \mu_*=2.3)$          & $+9.499$ & $+0.201$&$0.047$ & $+0.464$ & $5.964\rm{e}-05$ \\
$\checkmark$ & $\log Z_*$        &                          $\log M_*$&Center ($\log \mu_*=4.0)$         & $-0.369$ & $+0.059$&$0.034$ & $+0.410$ & $8.364\rm{e}-04$ \\
$\checkmark$ & $\log Z_*$        &                          $\log M_*$&Mid ($\log \mu_*=3.1)$            & $-0.491$ & $+0.058$&$0.043$ & $+0.467$ & $5.201\rm{e}-05$ \\
$\checkmark$ & $\log Z_*$        &                          $\log M_*$&Outer ($\log \mu_*=2.3)$          & $-1.637$ & $+0.143$&$0.071$ & $+0.518$ & $5.247\rm{e}-06$ \\
             & $\log\mli{Age}_*$ &                          $\log M_*$&Center ($\log \mu_*=4.0)$         & $+8.841$ & $+0.097$&$0.099$ & $+0.164$ & $1.997\rm{e}-01$ \\
             & $\log\mli{Age}_*$ &                          $\log M_*$&Mid ($\log \mu_*=3.1)$            & $+9.008$ & $+0.074$&$0.083$ & $+0.272$ & $2.395\rm{e}-02$ \\
             & $\log\mli{Age}_*$ &                          $\log M_*$&Outer ($\log \mu_*=2.3)$          & $+9.795$ & $+0.015$&$0.051$ & $+0.160$ & $1.901\rm{e}-01$ \\
             & $\log Z_*$        &      $\log \mli{Age}_*(R_\text{e})$&Center ($\log \mu_*=4.0)$         & $+0.398$ & $-0.011$&$0.037$ & $+0.170$ & $1.841\rm{e}-01$ \\
$\checkmark$ & $\log Z_*$        &      $\log \mli{Age}_*(R_\text{e})$&Mid ($\log \mu_*=3.1)$            & $-1.112$ & $+0.128$&$0.046$ & $+0.388$ & $1.077\rm{e}-03$ \\
$\checkmark$ & $\log Z_*$        &      $\log \mli{Age}_*(R_\text{e})$&Outer ($\log \mu_*=2.3)$          & $-4.812$ & $+0.483$&$0.074$ & $+0.494$ & $1.835\rm{e}-05$ \\
$\checkmark$ & $\log\mli{Age}_*$ &      $\log \mli{Age}_*(R_\text{e})$&Center ($\log \mu_*=4.0)$         & $+1.676$ & $+0.830$&$0.079$ & $+0.599$ & $2.157\rm{e}-07$ \\
$\checkmark$ & $\log\mli{Age}_*$ &      $\log \mli{Age}_*(R_\text{e})$&Mid ($\log \mu_*=3.1)$            & $+1.334$ & $+0.859$&$0.055$ & $+0.756$ & $9.169\rm{e}-14$ \\
$\checkmark$ & $\log\mli{Age}_*$ &      $\log \mli{Age}_*(R_\text{e})$&Outer ($\log \mu_*=2.3)$          & $+5.775$ & $+0.422$&$0.037$ & $+0.690$ & $7.367\rm{e}-11$ \\
             & $\log Z_*$        &              $\log Z_*(R_\text{e})$&Center ($\log \mu_*=4.0)$         & $+0.281$ & $+0.099$&$0.036$ & $+0.166$ & $1.931\rm{e}-01$ \\
$\checkmark$ & $\log Z_*$        &              $\log Z_*(R_\text{e})$&Mid ($\log \mu_*=3.1)$            & $+0.158$ & $+0.278$&$0.044$ & $+0.381$ & $1.341\rm{e}-03$ \\
$\checkmark$ & $\log Z_*$        &              $\log Z_*(R_\text{e})$&Outer ($\log \mu_*=2.3)$          & $-0.058$ & $+0.693$&$0.067$ & $+0.499$ & $1.494\rm{e}-05$ \\
             & $\log\mli{Age}_*$ &              $\log Z_*(R_\text{e})$&Center ($\log \mu_*=4.0)$         & $+9.911$ & $+0.038$&$0.104$ & $+0.084$ & $5.144\rm{e}-01$ \\
             & $\log\mli{Age}_*$ &              $\log Z_*(R_\text{e})$&Mid ($\log \mu_*=3.1)$            & $+9.832$ & $+0.252$&$0.085$ & $+0.180$ & $1.421\rm{e}-01$ \\
$\checkmark$ & $\log\mli{Age}_*$ &              $\log Z_*(R_\text{e})$&Outer ($\log \mu_*=2.3)$          & $+9.959$ & $+0.227$&$0.049$ & $+0.351$ & $3.386\rm{e}-03$ \\
             & $\log Z_*$        &                            $C_{31}$&Center ($\log \mu_*=4.0)$         & $+0.236$ & $+0.009$&$0.036$ & $+0.155$ & $2.240\rm{e}-01$ \\
             & $\log Z_*$        &                            $C_{31}$&Mid ($\log \mu_*=3.1)$            & $+0.253$ & $-0.015$&$0.048$ & $-0.129$ & $2.908\rm{e}-01$ \\
             & $\log Z_*$        &                            $C_{31}$&Outer ($\log \mu_*=2.3)$          & $-0.129$ & $+0.012$&$0.083$ & $-0.008$ & $9.510\rm{e}-01$ \\
             & $\log\mli{Age}_*$ &                            $C_{31}$&Center ($\log \mu_*=4.0)$         & $+9.936$ & $-0.004$&$0.104$ & $-0.024$ & $8.495\rm{e}-01$ \\
             & $\log\mli{Age}_*$ &                            $C_{31}$&Mid ($\log \mu_*=3.1)$            & $+9.751$ & $+0.013$&$0.085$ & $+0.068$ & $5.761\rm{e}-01$ \\
             & $\log\mli{Age}_*$ &                            $C_{31}$&Outer ($\log \mu_*=2.3)$          & $+9.889$ & $+0.011$&$0.051$ & $+0.069$ & $5.720\rm{e}-01$ \\
   \hline
 \end{tabular}
\end{table*}
\let\T\undefined       
\let\B\undefined 

\section{Tables of stellar population gradients}
\newcommand\T{\rule{0pt}{2.6ex}}       
\newcommand\B{\rule[-1.2ex]{0pt}{0pt}} 
\begin{landscape}
\begin{table*}
  \centering
 \caption{Radial gradients for different subsamples, as indicated in column (1) (see also Tab. 1 in the main paper).
 Columns (2) to (6) refer to the inner regions, columns (7) to (11) to the outer ones. For each quantity, $\zstar$ and $\agestar$,
 we provide both linear gradients in units of $\text{dex}\,\Reff^{-1}$ labeled as $\partial\mli{SMA}$ (see equation 3  in the main paper), 
 and logarithmic gradients in units of $\text{dex}\,\text{dex}^{-1}$,  labeled as $\partial\log\mli{SMA}$ (see equation 
 4 in the main paper). The $\pm$ range is defined by the $16^\text{th}-84^\text{th}$ percentile range. In column (6) and (11) are 
 the numbers of galaxies contributing to the statistics, that are determined by the sample size and the extension of the profiles.}
 \label{tab:grad_stelpop_sma}
 \begin{tabular}{crrrrc|rrrrc} 
\hline
Sample & \multicolumn{5}{c}{Inner ($0.2<\mli{SMA}/R_e<1.0$)}& \multicolumn{5}{c}{Outer ($1.0<\mli{SMA}/R_e<2.0$)}\\
&\multicolumn{2}{c}{$\nabla Z_*$}&\multicolumn{2}{c}{$\nabla \mli{Age}_*$}&$N$
&\multicolumn{2}{c}{$\nabla Z_*$}&\multicolumn{2}{c}{$\nabla \mli{Age}_*$}&$N$\\
&$\partial\mli{SMA}$&$\partial\log\mli{SMA}$&$\partial\mli{SMA}$&$\partial\log\mli{SMA}$&&$\partial\mli{SMA}$&$\partial\log\mli{SMA}$&$\partial\mli{SMA}$&$\partial\log\mli{SMA}$&\\
(1) & (2) & (3) & (4) & (5) & (6) & (7) & (8) & (9) & (10) & (11)\\ 
\hline
All ETGs &$-0.27^{+0.08}_{-0.08}$ & $-0.31^{+0.10}_{-0.10}$ & $ 0.08^{+0.09}_{-0.19}$ & $ 0.10^{+0.10}_{-0.22}$ & $65$ & $-0.12^{+0.06}_{-0.07}$ & $-0.40^{+0.21}_{-0.24}$ & $ 0.05^{+0.07}_{-0.04}$ & $ 0.18^{+0.23}_{-0.12}$ & $32$ \B \T \\
$\log M_*/\rm{M_\odot} < 10.9 $&$-0.28^{+0.05}_{-0.10}$ & $-0.32^{+0.06}_{-0.11}$ & $ 0.07^{+0.21}_{-0.20}$ & $ 0.08^{+0.24}_{-0.22}$ & $16$ & $-0.12^{+0.06}_{-0.08}$ & $-0.40^{+0.21}_{-0.27}$ & $ 0.09^{+0.03}_{-0.09}$ & $ 0.30^{+0.11}_{-0.28}$ & $11$ \B \T \\
$10.9 \leq \log M_*/\rm{M_\odot} < 11.3 $&$-0.27^{+0.09}_{-0.11}$ & $-0.31^{+0.10}_{-0.13}$ & $ 0.10^{+0.15}_{-0.24}$ & $ 0.12^{+0.17}_{-0.27}$ & $29$ & $-0.12^{+0.06}_{-0.05}$ & $-0.40^{+0.20}_{-0.16}$ & $ 0.05^{+0.04}_{-0.03}$ & $ 0.17^{+0.14}_{-0.09}$ & $14$ \B \T \\
$\log M_*/\rm{M_\odot} \geq 11.3 $&$-0.24^{+0.05}_{-0.09}$ & $-0.27^{+0.05}_{-0.10}$ & $ 0.08^{+0.06}_{-0.11}$ & $ 0.10^{+0.07}_{-0.12}$ & $20$ & $-0.07^{+0.05}_{-0.09}$ & $-0.25^{+0.15}_{-0.29}$ & $ 0.03^{+0.05}_{-0.02}$ & $ 0.11^{+0.18}_{-0.05}$ & $7$ \B \T \\
$\sigma_e/[\rm{km\,s}^{-1}] < 170 $&$-0.28^{+0.07}_{-0.11}$ & $-0.32^{+0.08}_{-0.13}$ & $ 0.13^{+0.15}_{-0.07}$ & $ 0.15^{+0.17}_{-0.08}$ & $21$ & $-0.12^{+0.08}_{-0.13}$ & $-0.40^{+0.28}_{-0.42}$ & $ 0.05^{+0.22}_{-0.05}$ & $ 0.18^{+0.75}_{-0.16}$ & $12$ \B \T \\
$170 \leq \sigma_e/[\rm{km\,s}^{-1}] < 210 $&$-0.27^{+0.18}_{-0.11}$ & $-0.31^{+0.20}_{-0.13}$ & $ 0.06^{+0.19}_{-0.19}$ & $ 0.07^{+0.21}_{-0.22}$ & $22$ & $-0.14^{+0.07}_{-0.06}$ & $-0.47^{+0.23}_{-0.20}$ & $ 0.09^{+0.01}_{-0.05}$ & $ 0.28^{+0.03}_{-0.18}$ & $10$ \B \T \\
$\sigma_e/[\rm{km\,s}^{-1}] \geq 210 $&$-0.23^{+0.04}_{-0.10}$ & $-0.26^{+0.05}_{-0.11}$ & $ 0.08^{+0.06}_{-0.20}$ & $ 0.09^{+0.07}_{-0.23}$ & $22$ & $-0.09^{+0.06}_{-0.07}$ & $-0.29^{+0.20}_{-0.25}$ & $ 0.05^{+0.04}_{-0.03}$ & $ 0.17^{+0.14}_{-0.11}$ & $10$ \B \T \\
All Es&$-0.28^{+0.07}_{-0.10}$ & $-0.32^{+0.08}_{-0.12}$ & $ 0.08^{+0.08}_{-0.12}$ & $ 0.10^{+0.10}_{-0.13}$ & $44$ & $-0.09^{+0.06}_{-0.08}$ & $-0.29^{+0.19}_{-0.26}$ & $ 0.04^{+0.05}_{-0.04}$ & $ 0.14^{+0.17}_{-0.13}$ & $18$ \B \T \\
S0 &$-0.23^{+0.15}_{-0.07}$ & $-0.26^{+0.17}_{-0.08}$ & $ 0.09^{+0.16}_{-0.22}$ & $ 0.10^{+0.19}_{-0.25}$ & $21$ & $-0.14^{+0.09}_{-0.09}$ & $-0.47^{+0.28}_{-0.29}$ & $ 0.09^{+0.04}_{-0.05}$ & $ 0.28^{+0.13}_{-0.17}$ & $14$ \B \T \\
Es, $\log M_*/\rm{M_\odot} < 11.3 $&$-0.28^{+0.07}_{-0.11}$ & $-0.33^{+0.08}_{-0.12}$ & $ 0.10^{+0.07}_{-0.16}$ & $ 0.12^{+0.08}_{-0.19}$ & $26$ & $-0.11^{+0.05}_{-0.08}$ & $-0.37^{+0.16}_{-0.27}$ & $ 0.08^{+0.14}_{-0.07}$ & $ 0.26^{+0.47}_{-0.25}$ & $12$ \B \T \\
Es, $\sigma_e/[\rm{km\,s}^{-1}] < 170 $&$-0.27^{+0.06}_{-0.19}$ & $-0.31^{+0.07}_{-0.22}$ & $ 0.13^{+0.23}_{-0.22}$ & $ 0.15^{+0.26}_{-0.25}$ & $11$ & $-0.12^{+0.04}_{-0.07}$ & $-0.40^{+0.13}_{-0.24}$ & $ 0.05^{+0.17}_{-0.05}$ & $ 0.18^{+0.55}_{-0.16}$ & $7$ \B \T \\
Es, $170 \leq \sigma_e/[\rm{km\,s}^{-1}] < 210 $&$-0.30^{+0.03}_{-0.11}$ & $-0.35^{+0.03}_{-0.13}$ & $ 0.06^{+0.09}_{-0.12}$ & $ 0.07^{+0.11}_{-0.14}$ & $13$ & $-0.08^{+0.02}_{-0.06}$ & $-0.28^{+0.08}_{-0.21}$ & $ 0.04^{+0.05}_{-0.03}$ & $ 0.14^{+0.17}_{-0.12}$ & $3$ \B \T \\
Es, $\sigma_e/[\rm{km\,s}^{-1}] \geq 210 $&$-0.24^{+0.05}_{-0.09}$ & $-0.27^{+0.05}_{-0.10}$ & $ 0.09^{+0.05}_{-0.12}$ & $ 0.11^{+0.06}_{-0.13}$ & $20$ & $-0.07^{+0.05}_{-0.04}$ & $-0.25^{+0.15}_{-0.12}$ & $ 0.04^{+0.05}_{-0.02}$ & $ 0.14^{+0.15}_{-0.08}$ & $8$ \B \T \\
\hline
 \end{tabular}
\end{table*}
\end{landscape}

\clearpage
\begin{landscape}
\begin{table*}
  \centering
 \caption{Gradients along $\mustar$ for different subsamples, as indicated in column (1) (see also Tab. 1 in the main paper).
 Columns (2) to (4) refer to the inner regions, columns (5) to (7) to the outer ones. For each quantity, $\zstar$ and $\agestar$,
 we provide gradients in units of $\text{dex}\,\text{dex}^{-1}$ (see equation 5 in the main paper). The $\pm$ range is defined by the $16^\text{th}-84^\text{th}$ percentile range. In column (4) and (7) are 
 the numbers of galaxies contributing to the statistics, that are determined by the sample size and the extension of the profiles.}
 \label{tab:grad_stelpop_mustar}
 \begin{tabular}{c|rrc|rrc} 
\hline
Sample & \multicolumn{3}{c}{Inner ($3.1 < \log \mu_*/[\rm{M_\odot pc^{-2}}] < 4  $)}& \multicolumn{3}{c}{Outer ($2.3 < \log \mu_*/[\rm{M_\odot pc^{-2}}] < 3.1$)}\\
&$\nabla Z_*$&$\nabla \mli{Age}_*$&$N$
&$\nabla Z_*$&$\nabla \mli{Age}_*$&$N$\\
(1) &(2) &(3) &(4) &(5) &(6) &(7)\\
\hline
All ETGs &$ 0.13^{+0.07}_{-0.08}$ & $ 0.06^{+0.10}_{-0.09}$ & $63$ & $ 0.27^{+0.11}_{-0.10}$ & $-0.14^{+0.10}_{-0.08}$ & $69$ \B \T \\
$\log M_*/\rm{M_\odot} < 10.9 $&$ 0.16^{+0.05}_{-0.11}$ & $ 0.07^{+0.09}_{-0.13}$ & $14$ & $ 0.29^{+0.09}_{-0.07}$ & $-0.19^{+0.09}_{-0.07}$ & $16$ \B \T \\
$10.9 \leq \log M_*/\rm{M_\odot} < 11.3 $&$ 0.14^{+0.09}_{-0.09}$ & $ 0.06^{+0.09}_{-0.12}$ & $28$ & $ 0.26^{+0.18}_{-0.10}$ & $-0.14^{+0.15}_{-0.11}$ & $31$ \B \T \\
$\log M_*/\rm{M_\odot} \geq 11.3 $&$ 0.11^{+0.06}_{-0.05}$ & $ 0.06^{+0.10}_{-0.08}$ & $21$ & $ 0.24^{+0.08}_{-0.09}$ & $-0.10^{+0.07}_{-0.08}$ & $22$ \B \T \\
$\sigma_e/[\rm{km\,s}^{-1}] < 170 $&$ 0.15^{+0.06}_{-0.12}$ & $ 0.03^{+0.14}_{-0.09}$ & $17$ & $ 0.29^{+0.13}_{-0.09}$ & $-0.19^{+0.06}_{-0.14}$ & $21$ \B \T \\
$170 \leq \sigma_e/[\rm{km\,s}^{-1}] < 210 $&$ 0.16^{+0.06}_{-0.09}$ & $ 0.06^{+0.09}_{-0.11}$ & $23$ & $ 0.28^{+0.09}_{-0.13}$ & $-0.12^{+0.13}_{-0.08}$ & $25$ \B \T \\
$\sigma_e/[\rm{km\,s}^{-1}] \geq 210 $&$ 0.11^{+0.06}_{-0.05}$ & $ 0.07^{+0.09}_{-0.09}$ & $23$ & $ 0.22^{+0.09}_{-0.06}$ & $-0.10^{+0.06}_{-0.09}$ & $23$ \B \T \\
All Es&$ 0.13^{+0.07}_{-0.08}$ & $ 0.06^{+0.10}_{-0.08}$ & $44$ & $ 0.27^{+0.09}_{-0.10}$ & $-0.12^{+0.09}_{-0.08}$ & $48$ \B \T \\
S0 &$ 0.16^{+0.06}_{-0.10}$ & $ 0.07^{+0.09}_{-0.13}$ & $19$ & $ 0.28^{+0.11}_{-0.11}$ & $-0.17^{+0.13}_{-0.12}$ & $21$ \B \T \\
Es, $\log M_*/\rm{M_\odot} < 11.3 $&$ 0.14^{+0.07}_{-0.09}$ & $ 0.04^{+0.13}_{-0.09}$ & $25$ & $ 0.27^{+0.11}_{-0.10}$ & $-0.13^{+0.10}_{-0.11}$ & $28$ \B \T \\
Es, $\sigma_e/[\rm{km\,s}^{-1}] < 170 $&$ 0.11^{+0.09}_{-0.08}$ & $ 0.03^{+0.25}_{-0.09}$ & $9$ & $ 0.29^{+0.20}_{-0.09}$ & $-0.19^{+0.07}_{-0.16}$ & $11$ \B \T \\
Es, $170 \leq \sigma_e/[\rm{km\,s}^{-1}] < 210 $&$ 0.17^{+0.05}_{-0.06}$ & $ 0.06^{+0.09}_{-0.03}$ & $14$ & $ 0.29^{+0.09}_{-0.13}$ & $-0.11^{+0.13}_{-0.07}$ & $16$ \B \T \\
Es, $\sigma_e/[\rm{km\,s}^{-1}] \geq 210 $&$ 0.11^{+0.06}_{-0.05}$ & $ 0.06^{+0.10}_{-0.08}$ & $21$ & $ 0.22^{+0.06}_{-0.06}$ & $-0.10^{+0.05}_{-0.08}$ & $21$ \B \T \\
\hline
 \end{tabular}
\end{table*}
\end{landscape}
\clearpage
\begin{table*}
  \centering
 \caption{Correlations between radial gradients of stellar population properties ($\nabla$) and global properties ($x$): $\nabla\equiv\frac{\partial y}{\partial u}$ vs $x$. (1) Checked if correlation is significant; (2)-(3)-(4) $u$, $y$ and $x$ variables; (5) region, i.e. radial range, in which stellar population gradients are computed; (6)-(7) coefficients of the linear fit
 $\nabla=a+b\,x$; (8) mean absolute deviation (MAD); (9) Spearman rank coefficient; (10) probability of null correlation.}
 \label{tab:grad_rad_corr}
 \begin{tabular}{ccccrrrrrrrr} 
\hline
Significant  & $u$        & $y$               &$x$                      &Region                           &$a$       &$b$       &MAD      &$C_{\rm{Spearman}}$ & $P_{\rm{null}}$ \\
(1)          & (2)        & (3)               &(4)                      &(5)                              &(6)       & (7)      &(8)      & (9)                & (10) \\
  \hline
  &               $\mli{SMA}$&                $\log Z_*$&               $\log\sigma_\text{e}$&   Inner ($0.2<\mli{SMA}/R_\text{e}<1.0$)& $  -0.745$ & $  +0.214$& $   0.077$& $  +0.187$ & $\rm{ 1.355e-01}$\\
  &               $\mli{SMA}$&                $\log Z_*$&               $\log\sigma_\text{e}$&   Outer ($1.0<\mli{SMA}/R_\text{e}<2.0$)& $  -0.344$ & $  +0.102$& $   0.058$& $  +0.192$ & $\rm{ 2.932e-01}$\\
  &           $\log\mli{SMA}$&                $\log Z_*$&               $\log\sigma_\text{e}$&   Inner ($0.2<\mli{SMA}/R_\text{e}<1.0$)& $  -0.853$ & $  +0.245$& $   0.088$& $  +0.187$ & $\rm{ 1.355e-01}$\\
  &           $\log\mli{SMA}$&                $\log Z_*$&               $\log\sigma_\text{e}$&   Outer ($1.0<\mli{SMA}/R_\text{e}<2.0$)& $  -1.142$ & $  +0.338$& $   0.192$& $  +0.192$ & $\rm{ 2.932e-01}$\\
  &           $    \mli{SMA}$&         $\log\mli{Age}_*$&               $\log\sigma_\text{e}$&   Inner ($0.2<\mli{SMA}/R_\text{e}<1.0$)& $  +0.819$ & $  -0.324$& $   0.111$& $  -0.223$ & $\rm{ 7.389e-02}$\\
  &           $    \mli{SMA}$&         $\log\mli{Age}_*$&               $\log\sigma_\text{e}$&   Outer ($1.0<\mli{SMA}/R_\text{e}<2.0$)& $  +0.462$ & $  -0.174$& $   0.050$& $  -0.264$ & $\rm{ 1.450e-01}$\\
  &           $\log\mli{SMA}$&         $\log\mli{Age}_*$&               $\log\sigma_\text{e}$&   Inner ($0.2<\mli{SMA}/R_\text{e}<1.0$)& $  +0.937$ & $  -0.371$& $   0.127$& $  -0.223$ & $\rm{ 7.389e-02}$\\
  &           $\log\mli{SMA}$&         $\log\mli{Age}_*$&               $\log\sigma_\text{e}$&   Outer ($1.0<\mli{SMA}/R_\text{e}<2.0$)& $  +1.534$ & $  -0.580$& $   0.167$& $  -0.264$ & $\rm{ 1.450e-01}$\\
  &               $\mli{SMA}$&                $\log Z_*$&                          $\log M_*$&   Inner ($0.2<\mli{SMA}/R_\text{e}<1.0$)& $  -0.688$ & $  +0.038$& $   0.077$& $  +0.129$ & $\rm{ 3.069e-01}$\\
  &               $\mli{SMA}$&                $\log Z_*$&                          $\log M_*$&   Outer ($1.0<\mli{SMA}/R_\text{e}<2.0$)& $  -0.470$ & $  +0.033$& $   0.059$& $  +0.086$ & $\rm{ 6.392e-01}$\\
  &           $\log\mli{SMA}$&                $\log Z_*$&                          $\log M_*$&   Inner ($0.2<\mli{SMA}/R_\text{e}<1.0$)& $  -0.787$ & $  +0.043$& $   0.088$& $  +0.129$ & $\rm{ 3.069e-01}$\\
  &           $\log\mli{SMA}$&                $\log Z_*$&                          $\log M_*$&   Outer ($1.0<\mli{SMA}/R_\text{e}<2.0$)& $  -1.563$ & $  +0.108$& $   0.195$& $  +0.086$ & $\rm{ 6.392e-01}$\\
  &           $    \mli{SMA}$&         $\log\mli{Age}_*$&                          $\log M_*$&   Inner ($0.2<\mli{SMA}/R_\text{e}<1.0$)& $  -0.167$ & $  +0.022$& $   0.114$& $  +0.067$ & $\rm{ 5.951e-01}$\\
  &           $    \mli{SMA}$&         $\log\mli{Age}_*$&                          $\log M_*$&   Outer ($1.0<\mli{SMA}/R_\text{e}<2.0$)& $  +0.619$ & $  -0.051$& $   0.049$& $  -0.278$ & $\rm{ 1.231e-01}$\\
  &           $\log\mli{SMA}$&         $\log\mli{Age}_*$&                          $\log M_*$&   Inner ($0.2<\mli{SMA}/R_\text{e}<1.0$)& $  -0.191$ & $  +0.025$& $   0.131$& $  +0.067$ & $\rm{ 5.951e-01}$\\
  &           $\log\mli{SMA}$&         $\log\mli{Age}_*$&                          $\log M_*$&   Outer ($1.0<\mli{SMA}/R_\text{e}<2.0$)& $  +2.055$ & $  -0.170$& $   0.162$& $  -0.278$ & $\rm{ 1.231e-01}$\\
  &               $\mli{SMA}$&                $\log Z_*$&      $\log \mli{Age}_*(R_\text{e})$&   Inner ($0.2<\mli{SMA}/R_\text{e}<1.0$)& $  +1.983$ & $  -0.227$& $   0.072$& $  -0.185$ & $\rm{ 1.399e-01}$\\
$\checkmark$  &               $\mli{SMA}$&                $\log Z_*$&      $\log \mli{Age}_*(R_\text{e})$&   Outer ($1.0<\mli{SMA}/R_\text{e}<2.0$)& $  -5.717$ & $  +0.568$& $   0.044$& $  +0.578$ & $\rm{ 5.354e-04}$\\
  &           $\log\mli{SMA}$&                $\log Z_*$&      $\log \mli{Age}_*(R_\text{e})$&   Inner ($0.2<\mli{SMA}/R_\text{e}<1.0$)& $  +2.269$ & $  -0.260$& $   0.083$& $  -0.185$ & $\rm{ 1.399e-01}$\\
$\checkmark$  &           $\log\mli{SMA}$&                $\log Z_*$&      $\log \mli{Age}_*(R_\text{e})$&   Outer ($1.0<\mli{SMA}/R_\text{e}<2.0$)& $ -18.992$ & $  +1.886$& $   0.146$& $  +0.578$ & $\rm{ 5.354e-04}$\\
  &           $    \mli{SMA}$&         $\log\mli{Age}_*$&      $\log \mli{Age}_*(R_\text{e})$&   Inner ($0.2<\mli{SMA}/R_\text{e}<1.0$)& $  -1.377$ & $  +0.148$& $   0.113$& $  +0.082$ & $\rm{ 5.157e-01}$\\
$\checkmark$  &           $    \mli{SMA}$&         $\log\mli{Age}_*$&      $\log \mli{Age}_*(R_\text{e})$&   Outer ($1.0<\mli{SMA}/R_\text{e}<2.0$)& $  +4.689$ & $  -0.468$& $   0.036$& $  -0.673$ & $\rm{ 2.437e-05}$\\
  &           $\log\mli{SMA}$&         $\log\mli{Age}_*$&      $\log \mli{Age}_*(R_\text{e})$&   Inner ($0.2<\mli{SMA}/R_\text{e}<1.0$)& $  -1.576$ & $  +0.169$& $   0.129$& $  +0.082$ & $\rm{ 5.157e-01}$\\
$\checkmark$  &           $\log\mli{SMA}$&         $\log\mli{Age}_*$&      $\log \mli{Age}_*(R_\text{e})$&   Outer ($1.0<\mli{SMA}/R_\text{e}<2.0$)& $ +15.575$ & $  -1.554$& $   0.121$& $  -0.673$ & $\rm{ 2.437e-05}$\\
$\checkmark$  &               $\mli{SMA}$&                $\log Z_*$&              $\log Z_*(R_\text{e})$&   Inner ($0.2<\mli{SMA}/R_\text{e}<1.0$)& $  -0.302$ & $  +0.770$& $   0.043$& $  +0.824$ & $\rm{ 3.213e-17}$\\
  &               $\mli{SMA}$&                $\log Z_*$&              $\log Z_*(R_\text{e})$&   Outer ($1.0<\mli{SMA}/R_\text{e}<2.0$)& $  -0.116$ & $  -0.186$& $   0.059$& $  -0.318$ & $\rm{ 7.594e-02}$\\
$\checkmark$  &           $\log\mli{SMA}$&                $\log Z_*$&              $\log Z_*(R_\text{e})$&   Inner ($0.2<\mli{SMA}/R_\text{e}<1.0$)& $  -0.345$ & $  +0.881$& $   0.049$& $  +0.824$ & $\rm{ 3.213e-17}$\\
  &           $\log\mli{SMA}$&                $\log Z_*$&              $\log Z_*(R_\text{e})$&   Outer ($1.0<\mli{SMA}/R_\text{e}<2.0$)& $  -0.384$ & $  -0.617$& $   0.195$& $  -0.318$ & $\rm{ 7.594e-02}$\\
$\checkmark$  &           $    \mli{SMA}$&         $\log\mli{Age}_*$&              $\log Z_*(R_\text{e})$&   Inner ($0.2<\mli{SMA}/R_\text{e}<1.0$)& $  +0.100$ & $  -0.314$& $   0.111$& $  -0.372$ & $\rm{ 2.277e-03}$\\
  &           $    \mli{SMA}$&         $\log\mli{Age}_*$&              $\log Z_*(R_\text{e})$&   Outer ($1.0<\mli{SMA}/R_\text{e}<2.0$)& $  +0.044$ & $  +0.245$& $   0.045$& $  +0.431$ & $\rm{ 1.386e-02}$\\
$\checkmark$  &           $\log\mli{SMA}$&         $\log\mli{Age}_*$&              $\log Z_*(R_\text{e})$&   Inner ($0.2<\mli{SMA}/R_\text{e}<1.0$)& $  +0.115$ & $  -0.359$& $   0.127$& $  -0.372$ & $\rm{ 2.277e-03}$\\
  &           $\log\mli{SMA}$&         $\log\mli{Age}_*$&              $\log Z_*(R_\text{e})$&   Outer ($1.0<\mli{SMA}/R_\text{e}<2.0$)& $  +0.146$ & $  +0.812$& $   0.151$& $  +0.431$ & $\rm{ 1.386e-02}$\\
  &               $\mli{SMA}$&                $\log Z_*$&                            $C_{31}$&   Inner ($0.2<\mli{SMA}/R_\text{e}<1.0$)& $  -0.039$ & $  -0.037$& $   0.075$& $  -0.160$ & $\rm{ 2.027e-01}$\\
  &               $\mli{SMA}$&                $\log Z_*$&                            $C_{31}$&   Outer ($1.0<\mli{SMA}/R_\text{e}<2.0$)& $  -0.060$ & $  -0.009$& $   0.059$& $  +0.124$ & $\rm{ 4.973e-01}$\\
  &           $\log\mli{SMA}$&                $\log Z_*$&                            $C_{31}$&   Inner ($0.2<\mli{SMA}/R_\text{e}<1.0$)& $  -0.044$ & $  -0.042$& $   0.086$& $  -0.160$ & $\rm{ 2.027e-01}$\\
  &           $\log\mli{SMA}$&                $\log Z_*$&                            $C_{31}$&   Outer ($1.0<\mli{SMA}/R_\text{e}<2.0$)& $  -0.199$ & $  -0.030$& $   0.195$& $  +0.124$ & $\rm{ 4.973e-01}$\\
  &           $    \mli{SMA}$&         $\log\mli{Age}_*$&                            $C_{31}$&   Inner ($0.2<\mli{SMA}/R_\text{e}<1.0$)& $  +0.057$ & $  +0.005$& $   0.115$& $  +0.005$ & $\rm{ 9.694e-01}$\\
  &           $    \mli{SMA}$&         $\log\mli{Age}_*$&                            $C_{31}$&   Outer ($1.0<\mli{SMA}/R_\text{e}<2.0$)& $  +0.156$ & $  -0.016$& $   0.047$& $  -0.327$ & $\rm{ 6.786e-02}$\\
  &           $\log\mli{SMA}$&         $\log\mli{Age}_*$&                            $C_{31}$&   Inner ($0.2<\mli{SMA}/R_\text{e}<1.0$)& $  +0.066$ & $  +0.005$& $   0.131$& $  +0.005$ & $\rm{ 9.694e-01}$\\
  &           $\log\mli{SMA}$&         $\log\mli{Age}_*$&                            $C_{31}$&   Outer ($1.0<\mli{SMA}/R_\text{e}<2.0$)& $  +0.518$ & $  -0.052$& $   0.157$& $  -0.327$ & $\rm{ 6.786e-02}$\\
\hline
 \end{tabular}
\end{table*}
\clearpage
\begin{table*}
  \centering
 \caption{Correlations between gradients of stellar population properties along $\log \mu_*$ profiles ($\nabla$) and global properties ($x$): $\nabla\equiv\frac{\partial y}{\partial (\log \mu_*)}$ vs $x$. (1) Checked if correlation is significant; (2)-(3) $y$ and $x$ variables; (4) region, i.e. $\mu_*$ range, in which stellar population gradients are computed; (5)-(6) coefficients of the linear fit
 $\nabla=a+b\,x$; (7) mean absolute deviation (MAD); (8) Spearman rank coefficient; (9) probability of null correlation.}
 \label{tab:grad_mustar_corr}
 \begin{tabular}{ccccrrrrrrrr} 
\hline
Significant  & $y$               &$x$                      &Region                           &$a$       &$b$       &MAD      &$C_{\rm{Spearman}}$ & $P_{\rm{null}}$ \\
(1)          & (2)               &(3)                      &(4)                              &(5)       & (6)      &(7)      & (8)                & (9) \\
  \hline
             & $\log Z_*$        & $\log\sigma_e$          & Inner ($3.1 < \log \mu_*/[\rm{M_\odot pc^{-2}}] < 4  $) & $+0.647$ & $-0.228$ & $0.055$ & $-0.141$ & $2.711\rm{e}-01$\\
$\checkmark$ & $\log Z_*$        & $\log\sigma_e$          & Outer ($2.3 < \log \mu_*/[\rm{M_\odot pc^{-2}}] < 3.1$) & $+1.162$ & $-0.393$ & $0.081$ & $-0.380$ & $1.267\rm{e}-03$\\
             & $\log\mli{Age}_*$ & $\log\sigma_e$          & Inner ($3.1 < \log \mu_*/[\rm{M_\odot pc^{-2}}] < 4  $) & $-0.435$ & $+0.213$ & $0.080$ & $+0.090$ & $4.806\rm{e}-01$\\
$\checkmark$ & $\log\mli{Age}_*$ & $\log\sigma_e$          & Outer ($2.3 < \log \mu_*/[\rm{M_\odot pc^{-2}}] < 3.1$) & $-0.999$ & $+0.375$ & $0.075$ & $+0.431$ & $2.204\rm{e}-04$\\
             & $\log Z_*$        & $\log M_*$              & Inner ($3.1 < \log \mu_*/[\rm{M_\odot pc^{-2}}] < 4  $) & $+0.620$ & $-0.045$ & $0.055$ & $-0.177$ & $1.662\rm{e}-01$\\
             & $\log Z_*$        & $\log M_*$              & Outer ($2.3 < \log \mu_*/[\rm{M_\odot pc^{-2}}] < 3.1$) & $+1.224$ & $-0.087$ & $0.085$ & $-0.229$ & $5.824\rm{e}-02$\\
             & $\log\mli{Age}_*$ & $\log M_*$              & Inner ($3.1 < \log \mu_*/[\rm{M_\odot pc^{-2}}] < 4  $) & $-0.286$ & $+0.030$ & $0.081$ & $-0.027$ & $8.314\rm{e}-01$\\
             & $\log\mli{Age}_*$ & $\log M_*$              & Outer ($2.3 < \log \mu_*/[\rm{M_\odot pc^{-2}}] < 3.1$) & $-1.154$ & $+0.090$ & $0.081$ & $+0.270$ & $2.484\rm{e}-02$\\
             & $\log Z_*$        & $\log \mli{Age}_*(R_e)$ & Inner ($3.1 < \log \mu_*/[\rm{M_\odot pc^{-2}}] < 4  $) & $+2.669$ & $-0.257$ & $0.053$ & $-0.246$ & $5.185\rm{e}-02$\\
             & $\log Z_*$        & $\log \mli{Age}_*(R_e)$ & Outer ($2.3 < \log \mu_*/[\rm{M_\odot pc^{-2}}] < 3.1$) & $+3.213$ & $-0.300$ & $0.082$ & $-0.308$ & $1.060\rm{e}-02$\\
             & $\log\mli{Age}_*$ & $\log \mli{Age}_*(R_e)$ & Inner ($3.1 < \log \mu_*/[\rm{M_\odot pc^{-2}}] < 4  $) & $-1.899$ & $+0.197$ & $0.080$ & $+0.076$ & $5.554\rm{e}-01$\\
$\checkmark$ & $\log\mli{Age}_*$ & $\log \mli{Age}_*(R_e)$ & Outer ($2.3 < \log \mu_*/[\rm{M_\odot pc^{-2}}] < 3.1$) & $-5.611$ & $+0.552$ & $0.070$ & $+0.461$ & $7.529\rm{e}-05$\\
             & $\log Z_*$        & $\log Z_*(R_e)$         & Inner ($3.1 < \log \mu_*/[\rm{M_\odot pc^{-2}}] < 4  $) & $+0.134$ & $-0.145$ & $0.054$ & $-0.178$ & $1.634\rm{e}-01$\\
$\checkmark$ & $\log Z_*$        & $\log Z_*(R_e)$         & Outer ($2.3 < \log \mu_*/[\rm{M_\odot pc^{-2}}] < 3.1$) & $+0.263$ & $-0.430$ & $0.079$ & $-0.354$ & $3.039\rm{e}-03$\\
             & $\log\mli{Age}_*$ & $\log Z_*(R_e)$         & Inner ($3.1 < \log \mu_*/[\rm{M_\odot pc^{-2}}] < 4  $) & $+0.049$ & $+0.170$ & $0.080$ & $+0.050$ & $6.957\rm{e}-01$\\
             & $\log\mli{Age}_*$ & $\log Z_*(R_e)$         & Outer ($2.3 < \log \mu_*/[\rm{M_\odot pc^{-2}}] < 3.1$) & $-0.132$ & $+0.107$ & $0.083$ & $+0.043$ & $7.257\rm{e}-01$\\
             & $\log Z_*$        & $C_{31}$                & Inner ($3.1 < \log \mu_*/[\rm{M_\odot pc^{-2}}] < 4  $) & $-0.061$ & $+0.032$ & $0.053$ & $+0.251$ & $4.762\rm{e}-02$\\
             & $\log Z_*$        & $C_{31}$                & Outer ($2.3 < \log \mu_*/[\rm{M_\odot pc^{-2}}] < 3.1$) & $+0.327$ & $-0.010$ & $0.087$ & $-0.046$ & $7.063\rm{e}-01$\\
             & $\log\mli{Age}_*$ & $C_{31}$                & Inner ($3.1 < \log \mu_*/[\rm{M_\odot pc^{-2}}] < 4  $) & $+0.136$ & $-0.012$ & $0.080$ & $-0.042$ & $7.429\rm{e}-01$\\
             & $\log\mli{Age}_*$ & $C_{31}$                & Outer ($2.3 < \log \mu_*/[\rm{M_\odot pc^{-2}}] < 3.1$) & $-0.367$ & $+0.035$ & $0.081$ & $+0.155$ & $2.027\rm{e}-01$\\
\hline
 \end{tabular}
\end{table*}
\let\T\undefined       
\let\B\undefined 
\clearpage

\section{Radial profiles of spectral indices and colours}\label{app:ind_col}

In this appendix we report the radial profiles of the ``raw'' observable quantities from
which the stellar population properties are estimated, i.e. the four stellar absorption indices,
$\mathrm{H\beta}$, $\mathrm{H\delta_A}+\mathrm{H\gamma_A}$, $[\mathrm{Mg_2Fe}]$ and $[\mathrm{MgFe}]^\prime$,
and the four broad band colours resulting from the five SDSS bands. These profiles are obtained with the
very same method used to extract the radial stellar population profiles presented in Fig. \ref{fig:stacked_profs}
and Sec. \ref{sec:SPprofiles}.

As obvious, a proper conversion of these profiles into stellar population properties can only be
performed with the aid of stellar population models, as done in the paper. However, it is instructive to
see how the slopes of the indices and of the $r-i$ colour change around $\sim 0.5\,\Reff$. This effect
is most likely at the origin of the inflection of the age profiles around that radius and of their overall
U-shape.

\begin{figure*}
\centerline{\large \textsf{All ETGs}}
\includegraphics[width=0.45\textwidth]{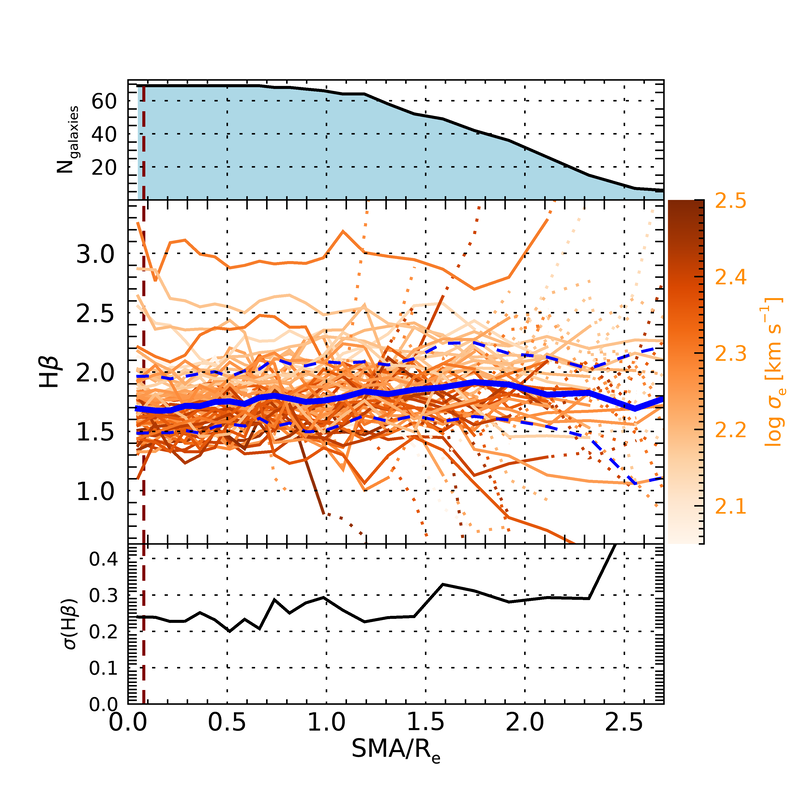}
\includegraphics[width=0.45\textwidth]{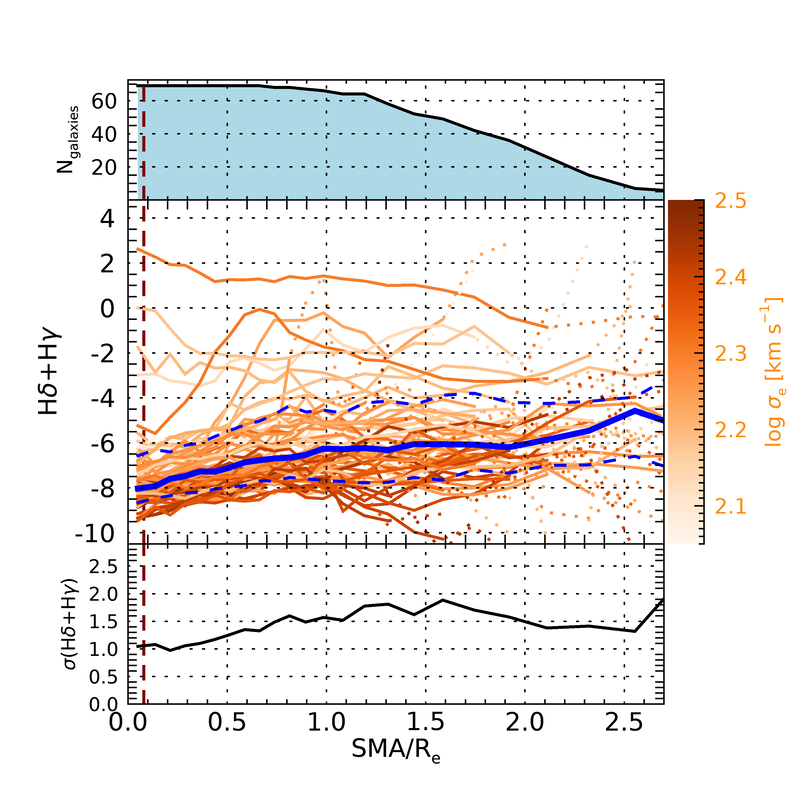}\\
\includegraphics[width=0.45\textwidth]{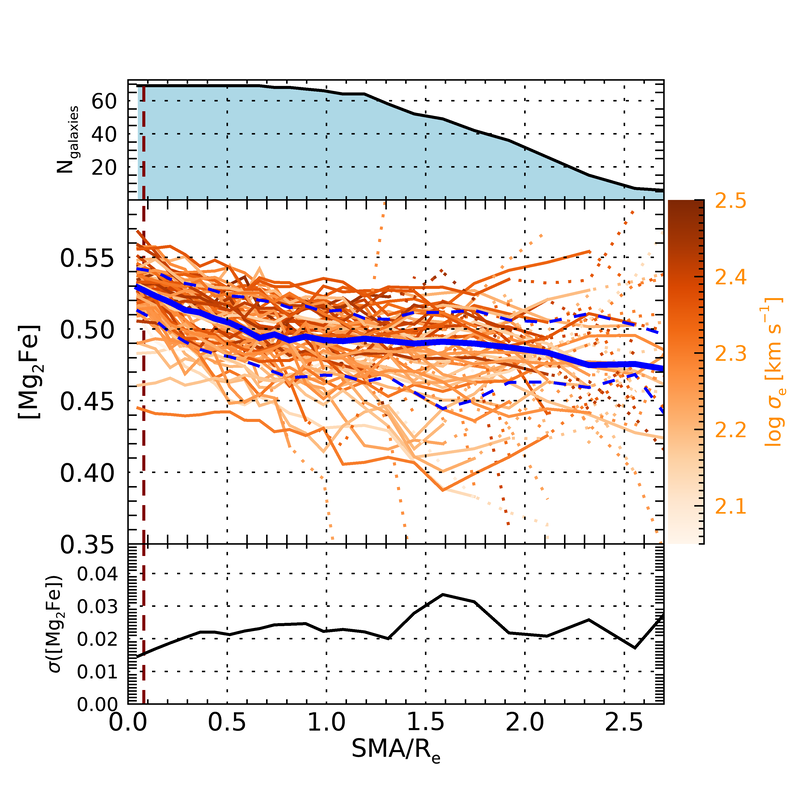}
\includegraphics[width=0.45\textwidth]{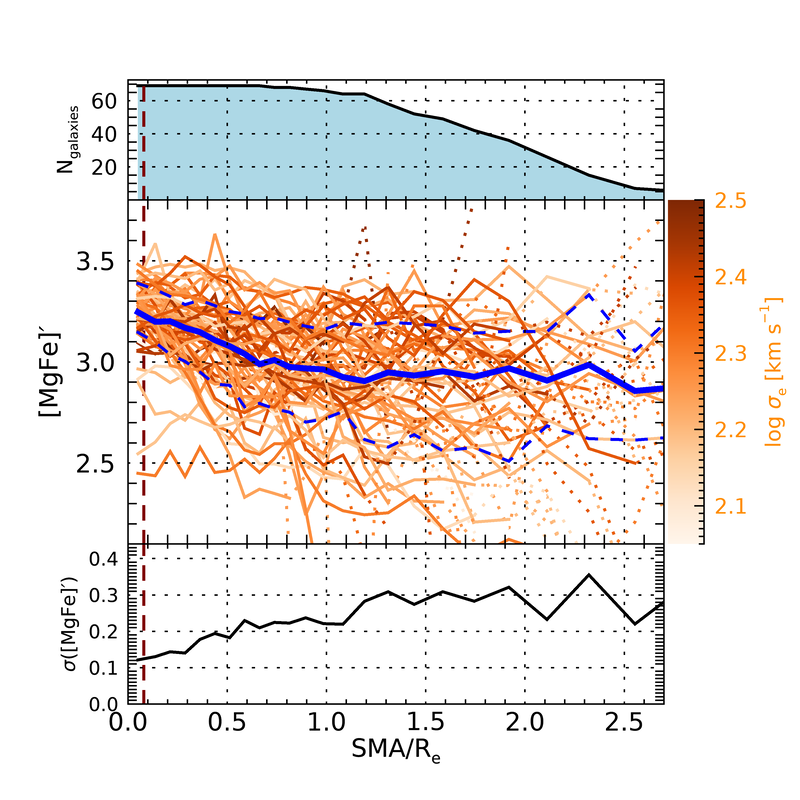}
\caption{Radial profiles of the four stellar absorption indices used to derive the stellar population parameters.
Each line is an individual galaxy, color coded by velocity dispersion. The blue line is the median, with the blue
dashed lines marking the $16^\mathrm{th}$ and $84^\mathrm{th}$ percentiles. 
The vertical dashed lines mark the median PSF radius (HWHM). See caption to Fig. \ref{fig:stacked_profs}
for the full description of lines and panels.}
    \label{fig:index_profiles}
\end{figure*}

\begin{figure*}
\centerline{\large \textsf{All ETGs}}
\includegraphics[width=0.45\textwidth]{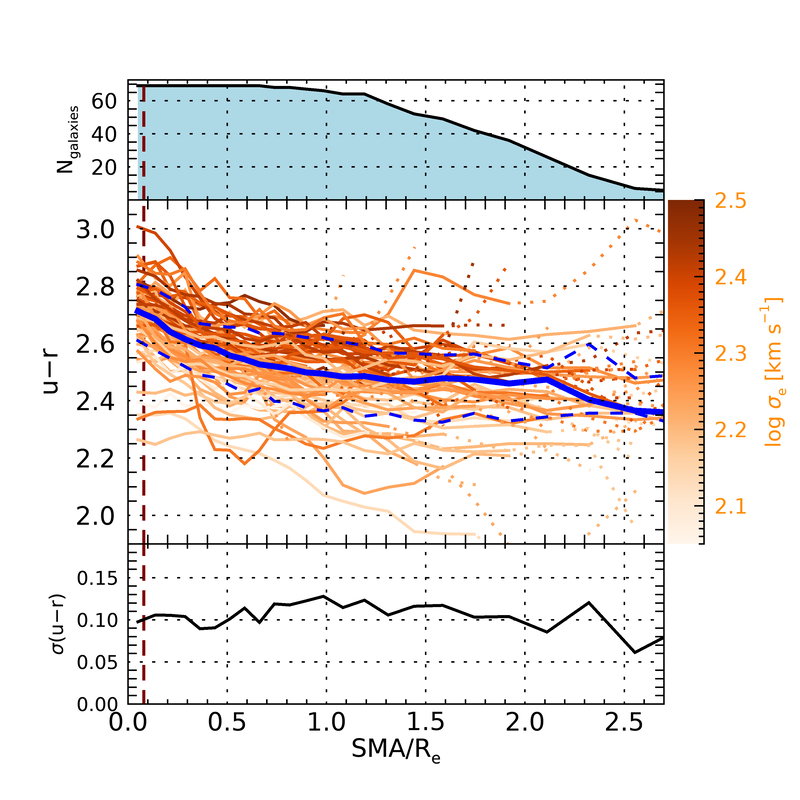}
\includegraphics[width=0.45\textwidth]{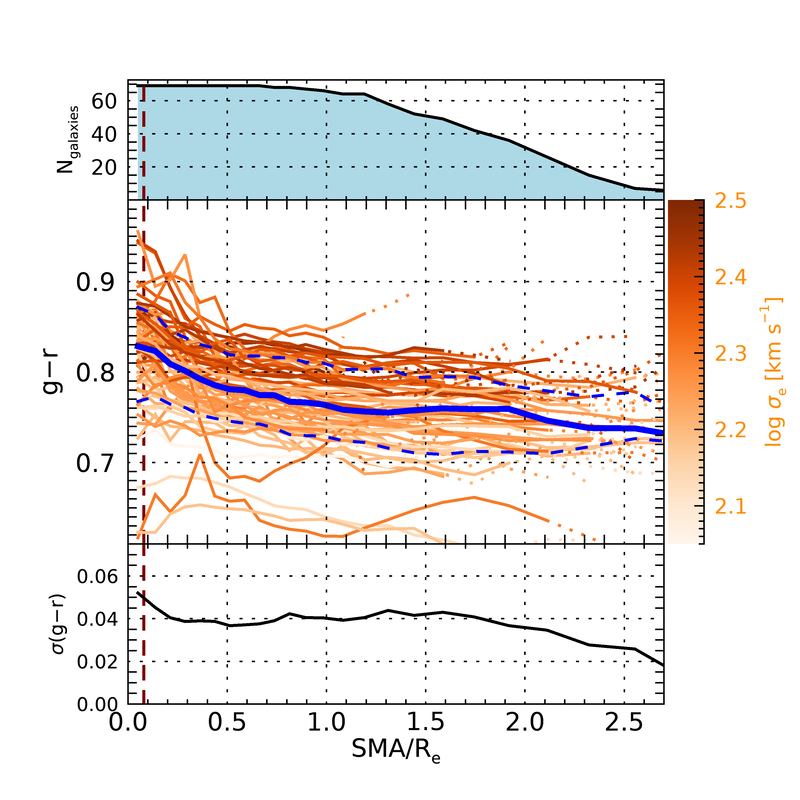}\\
\includegraphics[width=0.45\textwidth]{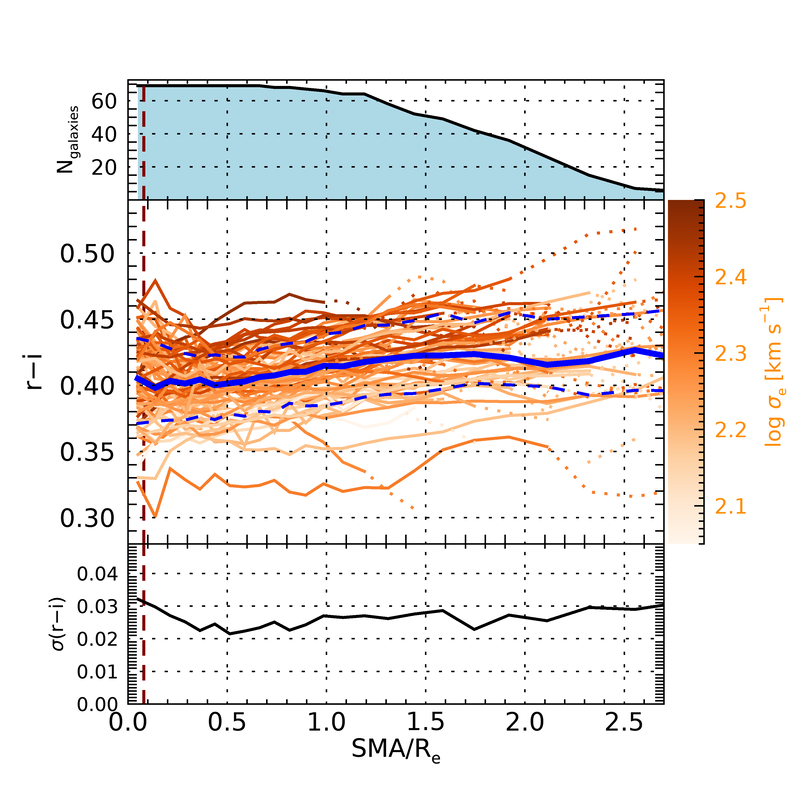}
\includegraphics[width=0.45\textwidth]{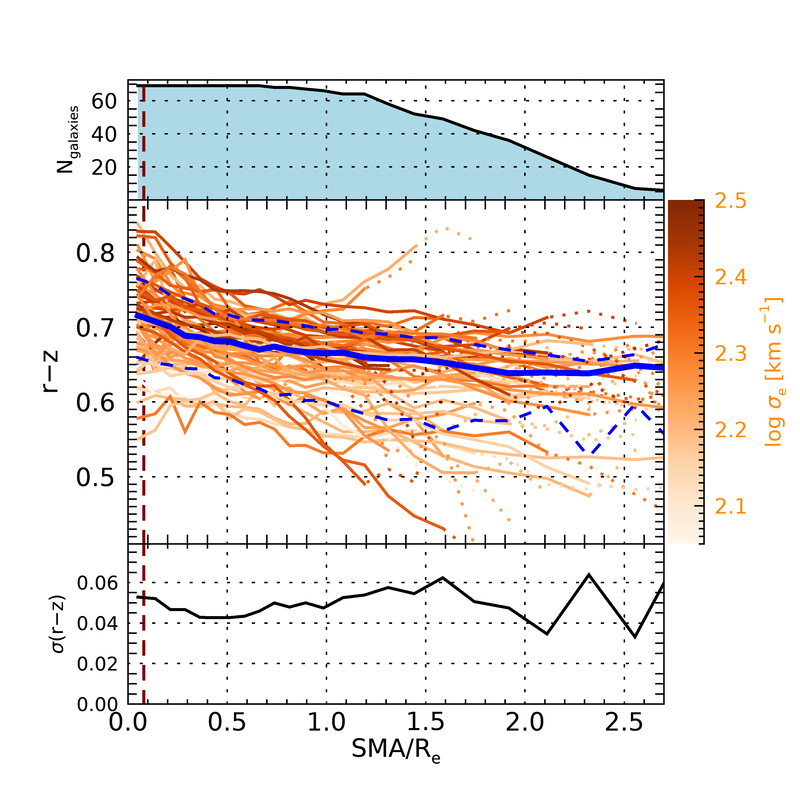}
\caption{Radial profiles of the four broad-band colours resulting from the five SDSS bands used to derive the 
stellar population parameters.
Each line is an individual galaxy, color coded by velocity dispersion. The blue line is the median, with the blue
dashed lines marking the $16^\mathrm{th}$ and $84^\mathrm{th}$ percentiles. 
The vertical dashed lines mark the median PSF radius (HWHM). See caption to Fig. \ref{fig:stacked_profs}
for the full description of lines and panels.}
    \label{fig:color_profiles}
\end{figure*}


\bsp	
\label{lastpage}
\end{document}